\documentclass[aps,preprint,superscriptaddress,nofootinbib,preprintnumbers]{revtex4-1}
 \pdfoutput=1

\usepackage{amsmath,amssymb,amsfonts}
\usepackage{graphicx}
\usepackage{xcolor}
\usepackage{hyperref}

\graphicspath{{./figs/}}
\newcommand{\mET}{{E\!\!\!\!/}_T}

\newcommand{\mpl}{m_\text{pl}}

\newcommand{\Tf}{T_f}

\newcommand{\beq}{\begin{eqnarray}}
\newcommand{\eeq}{\end{eqnarray}}

\newcommand{\centeron}[2]{{\setbox0=\hbox{#1}\setbox1=\hbox{#2}\ifdim
\wd1>\wd0\kern.5\wd1\kern-.5\wd0\fi \copy0
\kern-.5\wd0\kern-.5\wd1\copy1\ifdim\wd0>\wd1
                                  \kern.5\wd0\kern-.5\wd1\fi}}
\newcommand{\ltap}{\>\centeron{\raise.35ex\hbox{$<$}}
                          {\lower.65ex\hbox{$\sim$}}\>}
\newcommand{\gtap}{\>\centeron{\raise.35ex\hbox{$>$}}
                          {\lower.65ex\hbox{$\sim$}}\>}
\newcommand{\gsim}{\mathrel{\gtap}}

\newcommand\ZZ{\hbox{\zfont Z\kern-.4emZ}}
\font\zfont = cmss10 

\newcommand{\cref}[1]{Chapter \ref{c.#1}}

\def\beq{\begin{equation}}
\def\eeq{\end{equation}}
\newcommand{\ba}{\begin{array}}
\newcommand{\ea}{\end{array}}
\newcommand{\bea}{\begin{eqnarray}}
\newcommand{\eea}{\end{eqnarray} }
\newcommand{\bal}{\begin{align}}
\newcommand{\eal}{\end{align}}
\def\bi{\begin{itemize}}
\def\ei{\end{itemize}}
\def\ben{\begin{enumerate}}
\def\een{\end{enumerate}}
\def\beq{\begin{equation}}
\def\eeq{\end{equation}}
\def\bc{\begin{center}}
\def\ec{\end{center}}
\def\bt{\begin{table}}
\def\et{\end{table}}
\def\btb{\begin{tabular}}
\def\etb{\end{tabular}}

\newcommand{\XXb}[1]{#1^0-\bar{#1}^0}

\def\cC{{\mathcal C}}

\def\cL{{\mathcal L}}

\def\cO{{\mathcal O}}

\def\cH{{\mathcal H}}

\def\mpl{\, M_{\rm Pl}}

\def\mass2{mass${}^2$}

\def\mass{_{\text{\sc mass}}}
\def\Vckm{V_{\text{\sc ckm}}}
\newcommand{\ssc}[1]{\text{\scriptsize\sc #1}}
\newcommand{\fl}[1]{_{\text{\scriptsize\sc #1}}}
\newcommand{\ckm}{\left[\Vckm\right]}

\begin{document}
\pagestyle{plain}

\title{\boldmath Continuous Flavor Symmetries and the Stability of Asymmetric Dark Matter}

\def\Cincy{Department of Physics, University of Cincinnati, Cincinnati, Ohio 45221,USA}
\def\Fermilab{Theoretical Physics Department, Fermilab, P.O. Box 500, Batavia, IL 60510}

\author{Fady Bishara}
\email{bisharfy AT ucmail.uc.edu}
\affiliation{\Cincy}
\affiliation{\Fermilab}

\author{Jure Zupan}
 \email{zupanje AT ucmail.uc.edu}
\affiliation{\Cincy}

\begin{abstract}
  \vskip 3pt \noindent

Generically, the asymmetric interactions in asymmetric dark matter (ADM) models lead to decaying DM. We show that, for ADM that carries nonzero baryon number, the continuous flavor symmetries that generate the flavor structure in the quark sector also imply a looser lower bound on the mass scale of the asymmetric mediators  between the  dark and visible sectors. The mediators for $B=2$ ADM that can produce a signal in the future indirect dark matter searches can thus also be searched for at the LHC. For two examples of the mediator models, with either the MFV or Froggatt-Nielsen flavor breaking pattern, we derive the FCNC constraints and discuss the search strategies at the LHC.  
\end{abstract}

\preprint{FERMILAB-PUB-14-284-T}
\maketitle

\section{Introduction}

Dark matter (DM) is stable on cosmological time-scales.  A principal question about the nature of DM is what mechanism ensures its stability? Commonly this is assumed to be a result of an exact symmetry (for a concise review of proposed stabilization mechanisms see, e.g.,~\cite{Hambye2011}). One possibility is that the stability of DM is ensured by a gauge symmetry, mimicking the way QED gauge invariance ensures the stability of the electron in the standard model \cite{Ackerman:mha,Feng:2008mu,Feng:2009mn}.  A more frequent choice is to introduce  a $Z_2$ symmetry by hand. A prominent example is $R$-parity in the MSSM, which both stabilizes DM and ensures the stability of the proton \cite{Farrar:1978xj,Dimopoulos:1981dw,Farrar:1982te}. An exact $Z_2$ symmetry can be generated dynamically, e.g., as a remnant of a spontaneously broken U(1) gauge symmetry, such as $U(1)_{B-L}$ \cite{Kadastik:2009dj,Kadastik:2009cu,Frigerio:2009wf}.  An attractive possibility is that $Z_2$, and consequently the DM stability, is an accidental symmetry. Examples include minimal DM \cite{Cirelli:2009uv,Cirelli:2005uq}, hidden vector DM \cite{Hambye:2008bq}, and weakly interacting stable pions \cite{Bai:2010qg}.

In this paper we explore a possibility that the discrete $Z_2$ that ensures the stability of DM is {\it both accidental and approximate}. As a result DM is metastable with decay times potentially close to its present observational bound of $\tau \gtrsim 10^{26}$s. We focus on a particular subset of asymmetric DM models~\cite{Kaplan:2009ag} (for a recent review see \cite{Zurek:2013wia}), where DM carries baryon number. 
Our working assumptions are
\begin{itemize}
\item
Baryon number is a conserved quantum number (it could, for instance, be gauged at high scales).
\item
There is a sector that efficiently annihilates away the symmetric component. The exact form is not directly relevant for our discussion.
\item
The observed flavor structure in the quark sector is explained by flavor dynamics in the UV while DM is not charged under flavor. 
\end{itemize}

The flavor dynamics fixes the flavor structure of dark sector couplings to the visible sector in the same way that it fixes the structure of the SM Yukawa interactions. This has two important consequences. First, the exchange of DM in the loops does not generate dangerously large Flavor Changing Neutral Currents (FCNCs). Secondly, and most importantly, a flavor singlet DM is stable on cosmological timescales  even for TeV scale mediators between the dark and visible sectors. In this case, the nature of DM stability can even be probed directly at the LHC.

The underlying flavor symmetry is crucial for the stability of DM. We will demonstrate this for two realizations of flavor physics: the Minimal Flavor Violation (MFV) hypothesis and for abelian horizontal symmetries in the case where DM carries baryon number 2. In this case the mediators leading to the decay of DM can be at ${\mathcal O}(100{\rm GeV}$. In contrast, for completely anarchic flavor couplings where DM couples to all quark flavors with ${\mathcal O}(1)$ couplings, the indirect DM bounds would require the mediators to have masses in the ${\mathcal O}(10{\rm TeV})$ range. 
The implications of continuous flavor symmetries for DM interactions have also been explored in~\cite{Batell2011,Kile2011,Agrawal2012,Davoudiasl2011,Lopez-Honorez:2013wla,Agrawal:2014aoa,Batell:2013zwa,Agrawal:2014una,Haisch:2013fla,Kamenik:2011nb,Agrawal:2011ze,Merle:2011yv}. Our analysis differs from these studies in that we are assuming that DM is a flavor singlet (as is the case in most models of DM). This, along with its small mass and conserved baryon number, also ensures that DM is metastable in our setup. The stability of symmetry-less DM in the context of discrete flavor groups has been discussed in \cite{Kajiyama:2011gu} (for the potential relation of discrete flavor groups in the leptonic sector and the stability of DM, see also \cite{Lattanzi:2014mia,Hirsch:2010ru,Boucenna:2012qb}). The decaying DM in the context of ADM models was explored in \cite{Zhao:2014nsa,Feng:2013vva,Masina:2012hg,Masina:2011hu}.

The paper is structured as follows. In Sec. \ref{sec:DMmassADM} we review the relation between DM mass and relic abundance in asymmetric DM models. In Sec. \ref{metastability} we give two examples of flavor breaking models at the level of Effective Field Theory (EFT) analysis that can lead to metastable asymmetric DM. In Sec. \ref{sec:indir-detect} we derive the indirect detection bounds on the two EFT set-ups. In Sec. \ref{sec:uv-models} we give two examples of mediators that would lead to the EFT set-ups discussed in Sec.  \ref{sec:indir-detect}. The relevant bounds on the mediator masses and couplings, including collider signatures, are derived in Sec. \ref{sec:exp:signatures}. Conclusions are given in Sec. \ref{sec:conclusions}, while appendices contain technical details.

\section{Dark Matter Mass in Asymmetric Dark Matter models}
\label{sec:DMmassADM}

Asymmetric Dark Matter (ADM) models \cite{Kaplan:2009ag,Farrar:2005zd,Kitano:2004sv,Kitano:2008tk,Fujii:2002aj,Kaplan:1991ah,Nussinov:1985xr,Barr:1991qn,Barr:1990ca,Gudnason:2006ug} address the question of why the DM density, $\Omega_\chi$, and the baryon density  in the universe, $\Omega_B$, are so close to each other, $\Omega_\chi\simeq 5.3 \,\Omega_{\rm B}$~\cite{Beringer:1900zz}. In the standard weakly interacting massive particle (WIMP) models of DM this is to some extent pure coincidence. In this case DM is a thermal relic and 

\begin{equation}
\left(\frac{\Omega_\chi}{0.265} \right)\left(\frac{h}{0.673}\right)^2\sim  \frac{3\times 10^{-27}\,\text{cm}^3\text{s}^{-1}}{\langle{\sigma\,v}\rangle},
\end{equation}

with $\langle{\sigma\,v}\rangle$ the thermally averaged DM annihilation cross section. The coincidence $\Omega_\chi\sim \Omega_B$  then arises due to a fortuitous size of the annihilation cross section for a weakly coupled weak scale DM -- the WIMP miracle. 

\begin{figure}
	\includegraphics[scale=1]{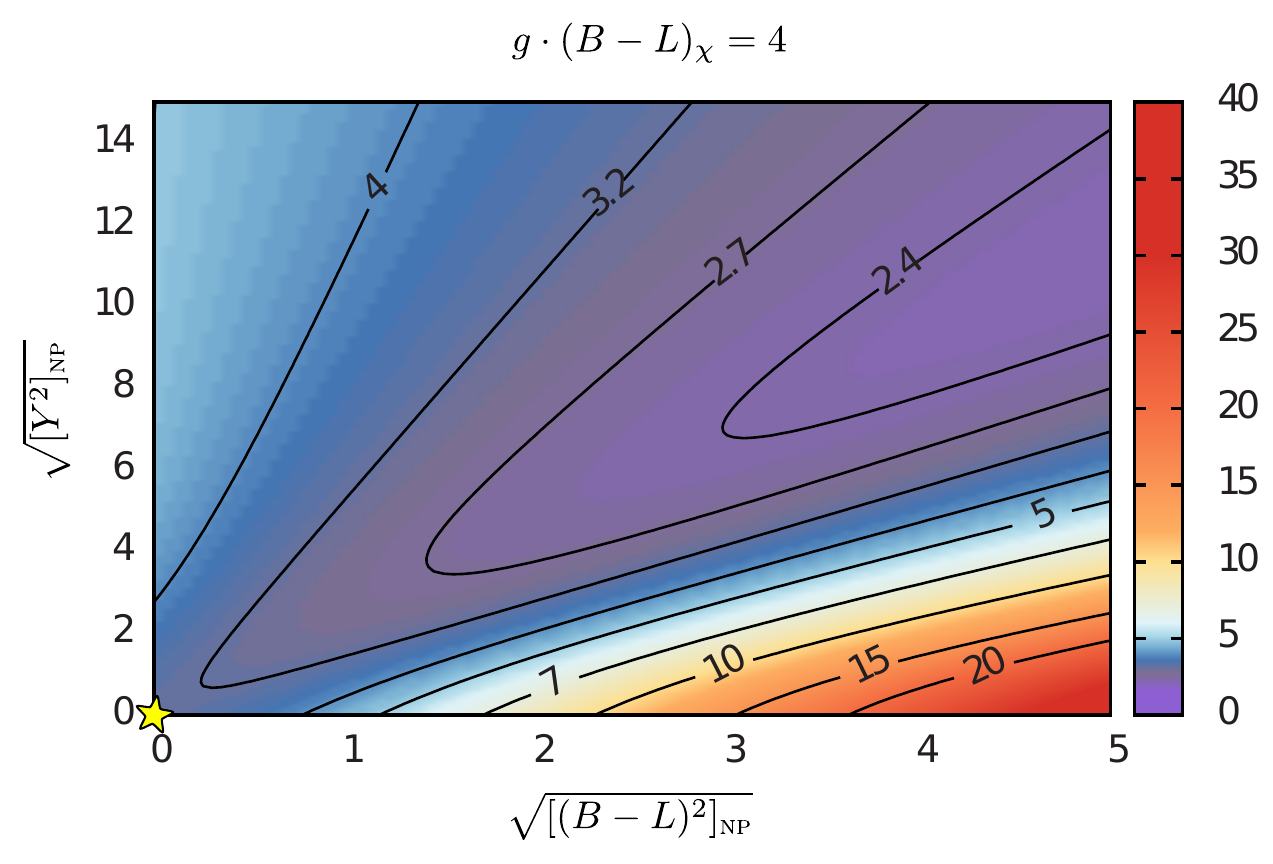}
	\caption{Contours of the ADM mass $m_\chi$ (in GeV) required to obtain the correct DM relic density as a function of  $[(B-L)^2]_{\rm NP}$, $[Y^2]_{NP}$, i.e. the $(B-L)^2$, and $Y^2$  summed over effective d.o.f. in visible NP sector, while keeping $[Y(B-L)]_{NP}=\sqrt{[(B-L)^2]_{\rm NP}[Y^2]_{NP}}$ and $g(B-L)_\chi=4$. The visible sector with only the SM, Eq.  \eqref{mchi:eq:init},  is denoted by a star. }
	\label{fig:adm-mass}
\end{figure}

In contrast, in ADM models the observed DM is not a thermal relic. Its relic abundance reflects the asymmetry in DM, $\chi$, and anti-DM, $\chi^\dagger$, densities in the early universe. The $\chi$ and $\chi^\dagger$ annihilate away, and only the asymmetric component remains. The coincidence of $\Omega_{\chi}$ and $\Omega_B$ is then due to the fact that the DM relic abundance has the same origin as the baryon asymmetry. 
The difference between $\Omega_{\rm \chi}$ and  $\Omega_{\rm B}$ is simply due to the fact that the DM particle is more massive than a proton by a factor of a few. More precisely, to explain the observed $\Omega_{\chi}$ the DM's mass needs to be (see Appendix~\ref{app:mass})
\beq
m_\chi= N_0 m_p \frac{\Omega_{\chi}}{\Omega_{B}} \frac{1}{ (B-L)_\chi},
\label{mchi:eq:init}
\eeq
where  $m_p$ is the proton mass. Here $(B-L)_\chi$ is the $B-L$ charge of the $\chi$ field. The exact value of numerical prefactor $N_0\simeq {\mathcal O}(1)$ depends on when the operators transferring the baryon asymmetry between the  visible and the dark sector decouple. For decoupling temperature above electroweak phase transition, and assuming that there are only the SM fields in the visible sector, gives $N_0=1.255$ for DM that is a complex scalar or a Dirac fermion. In this case the required DM mass is
\beq
m_\chi=(6.2\pm0.4){\rm GeV}\frac{1}{(B-L)_\chi},
\eeq
where the error reflects the errors on $\Omega_{\chi}=0.265\pm0.011$ and $\Omega_{B}=0.0499\pm0.0022$~\cite{Beringer:1900zz,Ade2013a}. We thus have
\beq\label{mchiSM}
m_\chi=\{6.2,3.1,2.1\}{\rm GeV},\qquad{\rm for}\quad(B-L)_\chi=\{1,2,3\},
\eeq
where we only quote the central values. Deviations from the above relations are possible if for instance the visible sector contains additional degrees of freedom beyond the SM. In that case, $m_\chi$ in \eqref{mchi:eq:init} is a functions of $[(B-L)^2]_{\rm NP}$, $[Y^2]_{NP}$, and $[Y(B-L)]_{NP}$, i.e., the $(B-L)^2$, $Y^2$ and $Y(B-L)$ summed over effective degrees of freedom in the visible NP sector, cf. Eq. \eqref{omegaBmoegachiNP}. The $m_\chi$ required to obtain the correct relic abundance is shown in Fig. \ref{fig:adm-mass}. For illustration we set $[Y(B-L)]_{NP}=\sqrt{[(B-L)^2]_{\rm NP}[Y^2]_{NP}}$ in the plot and assume that DM is a complex DM scalar with $(B-L)_\chi=2$. We see that for $[(B-L)^2]_{\rm NP} \sim [Y^2]_{NP}\sim [Y(B-L)]_{NP}$ the deviations from  \eqref{mchiSM} are modest, of ${\mathcal O}(1)$. Further deviations from Eqs.~\eqref{mchi:eq:init},  \eqref{mchiSM} are possible in more general frameworks such as ADM from leptogenesis \cite{Falkowski:2011xh} or dynamically induced mass mixing \cite{Cui:2011qe}. Henceforth, we will assume that $m_\chi$ is given by Eqs. \eqref{mchi:eq:init}, \eqref{mchiSM}. Our results can be trivially adjusted if this is not the case.

The thermal history of the universe in ADM models has several distinct epochs. At high temperatures a $B-L$ asymmetry is generated, e.g., via GUT-like baryogenesis \cite{Kaplan:2009ag} or via leptogenesis \cite{Falkowski:2011xh}. The $B-L$ asymmetry is efficiently transferred between the visible and the DM sectors through asymmetric
interactions. We do not require a discrete $Z_n$ symmetry in the dark sector so that unlike \cite{Kaplan:2009ag} the asymmetric interactions can involve just a single $\chi$ field. At low energies they  have a schematic form, 
\beq\label{Oschematic}
{\mathcal O}_{\rm asymm.}\sim \frac{C}{\Lambda^6} \chi (qq)^3,
\eeq
taking $(B-L)_\chi=2$ complex scalar DM as an example.  Here, $C$ is a flavor-dependent coefficient.
The asymmetric interactions freeze out at temperature $ T_f\sim \Lambda\gg m_\chi $, below which the $B-L$ asymmetries in the visible and dark sectors are separately conserved.
If the flavor breaking is due to a spontaneously broken horizontal symmetry (see Sec.~\ref{sec:sbhs},) the freeze out temperature for the above dimension 10 operator in Eq. \eqref{Oschematic} is, using Naive Dimensional Analysis (NDA), 

\begin{equation}
    \Tf\sim \left(1.66\times\sqrt{g_*}\, (16\pi^2)^{3} \frac{8\pi}{C^2}\frac{\Lambda^{12}}{\mpl\,}\right)^{1/11} \simeq  450~\text{GeV}.
\end{equation}

In the numerical evaluation, we used the lower bound $\Lambda=\Lambda_*=1.8$~TeV from indirect detection Eq. \eqref{LambdaFN}, taken the effective number of relativistic d.o.f. to be $g_*=108.75$, corresponding to the SM with a complex scalar DM,
and set $C=1$ which is appropriate for the $\chi b\to b sc tb$ transition dominance (with any permutation of the flavors). Note that $T_f$ is above the electroweak phase transition temperature $T_{\rm ew}\sim 170$ GeV. It is also well below $\Lambda$ so that the use of EFT is justified. If the mediator scale were too low, $\Lambda \lesssim 730$ GeV (or $\Lambda \lesssim 400$ GeV for MFV breaking), the asymmetric operator would not freeze out before electroweak phase transition started. Consequently, the DM quantum number would not be conserved and the DM density would be washed out. This places a lower bound on the asymmetric mediator masses to be above a few hundred GeV.

 Finally, at temperatures below DM mass the bulk of the DM efficiently annihilates back to the visible sector through symmetric interactions leaving only the small asymmetric component. We have nothing new to say about this mechanism and refer the reader to a set of model building ideas already present in the literature \cite{Zurek:2013wia,Lin:2011gj,Cohen:2010kn,Blennow:2012de}.

\section{Metastability and flavor breaking}
 \label{metastability}
We show next that the DM in ADM models can be stable on cosmological time-scales without invoking discrete $Z_n$ symmetries. We assume that the SM quark flavor structure is explained by a continuous flavor group and that the DM carries nonzero baryon number. This is a crucial ingredient in the argument. Since DM is not charged under the flavor group, while the SM fields are, there are no interactions between DM and the SM in the limit that the flavor group is unbroken (all flavor singlet interactions are forbidden by baryon number conservation). All the interactions between DM and the visible sector thus have to be flavor breaking and this leads to a significant suppression of the DM decay time.  

We show this explicitly for two examples of flavor breaking: i) the MFV ansatz, where all the flavor breaking is assumed to be due to the SM Yukawas, and ii) the spontaneously broken horizontal $U(1)$ symmetries. Integrating out the NP fields gives the effective DM decay Lagrangian
\beq\label{eqLeft}
	{\cal L}=\sum_i \frac{C_i}{\Lambda^{(D_i-4)}} {\cal O}_i.
\eeq
The size of $\mathcal{C}_i$ is fixed by the assumed flavor generating mechanism. The sum runs over the different forms of the local operators 
\begin{equation}
	\cO_i=\chi\,\left[u^c\right]^{n_u}\,\left[d^c\right]^{n_d}\,\left[q^*\right]^{n_q},
	\label{eq:OB}
\end{equation}
where $(n_u+n_d+n_q)\mod3=0$ in order for DM to be a color singlet. Note that the DM needs to carry integer baryon number in order not to forbid all interactions with the visible sector.  Here $u^c$, $d^c$ are the electroweak singlets and $q$ is the electroweak doublet left-handed quark fields in two component notation, see App.~\ref{app:spinors}. In the down-quark mass basis they are

\begin{equation}
	u^c\rightarrow u^c_{\text{\sc mass}}, \qquad d^c\rightarrow d^c_{\text{\sc mass}}, \qquad q=\binom{u}{d}\rightarrow\binom{V_{\text{\sc ckm}}\,u_{\text{\sc mass}}}{d_{\text{\sc mass}}}.
\label{eq:mass-basis-fields}
\end{equation}
The SM Yukawa matrices are then
\begin{equation}
Y_D\rightarrow Y_D^{\text{diag}},\quad Y_U\rightarrow \Vckm Y_U^{\text{diag}},
\label{eq:mass-basis-yukawa}
\end{equation}
with $Y_{D,U}^{\text{diag}}$ the diagonal Yukawa matrices. 

As an example, let us consider fermionic $B=1$ DM. Two distinct types of operators are allowed
\begin{equation}
\begin{split}
	\cO^{(B=1)}_{1}&=(\chi\,u^c)(d^c d^c)\rightarrow (\chi\,u^c\mass)(d^c\mass\,d^c\mass),\\
	\cO^{(B=1)}_{2}&=(\chi\,q^*_\rho)(d^c\, q^*_\sigma)\epsilon^{\rho\sigma}\rightarrow(\chi\,{u^*\mass} \Vckm)(d^c\mass d^*\mass),
\end{split}
\label{eq:B1}
\end{equation}
where $\rho,\sigma$ are $SU(2)_L$ indices while the $SU(3)_C$ and flavor indices are implicit and we have chosen one possible Lorentz contraction denoted by the parentheses.

\subsection{Minimal Flavor Violation}
The MFV assumption is that also in the NP sector the flavor is broken only by the SM Yukawas $Y_{U,D}$ \cite{Chivukula:1987py,D'Ambrosio:2002ex,Hall:1990ac,Buras:2003jf,Buras:2000dm}. The MFV assumption can be most succinctly cast in the spurion language \cite{D'Ambrosio:2002ex}. In 
the limit of vanishing quark masses the SM quark sector enjoys an enhanced flavor symmetry $G_F=SU(3)_Q\times SU(3)_U\times SU(3)_D$.
The Yukawa interactions $u^c Y_U^\dagger q H$, $d^c Y_D^\dagger q H^c$ are formally invariant under $G_F$, if $Y_{U,D}$ are promoted to spurions, i.e. if they are assumed to transform  under $G_F$ as 
$Y_U\rightarrow Y_U'=U_QY_UU_U^\dagger$, $Y_D\rightarrow Y_D'=U_QY_D U_D^\dagger$.
Here $U_{Q,U,D}$ are transformations from $SU(3)_{Q,U,D}$, respectively. 

 This means that also  low energy operators \eqref{eqLeft} need to be formally $G_F$ invariant. Keeping the minimal insertion of Yukawas the operators $\cO_{1,2}$ in Eq.~\eqref{eq:B1} for $B=1$ DM are
\begin{equation}
\begin{split}
	\cO^{(B=1)}_{1}=&\big(\chi\,u^c_{\alpha}Y^\dagger_UY_D\big)_K \big(d^c_{N\beta} d^c_{M\gamma}\big)\epsilon^{KNM}\epsilon^{\alpha\beta\gamma}\\
		&\quad\rightarrow \big(\chi\,u^c\mass Y_U^{\text{diag}\dagger}\Vckm^\dagger Y_D^\text{diag}\big)_{K\alpha}\big([d^c\mass]_{N\beta}\,[d^c\mass]_{M\gamma}\big)\epsilon^{KNM}\epsilon^{\alpha\beta\gamma},\\
	\cO^{(B=1)}_{2}=&(\chi\,q^*_{K\alpha i})([d^c_{\beta}Y^\dagger_D]_N q^*_{M\gamma j})\epsilon^{ij}\epsilon^{KNM}\epsilon^{\alpha\beta\gamma}\\
		&\quad\rightarrow
			\big(\chi\,u^*\mass \Vckm^\dagger \big)_{K\alpha}\big([d^c\mass Y^{\text{diag}\dagger}_D]_{N\beta} [d^*\mass]_{M\gamma}\big)\epsilon^{KNM}\epsilon^{\alpha\beta\gamma},
\end{split}
\end{equation}
where $\alpha, \beta, \gamma$ are the color indices, and $K,N,M$ run over the quark generations.

\begin{figure}[]
\centering
\includegraphics[scale=0.8]{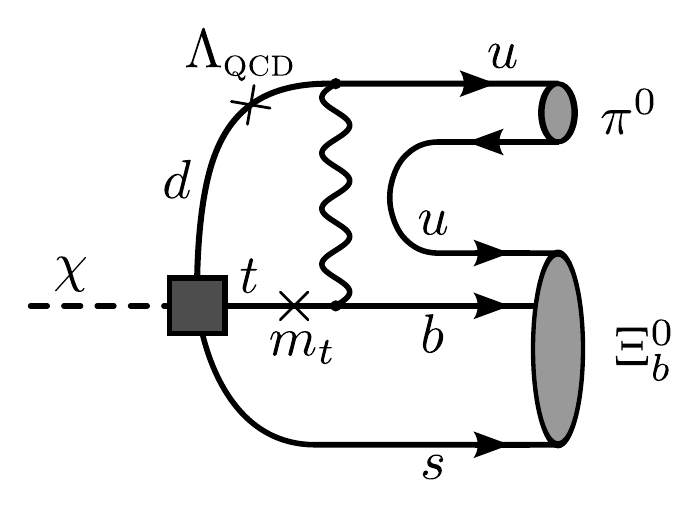}
\caption{Feynman diagram for the decay of DM with $B=1$ assuming MFV. This amplitude leads to the partial decay width $\Gamma_\chi^{(1)}$ in Eq.~\eqref{eq:b1-mfv-pdw}.}
\label{fig:b1-decay}
\end{figure}

The two operators lead to the $\chi\to bu s$ decay at the partonic level which is the least suppressed kinematically allowed transition. For the operator $\cO_1$, this transition arises at 1-loop and requires two chirality flips, see Fig.~\ref{fig:b1-decay}. The decay amplitude scales as $\sim y_t y_b$ with an extra loop factor and a chirality flip suppression $\sim m_t \Lambda_{\rm QCD}/m_W^2$. To be conservative, we count the chirality flip suppression due to the light $u,d,s$ quarks as proportional to $ \Lambda_{\rm QCD}$ and not to the much smaller quark masses. The operator $\cO_2$ leads to the decay $\chi\to bu s$ at tree level with the decay amplitude suppressed by $\sim y_b V_{ub}$.
  Once the quarks hadronize, the decays appear as $\chi\to \Xi_b\pi$, or $\chi\to \Lambda_bK$, with any number of pions. Using NDA to estimate the decay width gives (setting $V_{tb}\simeq V_{ud}\simeq 1$)
\beq
\begin{split}
\Gamma_\chi^{(1)}&\sim \frac{(y_ty_b)^2}{8\pi}\left(\frac{m_\chi}{\Lambda}\right)^4 \left(\frac{1}{16\pi^2}\frac{m_t \Lambda_{\rm QCD}}{m_W^2}\right)^2 \frac{m_\chi}{16\pi^2}=6.6 \cdot10^{-51} \text{GeV}\left(\frac{y_b}{0.024}\right)^2\left(\frac{4.0 \cdot 10^6\text{TeV}}{\Lambda}\right)^4,\\
\Gamma_\chi^{(2)}&\sim \frac{|y_b V_{ub}|^2}{8\pi}\left(\frac{m_\chi}{\Lambda}\right)^4 \frac{m_\chi}{16\pi^2}=6.6 \cdot10^{-51} \text{GeV}\left(\frac{y_b}{0.024}\right)^2\left(\frac{4.3 \cdot 10^7\text{TeV}}{\Lambda}\right)^4,
\end{split}
\label{eq:b1-mfv-pdw}
\eeq
for the case where $\cO_1$ and $\cO_2$ dominate the decay, respectively. The last $1/16\pi^2$ factor is due to three body final state and is required to obtain the correct estimate for the inclusive decay width as can be seen from the optical theorem and the use of OPE. In the numerics we use $m_t=173$ GeV, $m_\chi=6.2$ GeV, $|V_{ub}|=0.00415$. The numerical prefactor $6.6 \cdot 10^{-51}$ GeV $=1/(10^{26} s)$ is chosen to make contact with the bounds on the DM lifetime from indirect DM searches.

Note that MFV leads to two sources of suppression. First, there is the suppression of the Wilson coefficients due to Yukawa insertions, $y_b\sim 0.024$ for $\cO_1$ and $y_bV_{ub}\sim 10^{-4}$ for $\cO_2$. In addition, there is a loop suppression for $\cO_1$ where the decay has to proceed through an off-shell top quark. Without these additional suppressions the bounds from indirect DM detection would require about two orders larger NP scale, $\Lambda\gtrsim 4.3 \cdot 10^{9}$~TeV.

The suppression factors are much larger for $B=2$ DM, in which case DM is a scalar, and the asymmetric operators start at dimension 10. We investigate in detail the  operator 

\beq
\begin{split}\label{MFVop}
{\mathcal O}^{(B=2)}_1=\chi (d^c_{K\alpha} &d^c_{N\beta})([q^*Y_D]_{M\alpha'} q^*_{K'\beta'})(q^*_{N'\gamma'}q^*_{M'\gamma})\epsilon^{KNM}\epsilon^{K'N'M'}\epsilon^{\alpha\beta\gamma}\epsilon^{\alpha'\beta'\gamma'}\\
&\to \chi ([d^c\mass]_{K\alpha} [d^c\mass]_{N\beta})([u^*\mass {\Vckm^\dagger} Y_D^{\text{diag}}]_{M\alpha'} [d^*\mass]_{K'\beta'})\times\\
&\qquad \times([u^*\mass{\Vckm^\dagger}]_{N'\gamma'}[d^*\mass]_{M'\gamma})\epsilon^{KNM}\epsilon^{K'N'M'}\epsilon^{\alpha\beta\gamma}\epsilon^{\alpha'\beta'\gamma'},
\end{split}
\eeq
that gives the least suppressed decay amplitude.
Above, we chose one of the possible color contractions, implicitly assumed contractions of weak indices within brackets, and only kept the weak contraction leading to the largest decay rate in the second line.

The correct relic abundance requires DM mass of $m_\chi=3.1\pm0.2$ GeV, assuming the SM field content at the time of the decoupling of the asymmetric operators. We assume that $m_\chi<m_{\Lambda_c^+}+m_{\Sigma^-}=3.48$~GeV, and thus below the threshold  for the $\chi\to \Lambda_c^+\Sigma^-$ decay, kinematically forbidding the $\chi\to udc\, dds$ partonic transition. The least suppressed partonic level transition is therefore  $\chi\to uds\, uds$
resulting, after hadronization, in the decays $\chi\to \Lambda^0\Lambda^0,\Sigma^-\Sigma^+, \Xi^- p, \Xi^0 n, \dots$.
The NDA estimate of the $\chi$ decay width is then

\beq
\Gamma_\chi^{(1)}\sim \frac{|y_b V_{ub}^2|^2}{8\pi} \left(\frac{m_\chi}{\Lambda}\right)^{12}  \frac{m_\chi}{(16 \pi^2)^4}=6.6 \cdot10^{-51} \text{GeV}\left(\frac{y_b}{0.024}\right)^2\left(\frac{0.63~\text{TeV}}{\Lambda}\right)^{12}.
\eeq
The MFV assumption results in the $y_b V_{ub}^2$ suppression of the Wilson coefficient.
The $1/(16\pi^2)^4$ factor reflects the fact that, in the OPE, the leading contribution starts at 5 loops.  The use of the OPE may be suspect for such low $m_\chi$ masses and one could expect ${\mathcal O}(1)$ corrections to the above estimate from additional soft gluon loops. 

Indirect DM searches require the NP scale to be $\Lambda\gtrsim 0.49$~TeV.
This corresponds to the bounds on the masses of the mediators between the dark and the visible sectors, $m_{\rm mediator}\gtrsim 490$~GeV, $m_{\rm mediator}\gtrsim 210$~GeV, and  $m_{\rm mediator}\gtrsim 90$~GeV, if the  operator \eqref{MFVop} arises at tree level, 1-loop, or 2-loop, respectively.  The mediators can thus be searched for at the LHC as discussed in Sec. \ref{sec:collider}. Note that the flavor suppression was essential to have such a low bound on the NP
scale $\Lambda$. Without it, and taking the Wilson coefficient to be 1, the indirect bounds on the stability of DM would require $\Lambda\gtrsim 7.3$~TeV, implying that the mediators were most likely out of reach of the LHC.

\begin{table}[t!]\centering
\begin{tabular}{ccccccccc}
\hline\hline
\multicolumn{3}{c}{ADM model} & \multicolumn{3}{c}{MFV} & \multicolumn{3}{c}{FN}\\
~$\:B\:$~ & ~Dim.~ & ~~$m_\chi$ [GeV]~~ & ~~~~decay~~ & ~~$\tau$ [s]~~ & ~~$\Lambda$ [TeV]~~ & ~~~~decay~~ & ~~$\tau$ [s]~~ & ~~$\Lambda$ [TeV]~~ \\ 
\hline
1 & 6 & 6.2 & $\;\chi\to bus$ & $10^{26}$ & $4.0\times 10^{6}$ & $\chi\rightarrow bus$ & $10^{26}$ & $8.1\times10^{8}$ \\ 
2 & 10 & 3.1 & $\chi\to uds uds$ & $10^{26}$ & $0.63$ & $\chi\rightarrow uds uds$ & $10^{26}$ & $2.5$ \\ 
3 & 15 & 2.1 & forbidden & $\infty$ & -- & forbidden & $\infty$ & -- \\ 
\hline\hline
\end{tabular}
\caption{Leading decay modes for the $B=1,2,3$ ADM assuming MFV or FN flavor breaking. The dimensionality of the decaying operators are denoted in the 2nd column. With the suppression scales $\Lambda$ given in the 6th and 9th column the ADM decay time is $\tau\simeq 10^{26}$~s. 
The $B=3$ ADM decays to quarks are kinematically forbidden.}
\label{tab:O_i}
\end{table} 

The bound on the NP scale $\Lambda$ is quite sensitive to the actual value of $m_\chi$. For larger values of $m_\chi$, the $\chi$ can decay to top and bottom quarks 
reducing the loop and CKM suppression of the decay width. 
This is illustrated in Fig. \ref{fig:adm-life}, where the NP scale is fixed to $\Lambda_{\rm MFV}=1$ TeV and $m_\chi$ is varied. As the kinematic thresholds for the $\chi$ decays to $c$ or $b$ quarks are reached, this results in a change of several orders of magnitude in the predicted decay time.

\subsection{Spontaneously Broken Horizontal Symmetries}
\label{sec:sbhs}
The suppression we found above using the MFV ansatz is model dependent. To illustrate this point we turn to $U(1)$ Frogatt-Nielsen (FN) models of spontaneously broken horizontal symmetries \cite{Froggatt:1978nt}. The suppression of the Wilson coefficients in the effective Lagrangian \eqref{eqLeft} is then given by the horizontal charges of the quarks in the operators. For instance, for the two $B=1$ DM operators in \eqref{eqLeft} 
\newcommand{\Hchrg}[2]{|H(#1_{\text\sc #2})|}
\begin{equation}
\begin{split}
	\cO^{(B=1)}_{1}&=\left(\chi\,d^c_{K}\right)(u^c_{N} d^c_{M})\to \left(\chi\,[d^c\mass]_{K}\right)([u^c\mass]_{N} [d^c\mass]_{M}) , \\
	\cO^{(B=1)}_{2}&=(\chi\,q^*_{Ki})(d^c_{N} q^*_{Mj})\epsilon^{ij}\to
			\left(\chi\,[u^*\mass]_K\right)\left([d^c\mass]_{N}[d^*\mass]_{M}\right),
\end{split}\label{B1:FN}
\end{equation}
the Wilson coefficients are
\beq
C_1\sim \lambda^{|H(d^c_K)+H(u^c_N)+H(d^c_M)|}, \qquad C_2\sim \lambda^{|-H(q_K)+H(d^c_N)-H(q_M)|}.
\eeq
Here $H(u^c_K), \dots$, with $H(q^*_K)=-H(q_K)$, are the horizontal $U(1)$ charges of the quarks, and $\lambda\sim 0.2$ is the expansion parameter. The dependence of the operators and Wilson coefficients on the generational indices $KNM$ is implicit as are color, weak, and Lorentz contractions in \eqref{B1:FN}.

An example of a horizontal charge assignment that gives phenomenologically satisfactory quark masses and CKM matrix elements is~\cite{Leurer1994}, 
\begin{equation}
H(q, d^c, u^c)\Rightarrow \bordermatrix{
~   & 1 & 2 & 3\cr
q   & 3 & 2 & 0\cr
d^c & 3 & 2 & 2\cr
u^c & 3 & 1 & 0
},
\label{eq:Hi}
\end{equation}
\noindent where the column labels $\{1,2,3\}$ correspond to the first, second, and third generations of quarks.

 Since the heavier flavors carry smaller charges the DM preferentially decays into the heaviest accessible states. 
As in MFV the dominant decay is $\chi\to bus$, 
except that the $y_b V_{ub}\sim \lambda^5$ suppression gets replaced by a much more modest $\sim \lambda^{|-H(q_1)+H(s^c)-H(q_3)|}=\lambda$. This is the largest scaling allowed by FN charges. In concrete UV mediator models the suppression can, in fact, be much more severe as we will see explicitly in the next Section. 

For $B=2$  DM the least suppressed operator is
\beq\label{FNop}
\begin{split}
{\mathcal O}^{(B=2)}_1=\chi (d^c_{K} &d^c_{N})(q^*_Mq^*_{K'})(q^*_{N'}q^*_{M'})\\
&\to \chi ([d^c\mass]_{K} [d^c\mass]_{N})([u^*\mass]_{M} [d^*\mass ]_{K'})([u^*\mass]_{N'}[d^*\mass]_{M'}),
\end{split}
\eeq
suppressing again the color and weak contractions.
The corresponding Wilson coefficient is suppressed by
\beq\label{C_1:FN}
C_1\sim \lambda^{|H(d^c_K)+H(d^c_N)-H(q_M)-H(q_{K'})-H(q_{N'})-H(q_{M'})|}.
\eeq
At the partonic level, the dominant decay is $\chi\to u s s\, uds$ with a Wilson coefficient that is of parametric size $\sim \lambda^{|H(d^c)+H(s^c)-2 H(q_2)-2 H(q_1)|}=\lambda^5$. Note that in MFV this process proceeded through 2 loops so that the suppression was much more severe, $\sim V_{ts} V_{ub}/(16\pi^2)^2\sim \lambda^5/(16\pi^2)^2$ at the amplitude level. While the suppression in FN case is much less then in the MFV case, it is still nontrivial. It lowers the scale of NP allowed by indirect DM searches from $\Lambda\gtrsim 7.3$~TeV, in the case of no flavor structure, to $\Lambda\gtrsim2.5$~TeV in the FN case. Taking the bound from DM indirect detection searches gives $\Lambda\gtrsim 1.9$~TeV. If the operator arises at tree level, 1-loop or 2-loops, this corresponds to mediator masses, $m_{\rm mediator}\gtrsim1.9$~TeV, $m_{\rm mediator}\gtrsim 830$~GeV, and $m_{\rm mediator}\gtrsim 360$~GeV, respectively.

\section{Indirect detection}
\label{sec:indir-detect}
\begin{figure}[t!]
	\includegraphics[scale=1]{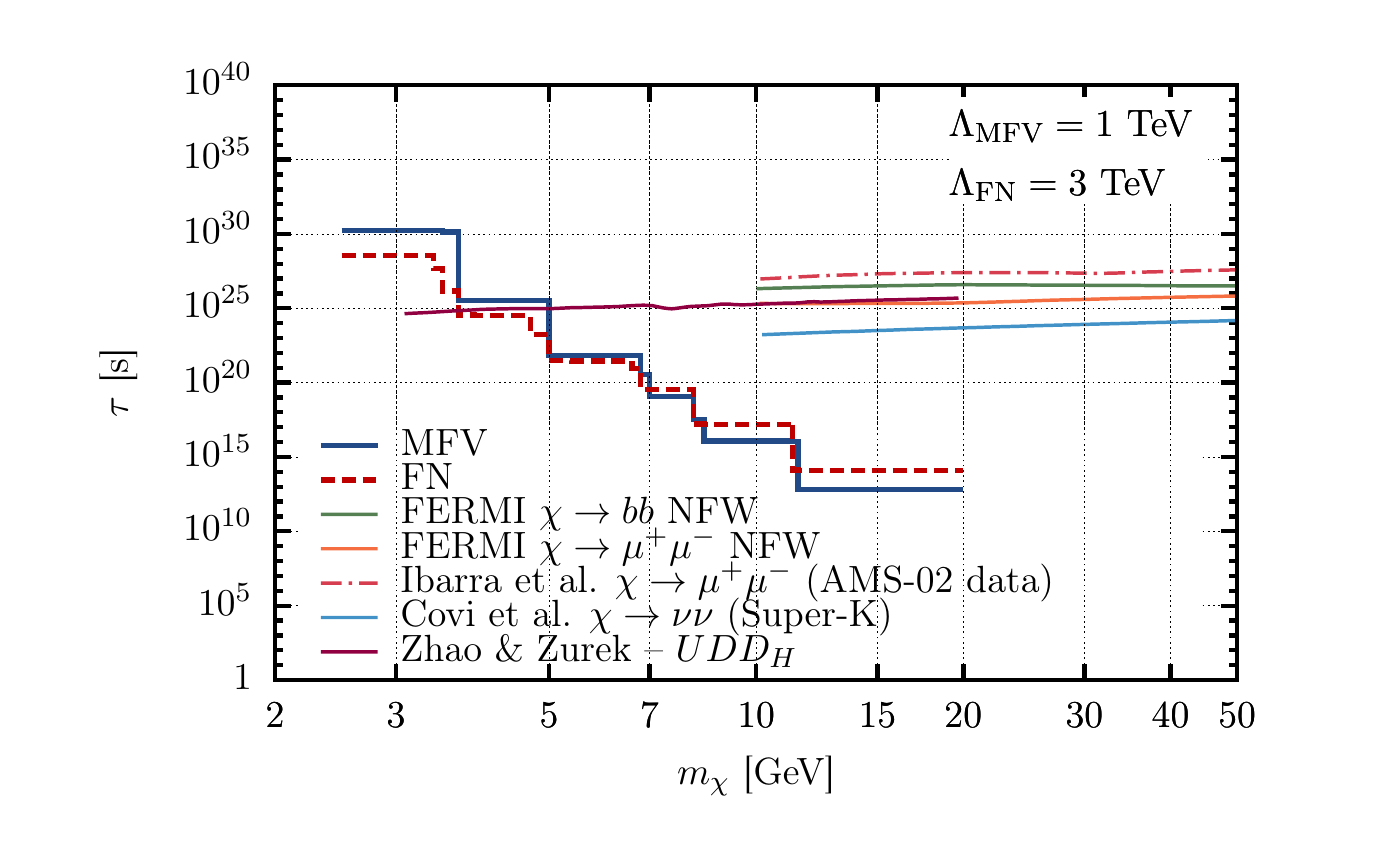}
	\caption{The solid blue (red dashed) line denotes the $B=2$ DM lifetime as a function of $m_\chi$ for the MFV (FN) case, fixing the NP scale to $\Lambda=1(3)$~TeV. Assuming the dominance of one decay mode, the green (orange) line shows the constraint on the decay time from FERMI-LAT~\cite{Ackermann2012} for $b\bar b$ ($\mu^+\mu^-$) final states using the NFW profile. The dash-dotted red line shows the AMS-02~\cite{Aguilar:2013qda} constraint on $\chi \to \mu^+\mu^-$ decay time derived in~\cite{Ibarra:2013zia}, while the light blue line shows the Super-Kamiokande~\cite{Desai2004} constraint on the $\chi \to \nu\bar \nu$ decay time obtained in~\cite{Covi2010}. The purple line shows the upper limit on $\chi\to uds$ and $\chi\to cbs$ decay times (indistinguishable at the scale of the figure) obtained in~\cite{Zhao:2014nsa}.
	}
	\label{fig:adm-life}
\end{figure}

\paragraph*{}
The asymmetric operators discussed in the previous section lead to a decaying DM which can be potentially seen in indirect DM searches.  In our models, the $\chi$ decays hadronicaly. The decay products thus contain a number of charged particles and photons. The flavor composition of the final state depends on the mass, $m_\chi$, and also on the assumed flavor breaking pattern. In Section \ref{metastability} we discussed in detail the case of 6.2 GeV $B=1$ DM, which decays through $\chi\to bus$ and a 3.1 GeV $B=2$ DM that decays through $\chi \to uds\, uds$. After hadronization these result in the decays $\chi\rightarrow\Xi^0_b\,\pi^0$ and $\chi\rightarrow\Lambda^0\Lambda^0$, respectively. The dominant decays for other DM masses, assuming the MFV or FN flavor breaking patterns, are given in Appendix \ref{app:adm-life}. The DM lifetime dependence on $m_\chi$ is shown in Fig. \ref{fig:adm-life} after fixing the NP scale to be $\Lambda=1(3)$ TeV for the MFV (FN) flavor breaking.

To guide the eye, we also show in Fig. \ref{fig:adm-life} the following bounds from indirect DM searches. The green (orange) line shows the constraint on the 
DM decay time from FERMI-LAT~\cite{Ackermann2012} for $\chi \to b\bar b(\mu^+\mu^-)$ decays using the NFW profile. The dash-dotted light red line shows the results of an analysis~\cite{Ibarra:2013zia} based on AMS-02~\cite{Aguilar:2013qda} and assuming $\chi \to \mu^+\mu^-$. The light blue line shows the result of an analysis~\cite{Covi2010} assuming $\chi\to \bar \nu \nu$ decay based on Super-Kamiokande~\cite{Desai2004} bounds. The purple line is an exclusion curve from~\cite{Zhao:2014nsa} based on galactic and extragalactic gamma ray flux measurements by  Fermi \cite{FermiLAT:2012aa,Ackermann:2012qk,Abdo:2010nz}. The authors in~\cite{Zhao:2014nsa} consider $\chi\to uds$ and $\chi\to cbs$ decays as two extreme choices for the flavor structure of the final states. The derived bounds on $\chi$ lifetime differ by less then a factor of $2$ such that the two bounds overlap on the scale of Fig. \ref{fig:adm-life}. The decays we consider fall between these two extreme choices with potentially weakened bounds in our cases above $m_\chi\gtrsim {\mathcal O}(10)$ GeV due to the increased multiplicity of final states. The bounds cross the expected $\chi$ decay times at $m_\chi\sim5$ GeV for  $\Lambda_{\rm MFV}=1$ TeV suppression scale in the case of MFV flavor breaking and at $m_\chi\sim4$ GeV for  $\Lambda_{\rm FN}=3$ TeV suppression scale in the case of FN flavor breaking. 

For the $3.1$ GeV $B=2$ DM we thus find that, for the MFV case, the indirect detection requires 
\beq
\Lambda_{\rm MFV}\gtrsim 0.49 {\rm~TeV},
\eeq
where the dominant operator is given in \eqref{MFVop}. For the FN case the bound is
\beq\label{LambdaFN}
\Lambda_{\rm FN}\gtrsim 1.9 {\rm~TeV},
\eeq
where the least suppressed operator is given in \eqref{FNop}.

\section{Mediator Models}
\label{sec:uv-models}
The EFT analysis of metastable ADM using asymmetric operators is an appropriate approach to derive the indirect DM detection signatures as we did in the previous section. However, for DM direct detection searches and the DM production at colliders, the dominant signals are due to either a single mediator exchange or from direct production of the mediators. To assess the reach of these DM searches, the UV completions to our models are therefore needed. 

We introduce two toy model UV completions that can generate the dimension 10 effective operators; that is, the operator in Eq. \eqref{MFVop} for the MFV case and the operator in Eq. \eqref{FNop} for the FN case. The EFT operators are generated when the $\sim$TeV mediators are integrated out. In our first model all the mediators are scalars, while in the second model there is also a fermionic mediator. The flavor structure in either of the two models could be of the MFV or of the FN type. For concreteness we fix the first model to have the MFV flavor breaking, and the second model to have the FN flavor breaking.

\subsection{MFV model with scalar mediators}
\begin{table}
\centering
\begin{tabular}{cccccc}
\hline \hline
~~Field~~ & ~$SU(3)_C$~ & ~$SU(2)_L$~ & ~$U(1)_Y$~ & ~~~~~$G_F$~~~~~ & $U(1)_{B-L}$~~\\ 
\hline 
$\phi_L$ & $\mathbf{\bar 3}$ & $\mathbf{1}$ & $1/3$ & $\mathbf{(6,1,1)}$ & $2/3$ \\ 
$\varphi_L$ & $\mathbf{6}$ & $\mathbf{1}$ & $1/3$ & $\mathbf{(\bar{3},1,1)}$ & $2/3$ \\ 
$\phi_R$ & $\mathbf{\bar 3}$ & $\mathbf{1}$ & $-2/3$ & $\mathbf{(\bar 3,1,1)}$ & $2/3$ \\
\hline\hline 
\end{tabular}
\caption{The gauge and global charge assignment for the three scalar mediators, $\phi_L$, $\varphi_L$ and $\phi_R$, in the first UV completion toy model for which we assume the MFV flavor breaking pattern.}
\label{tab:tl-charges:MFV}
\end{table}

The SM is extended by the DM, $\chi$, and three flavor multiplets of scalar mediators -- a color anti-triplet $\phi_L$ and a color sextet $\varphi_L$, both with  hypercharge $1/3$, and a color sextet $\phi_R$ with hypercharge  $-2/3$ (see Table \ref{tab:tl-charges:MFV}). They transform under the flavor group $G_F$ as $\mathbf{(6,1,1)}$, $\mathbf{(\bar 3,1,1)}$ and $\mathbf{(\bar 3,1,1)}$, respectively.
The interaction Lagrangian between mediators and the SM is thus given by
\begin{equation}
\begin{split}
\cL_\ssc{int}\supset\, 
	&\frac{\kappa_1}{2}\bar K_I^{AB} [\phi_L]^{I}_\gamma\left(q^{*}_{A,\alpha i} q^{*}_{B, \beta j}\right)\epsilon^{ij}\epsilon^{\alpha\beta\gamma}
  	+\frac{\kappa_2}{2} \bar K_\lambda^{\alpha\beta} [\varphi_L]^{\lambda}_A\left(q^{*}_{ B,\alpha i} q^{*}_{ C, \beta j}\right)\epsilon^{ij}\epsilon^{ABC}\\
     +& \frac{\kappa_3}{2} [Y_D]_X^A[\phi_R]_{A,\alpha}\left(d^c_{Y,\beta}\,d^c_{Z,\gamma}\right)\epsilon^{\alpha\beta\gamma}\epsilon^{XYZ}
	 +\kappa_4  \bar K_I^{AB} \bar K_\lambda^{\alpha\beta}\chi^\dagger [\phi_L]^{I}_\alpha[\varphi_L]^{\lambda}_A[\phi_R]_{B,\beta} + h.c.,
\end{split}
\label{eq:Lint-mfv}
\end{equation}
where the flavor indices $A,B,C$ belong to $SU(3)_Q$ and $X,Y,Z$ to $SU(3)_D$. The QCD indices are $\alpha\beta\gamma$, while the weak isospin indices are denoted by $i,j$. The flavor index $I$ and color index $\lambda$ run from $1$ to $6$. The matrices of the Clebsch-Gordan coefficients, $\bar K_I^{AB}$ and $\bar K_\lambda^{\alpha\beta}$, are the same as in~\cite{Han:2009ya} and satisfy the completeness relation $(\bar K_I^{AB})^* \bar K_I^{CD}=\tfrac{1}{2}(\delta_A^{D}\delta_B^C+\delta_A^C\delta_B^D)$, and similary for $\bar K_\lambda^{\alpha\beta}$. In the second line of \eqref{eq:Lint-mfv}, the down Yukawa insertions make the interaction term with right-handed down quarks formally invariant under $G_F$.

Integrating out the mediators $\phi_{L,R}, \varphi_L, $ gives the $\chi$ decay operator \eqref{MFVop}, with the Wilson coefficient
\beq
\frac{{\cal C}_1}{\Lambda^6}=-\frac{1}{8}\frac{\kappa_1 \kappa_2\kappa_3\kappa_4}{m_{\phi_L}^2 m_{\varphi_L}^2 m_{\phi_R}^2}.
\eeq

For $\kappa_1=\kappa_2=\kappa_3=\kappa_4=1$ the bounds from indirect DM searches thus require $m_{\phi_L, \phi_R, \varphi_L}\gtrsim 450$ GeV, if all the mediator masses are the same. This should be appropriately rescaled if either $\kappa_i$ have smaller values or if all masses are not the same. For instance, for $\kappa_i=0.3$ the mass degenerate case of the mediators is bounded from below by $m_{\phi_L, \phi_R, \varphi_L}\gtrsim 200$ GeV. Since the mediators carry color charges, they can be searched for at the LHC as discussed in Section \ref{sec:collider} below.

\begin{figure}\centering
  \includegraphics[scale=1.5]{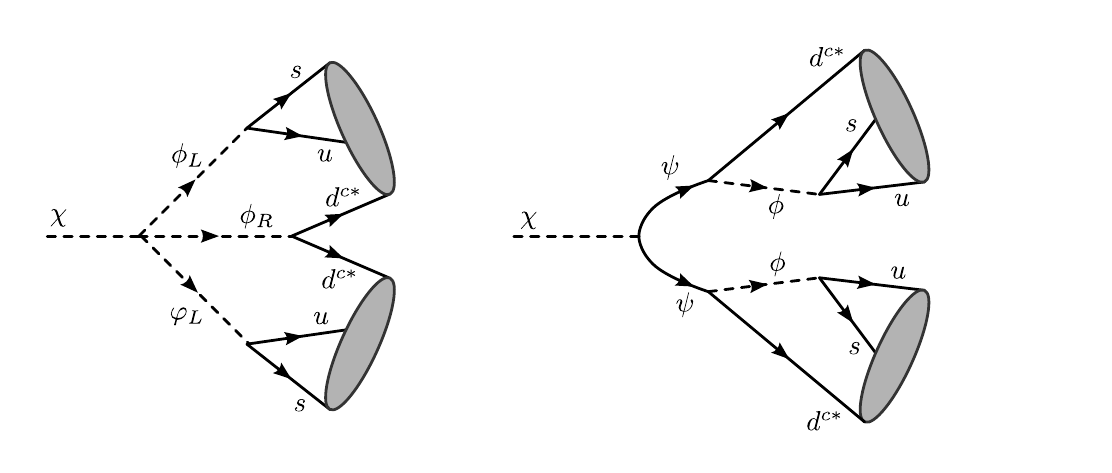}
  \caption{The $\chi$ decay in the MFV mediator model through the off-shell scalar mediators $\phi_{L,R}, \varphi_L$ (left), and through the off-shell fermion $\psi$ and scalar $\phi$ mediators in the FN model (right).}
  \label{fig:b2-uv-decay}
\end{figure}

\subsection{FN model with fermionic  and scalar mediators}
\label{subsec:oneloop}

\begin{table}
\centering
\begin{tabular}{ccccc}
\hline \hline
Field & $SU(3)_C$ & $SU(2)_L$ & $U(1)_Y$ & $U(1)_{B-L}$\\ 
\hline 
$\phi$ & $\mathbf{\bar{3}}$ & $\mathbf{1}$ & $1/3$ & $2/3$ \\ 
$\psi$ & $\mathbf{1}$ & $\mathbf{1}$ & 0 & 1 \\
\hline\hline 
\end{tabular}
\caption{Gauge and $B-L$ charges of the mediators $\phi$ and $\psi$ in the  second UV completion toy model. We also assume the FN flavor breaking pattern.}
\label{tab:tl-charges}
\end{table}

In the second model the SM is supplemented with DM scalar $\chi$,  a Dirac fermion $\psi$ and a complex scalar $\phi$ with SM gauge assignments as in Table \ref{tab:tl-charges}. The relevant terms in the baryon number conserving interaction Lagrangian are
\begin{equation}
\begin{split}\label{FN-model}
   \cL_\ssc{int}\supset\, &\frac{g_{q,AB}}{2}\phi_\gamma \left(q^{*j}_{A,\alpha i} q^{*k}_{B,\beta j}\right)\epsilon^{ij}\epsilon^{\alpha\beta\gamma}
    + g_{d,A} \phi^{*\alpha}\left(d^c_{A,\alpha}\,\psi\right)+ \frac{g_{\chi}}{2}\,\chi(\psi^c\,\psi^c) 
    + h.c.\,,
\end{split}
\end{equation}
where, for the couplings $g_{q}$, $g_{d}$, we also denote the flavor dependence. If the flavor breaking is of the FN type and the mediators do not carry a horizontal charge, then
\beq\label{FNmediatorscaling}
g_{q,AB}\sim g_q \lambda^{|H(q_A)+H(q_B)|}, \qquad g_{d,A}\sim g_d \lambda^{|H(d_A)|},
\eeq
where $g_{q,d}\sim {\mathcal O}(1)$.

Integrating out the mediators generates the operator \eqref{FNop} with the Wilson coefficient
\beq\label{CFNmodel}
\frac{\cal C}{\Lambda^6}\simeq \frac{1}{8\,m_\psi^2 m_\phi^4} g_{\chi} g_{q,M,K'} g_{q,N'M'} g_{d,K} g_{d,N}\sim \frac{1}{8\,m_\psi^2 m_\phi^4}\lambda^{|H(d_K^c)|+|H(d_N^c)|+|H(q_M)+H(q_{K'})|+|H(q_N')+H(q_{M'})|}.
\eeq

Note that the flavor suppression here is parametrically different than in \eqref{C_1:FN} which was obtained by assuming that the FN scale is close to the TeV scale and that the interactions of DM with the visible sector involve the FN fields. In the above model, however, the FN scale can be arbitrarily high and only fixes the flavor interactions between the mediator and the SM fields. Consequently, the leading decay is now $\chi\to uss uds$ where the suppression for the  amplitude is $\sim \lambda^{|H(d^c)|+|H(s^c)|+2|H(q_2)+H(q_{1})|}\sim \lambda^{15}$, to be compared with the $\lambda^4$ suppression in the more conservative case considered in Section \ref{sec:sbhs}~where the leading decay is $\chi\to uds uds$. The indirect detection bound \eqref{LambdaFN} thus translates in our toy mediator model to $m_{\phi, \psi}\gtrsim 130$ GeV for mass degenerate $\phi$ and $\psi$. However, since the coupling to the third generation quarks is $\cO(1)$, the scalar mediators should in fact be heavier than the top quark in order not to modify its total decay width.

The scaling \eqref{FNmediatorscaling} changes if the mediators carry nonzero horizontal charges. For instance, if the horizontal charge of $\phi$ is nonzero, $H(\phi)\ne 0$, one has
~$g_{q, AB}\sim \lambda^{|H(q_A)+H(q_B)-H(\phi)|}$, $g_{d,A}\sim  \lambda^{|H(d_A)+H(\phi)|}$. In this case, the indirect detection bounds need to be appropriately rescaled. For $-2\leq H(\phi)\leq 5$ the Wilson coefficient is still given by \eqref{CFNmodel} and thus $m_{\phi, \psi}\gtrsim 130$ GeV from indirect bounds as before. For other values of $H(\phi)$, the bound becomes even weaker.

\section{Experimental Signatures Of Mediators}
\label{sec:exp:signatures}
Now we turn to the experimental signatures of weak scale mediators, the flavor constraints, direct DM detection, and DM production at the LHC. 
\subsection{Flavor constraints}
\label{sec:flavor}
\begin{figure}
\includegraphics[scale=1.5]{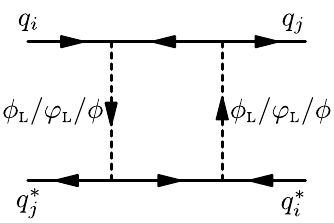}\includegraphics[scale=1.5]{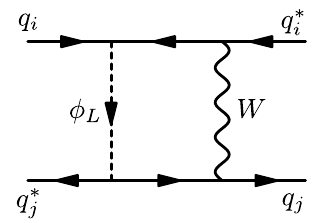}\includegraphics[scale=1.5]{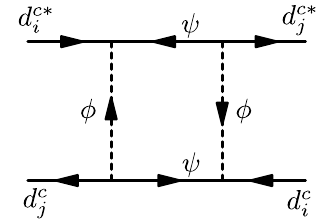}
\caption{Box diagrams contributing to the neutral meson mixing. In the MFV model, there is also a contribution with both $\phi_L$ and $\varphi_L$ in the loop, while $\phi_R$ contributions are suppressed and can be ignored.
\label{fig:one-loop-FCNC}
}
\end{figure}

The two mediator models from Sec.~\ref{sec:uv-models} do not lead to tree level flavor changing neutral currents (FCNCs). These are first generated at 1-loop, see Fig. \ref{fig:one-loop-FCNC}.  For real couplings $\kappa_i$ and $g_{q/d}$ in Eqs.~\eqref{eq:Lint-mfv}, \eqref{FN-model}, the constraints from $K^0-\bar K^0, D^0-\bar D^0$ and $B_{(s)}^0-\bar B_{(s)}^0$  mixing require the mediators masses to be generically above several hundred GeV, as we show below. 
For related analyses of flavor constraints on diquarks, see, e.g.,~\cite{Giudice:2011ak,Grinstein:2011dz}.

The $\Delta F=2$ effective weak Hamiltonian is
\beq\label{Heff}
\cH_{\text{eff}}^{\Delta F=2}=\sum_i \cC_i\cO_i+\sum_i \tilde\cC_i\tilde\cO_i,
\eeq
where $i=1,\dots,5$ runs over the dimension six operators (we use the notation in \cite{Bona:2007vi}). Integrating out the mediators and the $W$ at the weak scale
gives at leading order a nonzero Wilson coefficient for the operator
\beq
\cO_1=(\bar S^\alpha \gamma_\mu P_L D^\alpha)(\bar S^\beta \gamma^\mu P_L D^\beta)=\left(s^{*\alpha}\overline{\sigma}^\mu d^\alpha\right)\left(s^{*\beta}\overline{\sigma}_\mu d^\beta\right),
\eeq
in the case of the MFV model, and for both $\cO_1$ and its parity conjugate operator
\beq
\tilde\cO_1=(\bar S^\alpha \gamma_\mu P_R D^\alpha)(\bar S^\beta \gamma^\mu P_R D^\beta)=\left(s^{c\alpha}\overline{\sigma}^\mu d^{c\alpha*}\right)\left(s^{c\beta}\overline{\sigma}_\mu d^{c\beta*}\right),
\eeq
in the case of the FN model. Above we first give the operators in the 4-component notation and then also in the 2-component notation (for our notation see Appendix \ref{app:spinors}).

In the matching there are two types of contributions, with only the mediators running in the loop, or with both the scalar mediator and the $W$ boson running in the loop, see Fig. ~\ref{fig:one-loop-FCNC}.  
For MFV model these give for the $\XXb{K}$, $\XXb{D}$, and ${B_{(s)}}-\bar B_{(s)}$ mixing
\begin{align}
\begin{split}
\cC_{1K}^\text{MFV} =~&\frac{1}{64\pi^2\,m_\phi^2}\Big\{
  (V_{cs}V_{cd}^*)^2 \left[
  \left(\kappa_1^4 + 3\kappa_2^4 - 2 \kappa_1^2\kappa_2^2  \right)F(x_c)
   +4g_w^2 \kappa_2^2 \,G(x_w,x_c)\right] 
  + c\rightarrow t\\
  &-2 V_{cs}V_{cd}^*V_{ts}V_{td}^*
  \Big[
  \left(\kappa_1^4 + 3\kappa_2^4 - 2 \kappa_1^2\kappa_2^2  \right)F^F(x_c,x_t)
   +4g_w^2 \kappa_2^2 \,G^F(x_w,x_c,x_t)\Big] \Big\},
   \label{eq:C1-MFV-K}
\end{split}
\\
\begin{split}
\cC_{1D}^\text{MFV} =~&\frac{1}{64\pi^2\,m_\phi^2}\Big\{
  (V_{us}V_{cs}^*)^2 \left[
  \left(\kappa_1^4 + 3\kappa_2^4 - 2 \kappa_1^2\kappa_2^2  \right)F(x_s)
   +4g_w^2 \kappa_2^2 \,G(x_w,x_s)\right] 
  + s\rightarrow b\\
  &-2V_{us}V_{cs}^*V_{ub}V_{cb}^*
  \Big[
  \left(\kappa_1^4 + 3\kappa_2^4 - 2 \kappa_1^2\kappa_2^2  \right)F^F(x_s,x_b)
   +4g_w^2 \kappa_2^2 \,G^F(x_w,x_s,x_b)\Big] \Big\},
   \label{eq:C1-MFV-D}
\end{split}
\\
\begin{split}
\cC_{1B_q}^\text{MFV} =~&\frac{1}{64\pi^2\,m_\phi^2}
  (V_{tb}V_{tq}^*)^2 \left[
  \left(\kappa_1^4 + 3\kappa_2^4 - 2 \kappa_1^2\kappa_2^2  \right)F(x_t)
   +4g_w^2 \kappa_2^2 \,G(x_w,x_t)\right],
\end{split}
\label{eq:C1-MFV-Bd}
\end{align}
where $q=d,s$, $x_i = (m_i/m_\phi)^2$, and we have set $m_u=m_d=0$ and taken for simplicity that the $\phi$ and $\varphi$ are mass degenerate. The loop functions $F(x)$, $F^F(x_1,x_2)$, $G(x_1,x_2)$, $G^F(x_1,x_2,x_3)$ are given in Appendix \ref{app:meson-mixing}.
As in the SM also here the largest contribution to the $\XXb{K}$ mixing is due to the charm-charm loop, while for $\XXb{B_q}$ mixing the top  loop dominates, as expected.

For the FN model the Wilson coefficients are given by
\begin{align}
\begin{split}
\cC_{1K}^\text{FN} &\sim \frac{\lambda^{10} g_q^4}{16\pi^2 m_{\phi}^2}\Big[
	\,H(x_t)
	+2\lambda^{4} \,H^F(x_c,x_t)
\Big], \qquad 
\tilde\cC_{1K}^\text{FN} \sim \frac{\lambda^{10}g_d^4}{16\pi^2 m_{\phi}^2}
	\,H(x_\psi),
\end{split}
\\
\begin{split}
\cC_{1B_d}^\text{FN} &\sim \frac{\lambda^{6}g_q^4}{16\pi^2 m_{\phi}^2}\Big[
	\,H(x_t)
	+2\lambda^{4}\,H^F(x_c,x_t)
\Big], \qquad
\tilde\cC_{1B_d}^\text{FN} \sim \frac{\lambda^{10}g_d^4}{16\pi^2 m_{\phi}^2}
	\,H(x_\psi),
\end{split}
\\
\begin{split}
\cC_{1B_s}^\text{FN} &\sim \frac{\lambda^{4}g_q^4}{16\pi^2 m_{\phi}^2}\Big[
	\,H(x_t)
	+2\lambda^{4}\,H^F(x_c,x_t)
\Big], \qquad
\tilde\cC_{1B_d}^\text{FN} \sim \frac{\lambda^{8}g_d^4}{16\pi^2 m_{\phi}^2}
	\,H(x_\psi),
\end{split}
\end{align}
and $\cC_{1D}^\text{FN}=\cC_{1K}^\text{FN}$, $\tilde\cC_{1D}^\text{FN}=0$. Above we have indicated the scaling of different contributions to the Wilson coefficient in terms of $\lambda=0.2$, c.f., Sec.~\ref{sec:sbhs}. In the numerics we use the equality sign. The loop functions $H(x)$ and $H^F(x_1,x_2)$ are given in Appendix \ref{app:meson-mixing}.

\begin{table}
\begin{tabular}{c  c c c c }
\hline\hline
\multicolumn{1}{c}{}&\multicolumn{2}{c}{MFV}&\multicolumn{2}{c}{~~FN~~}\\
 & $~~~\kappa_{1,2}<~~~$ & $~m_{\phi_L,\varphi_L}>~$ & $~~~g_{q,d} <~~~$ & $~m_{\phi}>~$ \\ 
 \hline
 $K^0-\bar K^0$ ~& 0.33 & 2.9 TeV & 0.63 & 570 GeV\\
 $B_d-\bar B_d$~& 1.3 & 710 GeV & 0.54 & 1 TeV \\ 
 $B_s-\bar B_s$~& 1.3 & 780 GeV & 0.59 & 840 GeV \\
 $D^0-\bar D^0$ ~& 30 & 34 GeV & 4.3 & 56 GeV \\ 
\hline\hline
\end{tabular} 
\caption{The 95 \% C.L. bounds on the MFV and FN mediator models from meson mixing. Taking $m_{\phi_L}=m_{\varphi_L}=m_\phi=1$TeV and $\kappa_1=\kappa_2$ gives the upper bounds on the couplings in the 2nd column,  and in 4th column for $g_q=g_d$. Taking in turn $\kappa_{1,2}=g_{q,d}=1$ gives lower bounds on the mediator masses in 3rd and 5th columns. The mass of the fermion in the FN model is fixed to $m_\psi=20$ GeV, see also Sec.~\ref{sec:collider}. The bounds are not very sensitive to $m_\psi$.}
\label{tab:mmbar}
\end{table}

Note that the above Wilson coefficients contain $\log(m_i/m_\phi)$ that can become large for  $m_\phi \gg m_i$. We do not attempt to resume these logarithms, which also means that we treat all the NP contributions as local.  We expect that our numerical results can receive ${\mathcal O}(1)$ corrections due to neglected terms, which is within precision required  for our analysis. We do include, though, the usual RGE effects due to the NLO QCD running of the effective weak Hamiltonian from the weak scale to the  low energy.  For constraints from $K^0-\bar K^0$ and  $B_{(s)}-\bar B_{(s)}$ mixing we use the recent results of a fit to the mixing parameters in \cite{Charles:2013aka}. The constraints from $D^-\bar D^0$ mixing are obtained by  that the NP contribution saturates $\Delta m_D$, so that in the equation $x_D=2 \left|\langle \bar D^0|\cH_\text{eff}^{\Delta C=2}|D^0\rangle\right|/\Gamma_D$, valid in the limit of no CP violation, we only include the NP contribution~\cite{Bona:2007vi}. The resulting bounds on couplings and masses are shown in Table \ref{tab:mmbar}. In the case of MFV model the most severe bound comes from $K^0-\bar K^0$ and is due to $\epsilon_K$. Since we assume that all the $\kappa_i$ in \eqref{eq:Lint-mfv} are real, the NP contribution does carry a weak phase due to the $V_{ts}V_{td}^*$ CKM factors and does contribute to $\epsilon_K$. In contrast for FN model the NP contribution to the mixing do not carry a weak phase, and thus to not have an effect on $\epsilon_K$, and thus the bounds from $K^0-\bar K^0$ mixing are much less severe.

\begin{figure}[]
\includegraphics[width=0.5\textwidth]{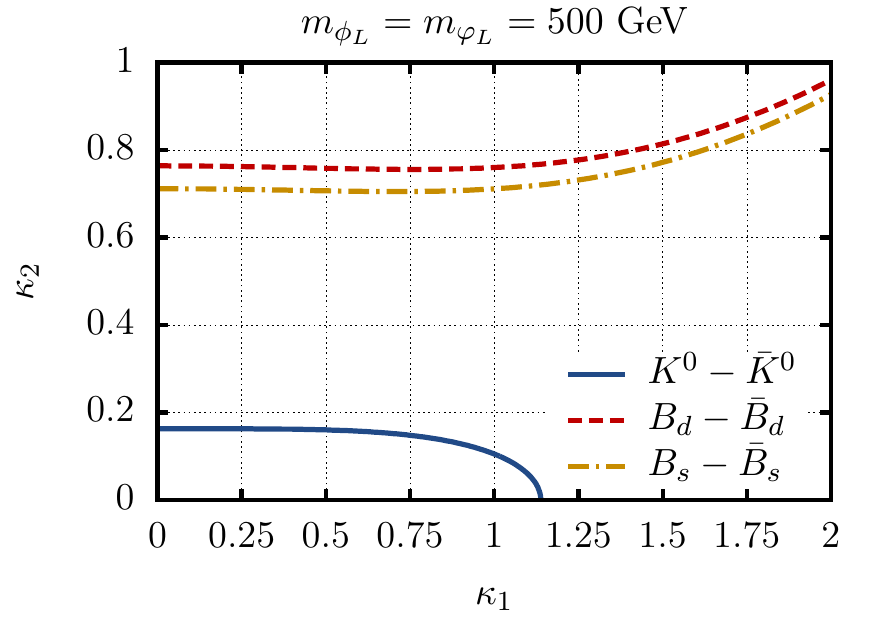}\includegraphics[width=0.5\textwidth]{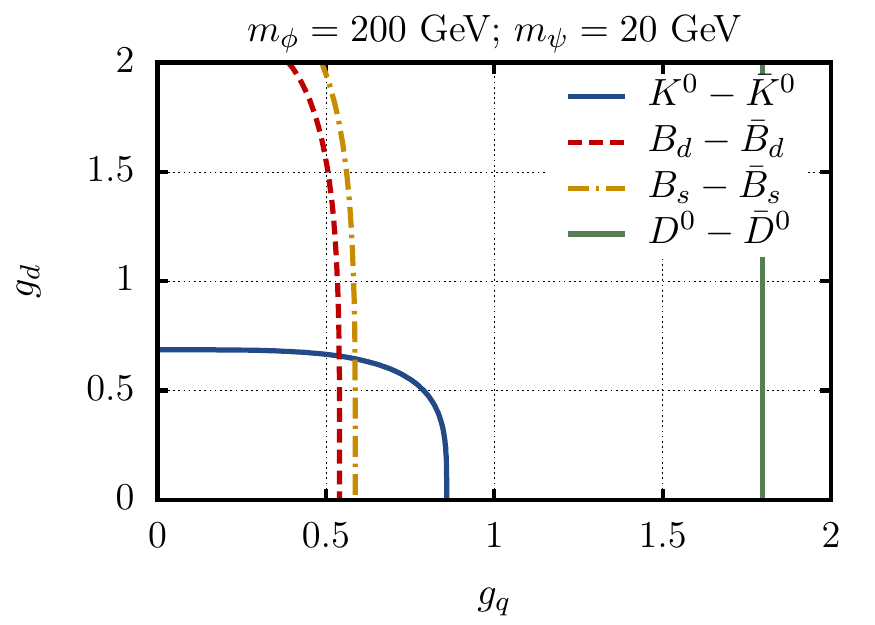}
\caption{Meson mixing constraints on the couplings $\kappa_{1,2}$ in the MFV mediator model (left) and $g_{q,d}$ in the FN model (right), taking $m_{\phi_L}=m_{\varphi_L}=500$ GeV and $m_\phi=200$ GeV, $m_\psi=20$ GeV respectively. The regions excluded are above and to the right of the curves. }
\label{fig:mmbarmfv}
\end{figure}

In Fig.~\ref{fig:mmbarmfv} we show the constraint on the couplings $\kappa_{1,2}$ in the MFV model, fixing $m_{\phi_L}=m_{\varphi_L}=500$GeV (left figure), and the constraints on $g_{q,d}$ in the FN model, fixing $m_\phi=200$GeV, $m_\psi=20$GeV (right figure). Since in the case of MFV the largest contribution to $K^0-\bar K^0$ is from the mediator-$W$ loop, the $\epsilon_K$ bound places a stringent constraint on $\kappa_2$. For $\kappa_2\gg 1$, however, $\kappa_1$ can be ${\mathcal O}(1)$. Since the NP contributions to the meson mixing were assumed to be CP conserving in the case of the FN model, the couplings $g_{d,q}\sim {\mathcal O}(1)$ are allowed even for $m_\phi$ as low as 200 GeV.

\subsection{Relic abundance and direct detection}
\label{sec:dir-detect}
We note in passing that the virtual exchanges of the mediators generate contact operators of the schematic form $\chi^\dagger \chi \bar q q$ that contribute to the $\chi\chi^\dagger$ annihilation cross section and to the cross section for DM scattering on nuclei. The symmetric couplings of DM and the mediators, of schematic form $\chi\chi^\dagger \phi \phi^\dagger$, 
do not suffice to create large enough annihilation cross sections that would annihilate away the symmetric component of DM relic abundance. 

As an example consider the MFV model with scalar mediators, Eq. \eqref{eq:Lint-mfv}, and assume that the lightest mediator is $\phi_L$. It can have a symmetric coupling to DM of the form
\beq
{\cal L} \supset \kappa' [\phi_L]^{I}_{\gamma}  [\phi_L^\dagger]^{I}_{\gamma} \chi^\dagger \chi.
\eeq
At 1-loop this generates a contact interaction $\chi^\dagger\partial_\mu \chi \bar q \gamma^\mu q$, which leads to annihilation cross section $\langle \sigma v\rangle\sim {\mathcal O}(10^{-28}cm^3/s) (100{\rm GeV}/m_{\phi_L})^4$ for ${\mathcal O}(1)$ couplings. This annihilation cross section is more than three orders of magnitude too small to obtain the observed relic density and satisfy CMB constraints for $s$-wave annihilation \cite{Lin:2011gj}. 

The symmetric component of DM thus needs to annihilate away through a different mechanism. An attractive possibility is that $\chi$ is charged under dark force which leads to large enough annihilation cross section \cite{Lin:2011gj,Cohen:2010kn,Blennow:2012de}. The dark forces will then also lead to the dominant contribution to the direct detection cross section, for details see, e.g.,  \cite{Lin:2011gj}.


\subsection{Collider signatures}
\label{sec:collider}

In both the MFV and FN flavor breaking scenarios the mediator models involve colored scalars. These can be searched for at the LHC through the gluon initiated pair production or through a single production. To estimate the LHC reach we use our two mediator models. The MFV mediator model, Eq. \eqref{eq:Lint-mfv},
contains three colored scalars that are either triplets or sextets of color and flavor group, see Table~\ref{tab:tl-charges:MFV}. The FN model, Eq. \eqref{FN-model},  contains a colored scalar and a neutral fermion, see Table~\ref{tab:tl-charges}.

\begin{figure}
\includegraphics[scale=1]{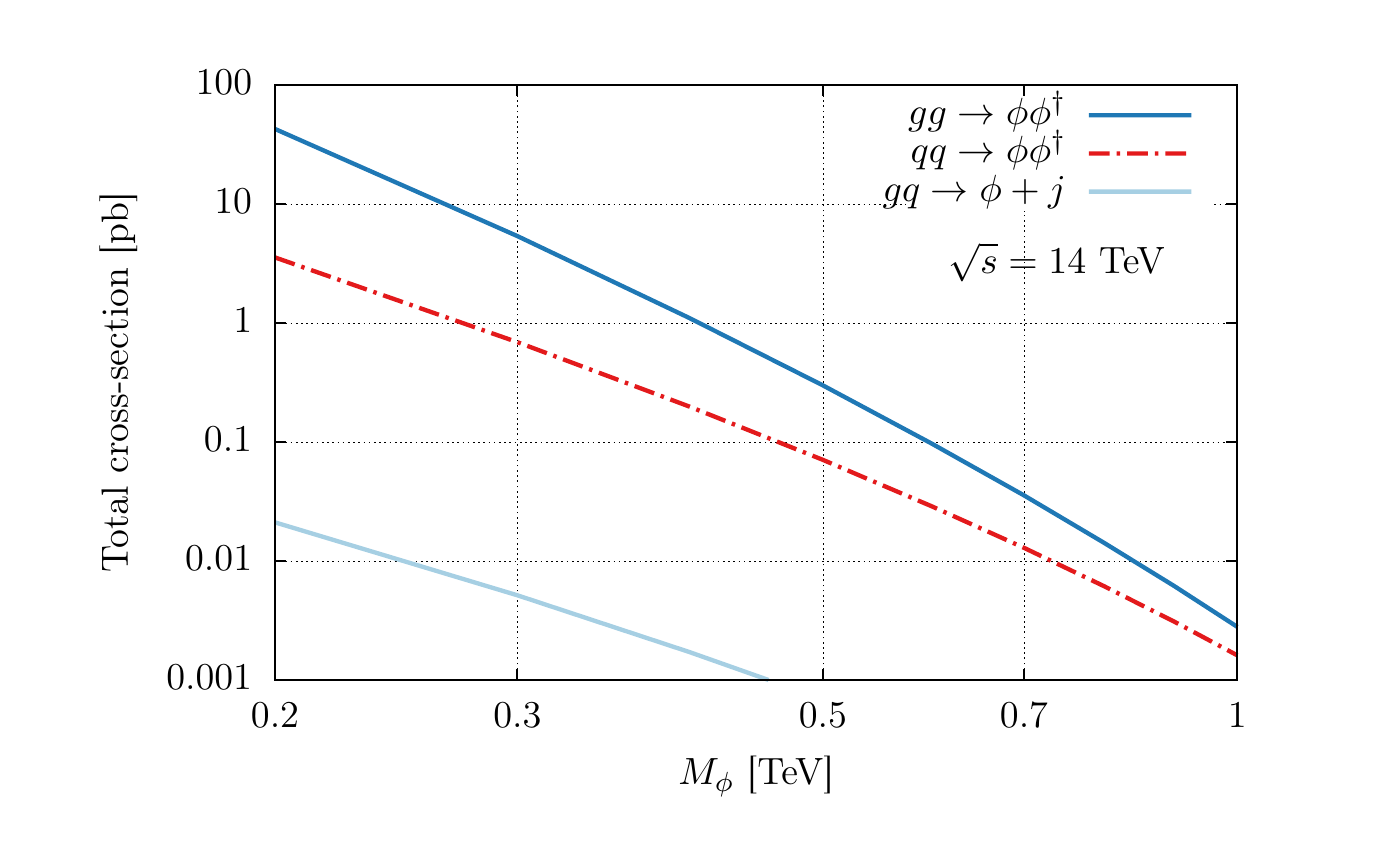}
\caption{The $gg\to \phi\phi^\dagger$ (solid blue), $qq\to \phi\phi^\dagger$ (dot-dashed red) and $gq\to \phi j$ (solid light blue) contributions to the pair-production and single-production cross-section at the LHC with $\sqrt{s}=14$ TeV as a function of a mass of a color triplet scalar $\phi$, a mediator in the FN model. }
\label{fig:fn-phi-prod}
\end{figure}

Pair production of colored scalars is the dominant production mechanism of the mediators for the masses of interest, below ${\mathcal O}({\rm TeV})$.
We illustrate this in Fig. \ref{fig:fn-phi-prod} for the color triplet $\phi$ in the FN model, where we compare the pair production cross section from gluon fusion and from quark-guon fusion, and the single production of $\phi$ in association with a jet. Gluon fusion clearly dominates in the mass range of interest.

The signatures of pair produced colored scalars depend on their decay modes. In our two models they decay either directly to two SM quarks or, alternatively, first to two lighter scalars that then in turn decay to two jets each. In the FN model the decay $\phi \to j\psi$ is also possible. The flavor composition of the jets depends on the flavor quantum numbers of the scalar. For instance, the states in $\phi_L$ flavor multiplet can decay either predominantly through $\phi_L\to tb$, through $\phi_L\to bj$, or through $\phi_L\to j j$, depending on the flavor numbers of $\phi_L$ (and similarly for $\varphi_L$), see Eq. \eqref{eq:Lint-mfv}. The scalars in $\phi_R$ flavor multiplet, on the other hand, decay through $\phi_R\to bj$, or through $\phi_R\to j j$, again depending on the flavor index carried by the $\phi_R$ state. 
In the FN model one needs to require $m_\phi>m_t$  in order not to modify the  total decay width of the top quark, see Sec.~\ref{sec:uv-models}. Then the dominant decay is either $\phi \to \bar b \psi$ or $\phi \to t b$, depending on the relative sizes of the two couplings, while the other decays are suppressed by additional powers of $\lambda$.

\begin{figure}
\includegraphics[scale=1]{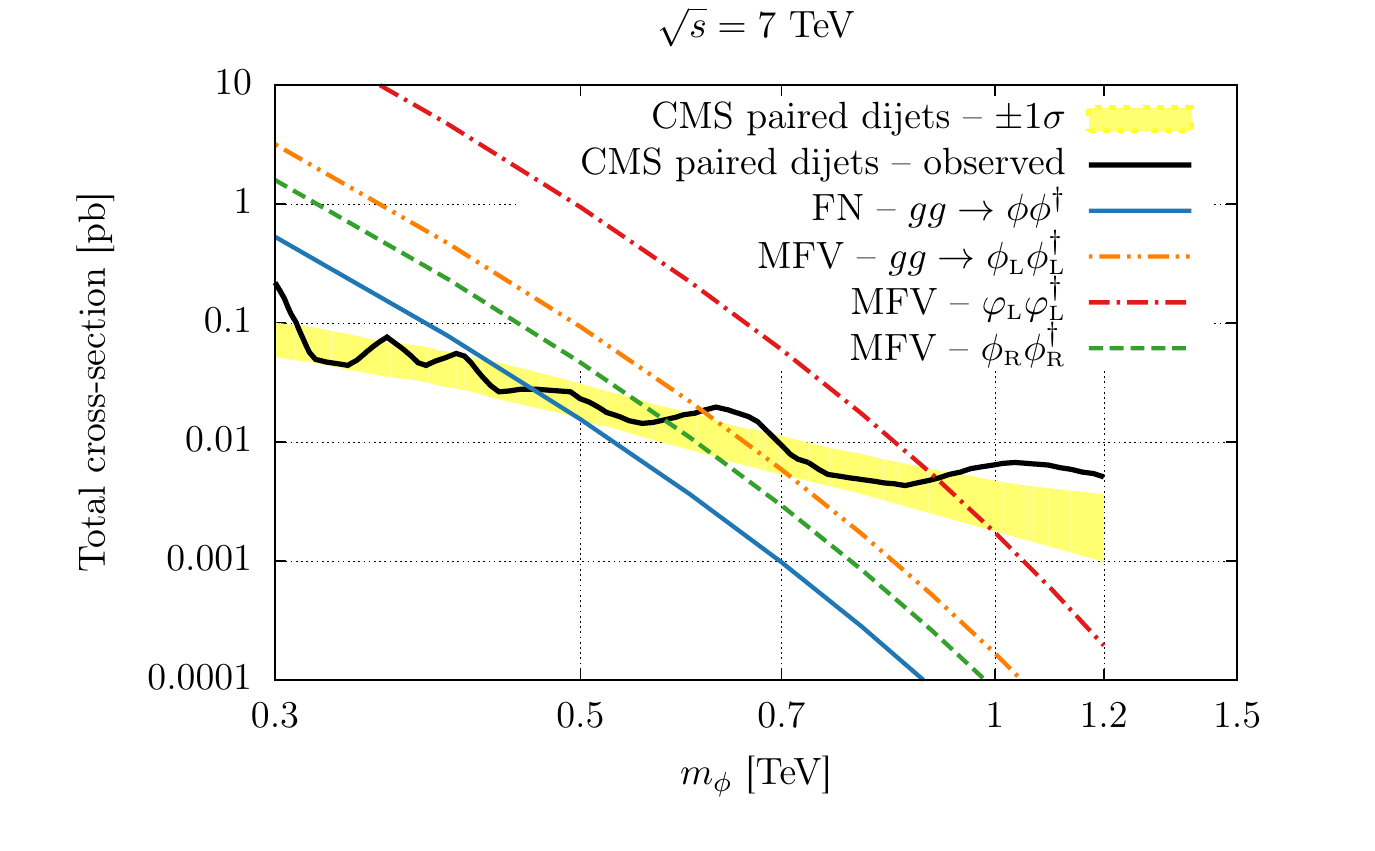}
\caption{Constraints on scalar mediator $\phi$ in the FN model, and $\phi_L$, $\varphi_L$, $\phi_R$  in the MFV model, that follow from the CMS search for pair-produced dijet-resonances \cite{Chatrchyan:2013izb}. The state in the same flavor and color multiplet are taken to be mass-degenerate.}
\label{fig:cms-paired-dijets}
\end{figure}

To get a rough estimate of the LHC sensitivity we treat all the decay modes as two-jet final states (this overestimates the reach slightly since for $tj$ final state the real efficiency is expected to be lower). The strongest constraint on pair-production of the lightest scalar mediators is then the search for pair-produced dijet resonances from CMS at 7 TeV LHC with integrated luminosity of 5 ${\rm fb}^{-1}$\cite{Chatrchyan:2013izb}. 
This places the bounds ${m_\phi} \gsim 470$ GeV in the case of FN model assuming  that $\phi\to \bar b\psi$ decay is negligible, and $m_{\phi_L} \gsim 620$ GeV,  $m_{\varphi_L} \gsim 910$ GeV,   $m_{\phi_R} \gsim 580$ GeV, in the case of MFV flavor breaking as shown in Fig.~\ref{fig:cms-paired-dijets}.

Note that when all three mediators are degenerate in mass, the color sextet scalar has the largest pair production cross section due to the large color factor.

For FN model a new experimental signature is obtained in the limit $g_{d}\gg \lambda^2 g_q$.  Then the dominant decay of $\phi$ is $\phi \to \bar b \psi$.
In order not to have fast decaying DM  $m_\psi>m_{\chi}/$. 
Using NDA the $\psi$ decay length is
\beq
c\tau({\psi\rightarrow b b c}) \sim \left(g_q^2 g_d^2\lambda^8\,\frac{1}{8\pi}\frac{1}{16\pi^2}\frac{m_\psi^5}{m_\phi^4}\right)^{-1} \sim 30\text{m}\left(\frac{20\,\text{GeV}}{m_\psi}\right)^5\left(\frac{m_\phi}{750\,\text{GeV}}\right)^4\left(\frac{0.03}{g_q g_d}\right)^2.
\label{eq:psideclen}
\eeq

\begin{figure}
\includegraphics[scale=1]{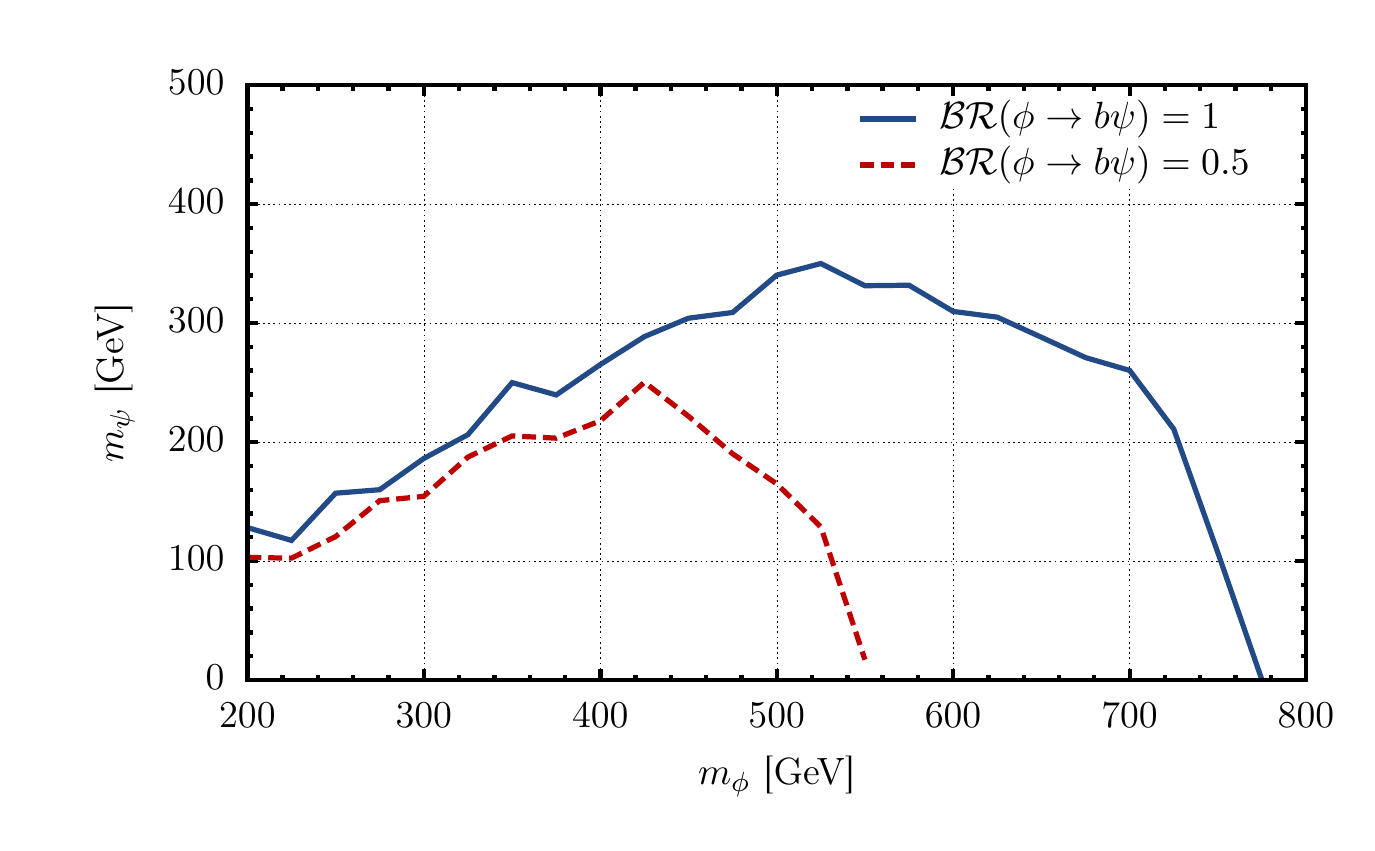}
\caption{The 95\% exclusion limit on $\phi\phi^\dagger$ production in the FN model for the $b\bar{b}\psi\bar{\psi}$ final state, where $\psi$ escapes the detector and sbottom search applies \cite{CMS:2014nia}.
The solid blue (dashed red) line is for  $\phi\rightarrow b\psi$  branching ratios of 50\% and 100\%.}
\label{fig:2b2psi}
\end{figure}

For light enough $m_\psi$ (or heavy enough $m_\phi$ ),
the fermion $\psi$ does not decay in the detector and appears as $\mET$. The $pp\to \phi \phi^\dagger$ pair production then results in $2j+\mET$ or $2b+\mET$ final state, and is bounded from sbottom searches as shown in Fig.~\ref{fig:2b2psi}. The resulting bound is $m_\phi>550(760)$ GeV for $m_\psi=20$ GeV and $\mathcal{BR}(\phi\rightarrow\psi b)=0.5(1.0)$. The choice $g_q g_d=0.03$ in Eq.~\eqref{eq:psideclen} gives $\mathcal{BR}(\phi\rightarrow b\psi)\approx\mathcal{BR}(\phi\rightarrow s\psi)=0.33$. For the same input parameters the single production of $\psi$ in association with $b$, $t$, or $\phi$ has a cross section $\sim 7 \cdot 10^{-2}$ fb while the pair production is dominated by the process $ss\rightarrow\psi\psi$ and has a negligible cross section of $\sim 4 \cdot 10^{-4}$ fb.

The single production of mediators, e.g.,  $u d \to \phi$, $u d \to \phi_L$,   $u d \to \varphi_L$, $d s \to \phi_R$, is suppressed due to the small couplings of the mediators to the first and the second generation quarks. Similarly, the single production from heavy quarks in the initial state suffers  the PDF suppression.

For single top production, the MFV model gives the largest contribution with a cross-section of $\sigma(ud\rightarrow\phi_L\rightarrow tb)_\text{\sc mfv} = 7.6\times 10^{-6}$ pb for $m_{\phi_L}=500$ GeV and $\sqrt{s}=8$ TeV. This is well below the SM production cross section. Thus, the ATLAS and CMS combined measurement  of the single top cross section at $\sqrt{s}=8$ TeV,  $85\pm12$ pb~\cite{ATLASandCMSCollaborations:2013ofa}, does not impose any limits on the mediator model.

The production of DM $\chi$ can occur from the decay of heavier mediators. For instance, for $\kappa_4\sim \kappa_3$ and $\phi_R$ heavy enough, the dominant decay mode of $\phi_3$ is $\phi_3\to \chi \varphi_L^\dagger \phi_L^\dagger$. Pair production $pp\to \phi_R\phi_R^\dagger$ would thus result in $8j+\mET$ signature, where paired dijets would reconstruct $\phi_L$ and $\varphi_L$ mass peaks (depending on the flavor assignments some of the jets can be replaced by $t$ of $b$ jets).

\section{Conclusions}
\label{sec:conclusions}
We showed that for asymmetric DM (ADM) models the stability of DM on cosmological time scales may be purely accidental. 
We do not require that the DM is charged under an ad-hoc conserved $Z_n$ symmetry. Rather, we assume that such a discrete symmetry is explicitly broken by the mediator interactions that transfer the $B-L$ between the DM sector and the visible sector in the early universe. Such asymmetric interactions are necessary in all models of ADM though they may be made to obey a $Z_4$ symmetry (i.e. one can demand that they involve only the $\chi\chi\to {\rm\it visible}$ or  $\chi^\dagger\chi^\dagger\to {\rm\it visible}$ transitions instead of $\chi\to {\rm\it visible}$ transitions as is in our case).

At low energies, the DM then carries a conserved $\chi$ charge that is broken only by the higher dimensional operators obtained by integrating out the mediators. Such operators also lead to DM decays. In this paper we explored the role of continuous flavor symmetries for the properties of such decaying DM focusing on the case where DM that carries nonzero baryon number. For $B=1$ DM, the direct detection bounds are evaded if the mediators are above $\sim 4\cdot 10^9$ TeV assuming ${\mathcal O}(1)$ couplings. However, if quark flavor breaking is of the MFV type, the mediators can be lighter by around two orders of magnitude.  For $B=2$ DM, the scale of mediators can be much lighter ${\mathcal O}(8{\rm TeV})$ for ${\mathcal O}(1)$ couplings. This is then lowered by an order of magnitude if quark flavor breaking is of the MFV or Froggatt-Nielsen type. The mediators that would lead to indirect DM signals in the next generation of experiments can thus be, at the same time, searched for at the LHC.

We have explored this possibility by constructing two mediator models, one with assumed MFV and one with a FN flavor breaking pattern. The MFV mediator model (Eq. \eqref{eq:Lint-mfv})
contains three colored scalars that are either triplets or sextets of the color and the flavor groups, see Table~\ref{tab:tl-charges:MFV}. The FN model (Eq. \eqref{FN-model}), on the other hand, contains one colored scalar and one neutral fermion, see Table~\ref{tab:tl-charges}. These mediators generate FCNCs at 1-loop. While this leads to nontrivial constraints on their masses and couplings, the mediators can still be as light as a few $\times 100$ GeV with ${\mathcal O}(1)$ couplings. Since the mediators are charged under QCD, they can be singly or pair-produced at the LHC with large cross sections. This means that the searches at the LHC can lead to interesting constraints or discoveries.  The signatures depend on how the mediators decay. In the FN model, for instance, the decay to heavy quarks, $\phi\to tb$, is favored. Modifying the paired dijet searches to the $pp\to \phi\phi \to tb \bar t \bar b$ signal could thus enhance the reach of the LHC in the search for these mediators.  In the MFV model, on the other hand, paired light dijets, paired $tb$ and paired $bj$ are possible. Other signatures are discussed in Sec. \ref{sec:collider}.

In conclusion, ADM can quite generically be metastable with a possibility of complementary  signals in indirect detection and at the LHC.

{\bf Acknowledgements:} J. Z. and F.B. are supported by the U.S. National Science Foundation under CAREER Grant PHY-1151392. F.B. is supported by the Fermilab Fellowship in Theoretical Physics.
Fermilab is operated by Fermi Research Alliance, LLC, under Contract
No.~DE-AC02-07CH11359 with the United States Department of Energy. J.Z. is grateful to the Mainz Institute for Theoretical Physics (MITP) for its hospitality and its partial support during the completion of this work. J.Z. thanks the Aspen Center for Physics, supported by the NSF Grant \#1066293, and the KITP, supported in part by the National Science Foundation under Grant No. NSF PHY11-25915, for their warm hospitality. F.B. thanks Prateek Agrawal, Roni Harnik, and Felix Yu for helpful discussions. Some of the cross-sections were computed using \verb|MadGraph5|~\cite{Alwall:2014hca} using a model file generated by \verb|FeynRules 2.0|~\cite{Alloul:2013bka}.

\appendix
\section{Operators in four component notation}
\label{app:spinors}
In the paper we are using a two-component notation, where the left-handed Weyl fermion fields $(q_i,u^c,d^c,l_i,e^c)$ have hypercharges $(+1/6,-2/3,+1/3,-1/2,+1)$ and $B-L$ charges $(1/3,-1/3,-1/3,-1,+1)$. The higgs doublet is denoted by $H$ and has $Y=+1/2$, while $
\tilde H=i \sigma_2 H^*$. The two Weyl spinors of the DM Dirac fermion are $\psi$ and $\psi^c$ with $B-L=-1$ and $+1$, respectively. Finally, $\phi$ is the complex scalar DM with $B-L=2$. Capital letters denote four-component spinors following the notation in \cite{Dreiner:2008tw}. The DM Dirac fermion $\Psi$ and its charge conjugate $\Psi^C$ are
\beq
\Psi=
\begin{pmatrix}
\psi_\alpha \\
\psi^{c\dagger\dot\alpha}
\end{pmatrix}, 
\quad
\Psi^C=
\begin{pmatrix}
\psi^c_\alpha \\
\psi^{\dagger\dot\alpha}
\end{pmatrix},
\eeq
while $\bar \Psi=(\psi^{c\alpha},\psi_{\dot \alpha})$. Writing for the two-component spinors $q_1=u_L$, $q_2=d_L$ and $\ell_1=\nu_L$, $\ell_2=e_L$, suppressing generation indices, we introduce
\beq
U=
\begin{pmatrix}
u_{L\alpha} \\
u^{c\dagger\dot\alpha}
\end{pmatrix}, 
\quad
D=
\begin{pmatrix}
d_{L\alpha} \\
d^{c\dagger\dot\alpha}
\end{pmatrix},
\quad
E=
\begin{pmatrix}
e_{L\alpha} \\
e^{c\dagger\dot\alpha}
\end{pmatrix},
\quad
N=
\begin{pmatrix}
\nu_{L\alpha} \\
\nu^{c\dagger\dot\alpha}
\end{pmatrix},
\eeq
where $\nu^c$ is the right-handed neutrino field introduced for completeness. If neutrino is Majorana, $\nu_L=\nu^c$. The weak doublets in the four-component notation are
\beq
Q_L=(U_L, D_L), \quad L_L=(N_L, E_L),
\eeq
with $U_L\equiv P_L U$, etc, and similarly $U_R\equiv P_R U, D_R\equiv P_R D, ...$. 
Some examples of the relevant asymmetric operators in the two- and four-component notations are given below.
\begin{align}
&\text{ \underline{\bf dim 6:}} &\quad &{\cal Q}_1^{(6)}=  
(q_i q^i)(d^{c*} \psi^{c*})=(\overline{Q_{Li}^C} Q_L^i) (\overline{D^C}P_R \Psi),\\
&\quad &\quad &{\cal Q}_2^{(6)}=  
(u^c d^c)(d^{c} \psi^c)=(\overline{U} P_L D^C) (\overline{D}P_L\Psi^C),\\
&\text{ \underline{\bf dim 10:}} &\quad &{\cal Q}_1^{(10)}=  \phi(d^c d^c)(q^*_i q^{i*})(q^*_j q^{j*})
	=\phi(\overline{D} P_L D^C) (\overline{Q_{Li}}P_R Q_L^{Ci}) (\overline{Q_{Li}}P_R Q_L^{Ci}).
\end{align}

\section{Asymmetric DM relic density}
\label{app:mass}
Here we review the relations between the DM relic density and the DM mass in ADM models.
We assume that the operator(s) transferring the $B-L$ asymmetry from the visible to the dark sector decouple above electroweak phase transition, $T_C>T_{\rm ew}\sim 170$ GeV \cite{D'Onofrio:2012ni}, as is the case for our ADM models, see Sec. \ref{sec:DMmassADM}. We first assume that  the visible sector consists below $T_C$  of only the SM fields (we will later relax this). 
The number density asymmetry  for relativistic particles is
\beq
(n-\bar n)_i=\frac{T^3}{6}\hat{g}_i \frac{\mu_i}{T}, 
\label{density}
\eeq
where  $n(\bar n)$ are the particle(anti-particle) number densities, $\mu_i$ is the chemical potential for species $i$, and 
$\hat{g}_i=g_i(g_i/2)$ for bosons (fermions) with 
$g_i$ internal degrees of freedom so that
$\hat g_i=1$ for a Weyl fermion, while $\hat g_i=2$ for a Dirac fermion or a complex scalar. 

All the SM particles are in chemical equilibrium,  so that the chemical potentials  
 are proportional to the conserved quantum numbers \cite{Harvey:1990qw}. Above the electroweak phase transition these are $B-L$, $Y$ and $SU(2)_L$, while $B+L$ is broken by  sphalerons. Thus (see also \cite{Feldstein2010})
\beq
\mu_i=(T_3)_i c_3 +Y_i c_Y+(B-L)_i c_{B-L},
\eeq
where the $c_i$ are constants that we determine from  net weak isospin, hypercharge and $B-L$ densities. The net weak isospin charge density in the universe normalized to entropy density is
\beq
T_3 \propto \sum_i\hat{g}_i\,(T_3)_i\,\mu_i=\sum_i\hat{g}_i\,(T_3)_i^2 c_3+0\cdot c_Y+0\cdot c_{B-L}=0.
\eeq
For the first equality we used that for each SU(2) multiplet $ \sum_i  (T_3)_i=0$, and in the second equality that the net $T_3$ charge is zero since $SU(2)_L$ is not explicitly broken. Thus $c_3=0$ and the  $SU(2)_L$ charge of a particle does not contribute to its chemical potential.

Flavor mixing ensures that the chemical potentials for SM Weyl fermions from different generations are the same. Similarly, $SU(2)_L$ interactions ensure that $\mu_{u_L}=\mu_{d_L}\equiv \mu_Q$, and $\mu_{\ell_L}=\mu_{\nu}\equiv \mu_L$. We thus have
\beq
\begin{matrix}
\mu_H=\frac{1}{2}c_Y,  &\mu_L=-\frac{1}{2}c_Y-c_{B-L}, &\mu_E=-c_Y-c_{B-L},\\
\mu_Q=\frac{1}{6}c_Y+\frac{1}{3}c_{B-L}, &\mu_U=\frac{2}{3}c_Y+\frac{1}{3}c_{B-L}, &\mu_D=-\frac{1}{3}c_Y+\frac{1}{3}c_{B-L},
\end{matrix}
\eeq
while for the gauge bosons $\mu_G=\mu_W=\mu_B=0$. The net hypercharge of the universe is thus
\beq
\begin{split}
Y&\propto \sum_i\hat{g}_i\,(Y)_i\,\mu_i=\frac{1}{2} 2\cdot 2 \mu_H +N_f\left[-\frac{1}{2} 2 \mu_L-\mu_E+N_c\left(\frac{1}{6}\cdot 2\mu_Q+\frac{2}{3}\mu_U-\frac{1}{3}\mu_D\right)\right]\\
&=2\mu_H+N_f\left(\mu_Q+2\mu_U-\mu_D-\mu_L-\mu_E\right)=11c_Y+8c_{B-L},
\end{split}
\label{Y_SM}
\eeq
where
$N_f=3$ is the number of generations and $N_c$ is the number of colors. 
Setting the net hypercharge density in the universe to zero, $Y=0$, gives
\beq
c_Y=-\frac{8}{11}c_{B-L}.
\eeq
The net $B-L$ number density in the visible sector (i.e. excluding the $B-L$ asymmetry carried by the $\chi$ fields in the dark sector) is then
\beq
B-L\propto
N_f\left(-2 \mu_L-\mu_E+2\mu_Q+\mu_U+\mu_D\right)= \frac{79}{11} c_{B-L}.
\label{eq:b-l_sm}
\eeq

There are two types of interactions between the dark and visible sector: the asymmetric interactions that involve a single $\chi$ field, and the symmetric interactions of the form $\chi^\dagger \chi$ times the SM fields. The symmetric operators keep the dark and the visible sectors in thermal equilibrium. The asymmetric interactions are suppressed, and decouple at temperatures well above the $\chi$ mass. At lower temperatures the $\chi$ number is thus effectively conserved. The chemical potential $\mu_\chi$ is the same as it was before the decoupling.
We thus have
\beq
\mu_\chi^i=(B-L)_{\chi}^i c_{B-L},
\eeq
where $(B-L)_{\chi}^i$ is the $B-L$ charge of the $\chi^i$ field. Here we allow for several $\chi^i$ fields in the dark sector and also define  the weighted $B-L$ charge of the dark sector fields as
\beq\label{B-Lsum}
(B-L)_\chi^{\rm sum}\equiv \sum_i \hat g_\chi^i (B-L)_{\chi}^i.
\eeq

The net $\chi$ number density normalized to entropy density we denote by $\Delta \chi$ and is
\beq
\begin{split}
\Delta \chi &\propto \sum_i \hat{g}_\chi^i\mu_\chi^i=(B-L)_\chi^{\rm sum}c_{B-L}.
\end{split}
\eeq
Since $B-L$ and $\chi$ are conserved quantum numbers below the decoupling temperature, each of the number densities scales as $R^{-3}$ as universe expands.
The ratio
\beq\label{ratiochiBL}
\frac{\Delta \chi}{B-L}=\frac{{\Delta \chi}}{ B-L}\Big|_{\rm decoup.}=\frac{11}{79}(B-L)_\chi^{\rm sum},
\eeq
thus stays fixed. 

Even if at the decoupling there are more $\chi_i$ dark sector states, we assume that DM is composed only from one state, $\chi$. We therefore have for the ratio of baryon and dark matter energy densities
\beq\label{omegaBmoegachi}
\frac{\Omega_B}{\Omega_{\chi}}=\frac{m_p}{m_\chi}\frac{B}{B-L}\frac{B-L}{\Delta \chi}.
\eeq
The ratio of net $B$ and $B-L$ numbers  $B/(B-L)=28/79=0.354$ just above the electroweak phase transition \cite{Harvey:1990qw}. This remains essentially unchanged even if sphaleron and top mass effects are taken into account, in which case using results from \cite{Burnier:2005hp,D'Onofrio:2012ni} one has $B/(B-L)=0.349$ for both scalar and fermionic DM. Using $(B-L)/\Delta \chi=79/(11 (B-L)_\chi^{\rm sum})$ from \eqref{ratiochiBL}  finally leads to 
\beq
m_\chi=2.509 m_p \frac{\Omega_{\chi}}{\Omega_{B}} \frac{1}{(B-L)_\chi^{\rm sum}}=(12.5\pm0.8) {\rm GeV} \frac{1}{(B-L)_\chi^{\rm sum}},
\eeq
where in the last equality we used $\Omega_{\chi}=0.265\pm0.011$ and $\Omega_{B}=0.0499\pm0.0022$~\cite{Beringer:1900zz}. Note that the error is dominated by the experimental determination of DM and baryon densities. For instance, the difference between $B/(B-L)$ determination with and without sphaleron effects leads to a smaller shift in $m_\chi$ than the above quoted error.

We turn next to the case of additional fields in the visible sector. An example would be that SM gets completed to the MSSM. The relation between $Y, B-L$ and the constants $c_{Y, B-L}$  can be written in the matrix form
\beq
\begin{split}
\begin{pmatrix}
Y\\
B-L
\end{pmatrix}
&=
\frac{15}{4\pi^2 g_*T}\,\begin{pmatrix}
\sum_i \hat g_i Y_i^2 & \sum_i \hat g_i Y_i (B-L)_i\\
\sum_i \hat g_i Y_i (B-L)_i& \sum_i \hat g_i(B-L)_i^2 \
\end{pmatrix}
\cdot
\begin{pmatrix}
c_Y\\
c_{B-L}
\end{pmatrix}\\
&=\frac{15}{4\pi^2 g_*T}\,
\begin{pmatrix}
11+[Y^2]_{\rm NP} & 8 + [Y(B-L)]_{\rm NP}\\
8 + [Y(B-L)]_{\rm NP}& 13+  [(B-L)^2]_{\rm NP} \
\end{pmatrix}
\cdot
\begin{pmatrix}
c_Y\\
c_{B-L}
\end{pmatrix}.
\end{split}
\eeq
Here we defined
\beq
[Y^2]_{\rm NP} =\sum_i \hat g_i Y_i^2, \quad[Y(B-L)]_{\rm NP}=\sum_i \hat g_i Y_i (B-L)_i, \quad  [(B-L)^2]_{\rm NP}= \sum_i \hat g_i (B-L)_i^2,
\eeq
where the sums run over the new states only. The solution for $B-L$ in terms of $c_{B-L}$ is obtained by solving the above matrix equation setting $Y=0$, from which
\beq
B-L=\frac{15\,c_{B-L}}{4\pi^2 g_*T}\left(13+ [(B-L)^2]_{\rm NP}-\frac{(8+ [Y(B-L)]_{\rm NP})^2}{11+[Y^2]_{\rm NP}}\right).
\eeq
The net $\chi$ charge is still given by Eq. \eqref{ratiochiBL}, while the ratio $\Omega_B/\Omega_\chi$ is given by \eqref{omegaBmoegachi} with $(B-L)/\Delta\chi$ fixed at the decoupling temperature and $B/(B-L)$ at the electroweak phase transition. We thus have
\beq\label{omegaBmoegachiNP}
{m_\chi}={m_p}\frac{\Omega_\chi}{\Omega_{B}}\frac{B}{B-L}\left(13+ [(B-L)^2]_{\rm NP}-\frac{(8+ [Y(B-L)]_{\rm NP})^2}{(11+[Y^2]_{\rm NP})^2}\right)\frac{1}{(B-L)_\chi^{\rm sum}},
\eeq
where $B/(B-L)=0.349$ and $(B-L)_\chi^{\rm sum}$ given in \eqref{B-Lsum}. 

\section{Calculation of the DM decay time}
\label{app:adm-life}

Here we give further details of the DM lifetime calculation in the MFV and FN models for  $B=2$ DM, Sec. \ref{metastability}, while also varying the DM mass. The results are shown  in Fig.~\ref{fig:adm-life}. There are three different types of dimension 10 operators that can lead to DM decay, of schematic form $\chi (d^c d^c) (d^c d^c) (u^c u^c)$, $\chi (d^c d^c) (d^c u^c) (q^{*} q_{}^*) $, and $\chi (q^{*} q_{}^*) (q^{*} q_{}^*)(d^c d^c) $. For the same NP suppression scale $\Lambda$ the last type of operators gives the shortest lifetime. The dominant effective decay Lagrangian is thus, schematically,
\begin{equation}
{\cal L}_{\rm dec}^{(B=2)}\supset\frac{\cC}{\Lambda^6}\chi(q^*q^*)(q^*q^*)(d^cd^c),
\label{eq:b2-op}
\end{equation}
where $\cC$ is a flavor-dependent Wilson coefficient, the brackets enclose Lorentz contracted pairs, and summation over different flavor, color and weak isospin contractions is understood.  

In Sec. \ref{metastability} we included the SM Yukawa insertions in the definition of the operators. To unify the notation we instead use in this appendix the convention that the Wilson coefficient $\cC$ encodes all the flavor suppressions. The effective decay Lagrangian is thus, going to the mass basis, and displaying the flavor indices only,
\begin{equation}
{\cal L}_{\rm dec.}^{(B=2)}\supset\frac{\cC^\ssc{ijbcef}}{\Lambda^6}\chi\,u^*\fl{i}u^*\fl{j}d^*\fl{b}d^*\fl{e}d^c\fl{c}d^c\fl{f},
\end{equation}
where the flavor dependent Wilson coefficients are
\begin{align}
\cC^{\ssc{ijbcef}}_{(\ssc{mfv})}
&\simeq\ckm\fl{ia} [Y_d^\text{\scriptsize diag}]\fl{b}\ckm\fl{jd}\varepsilon^\ssc{abc}\varepsilon^\ssc{def},\label{eq:wilson-mfv} \\
\cC^{\ssc{ijbcef}}_{(\ssc{fn})}
&\simeq \lambda^{|-H(q\fl{I})-H(q\fl{J})-H(q\fl{b})-H(q\fl{e})+H(d^c\fl{c})+H(d^c\fl{f})|},\label{eq:wilson-sbhs}
\end{align}
The partial decay width for $\chi \to q q q q d d$ transition is then, using NDA, 
\begin{equation}
\Gamma_\chi\simeq \frac{\cC^2}{8\pi}\frac{1}{\left(16\pi^2\right)^4}\left(\frac{m_\chi}{\Lambda}\right)^{12}m_\chi.
\end{equation}
The factor $1/(8\pi)\times 1/(16\pi^2)^4$ results from integrating over the 6-body phase space. 

\begin{figure}[]\centering
\includegraphics[scale=0.8]{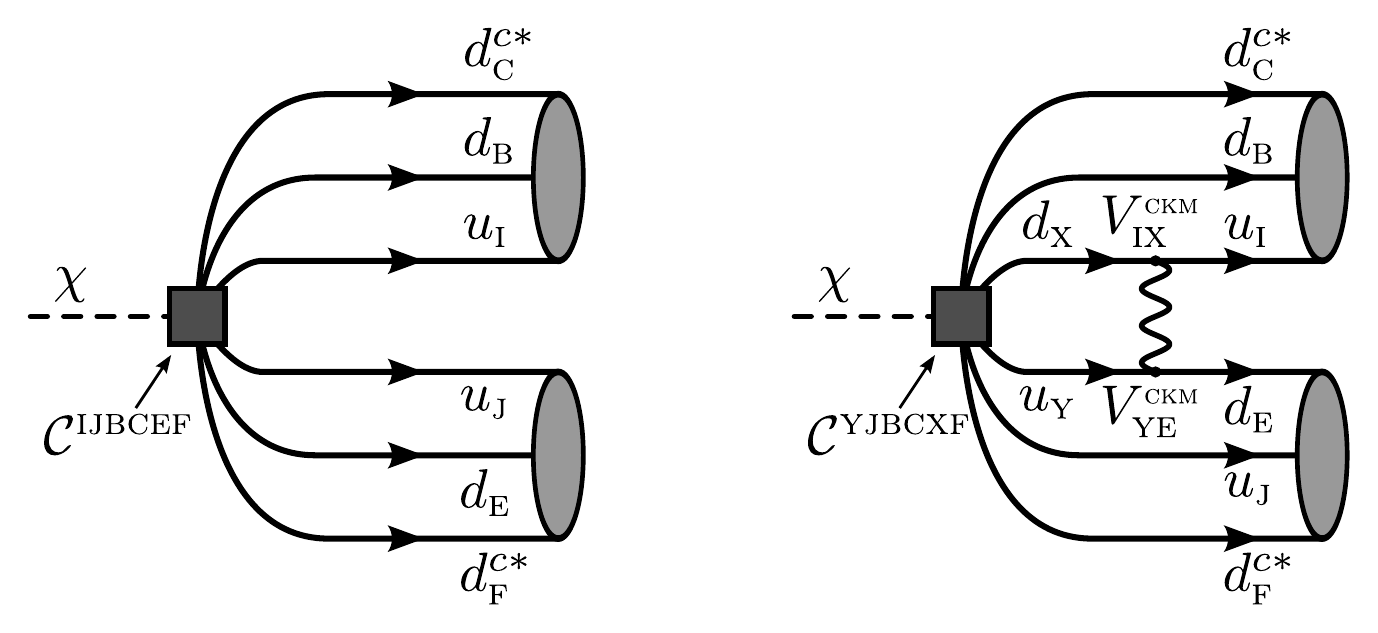}
\caption{Example Feynman diagrams for the decay of $B=2$ DM. The diagram on the left shows the tree level decay whereas the one on the right shows the loop-induced decay.}
\label{fig:b2-decay}
\end{figure}

For the MFV flavor breaking case there are several subtleties when calculating the decay width. For instance, the Levi-Civita tensor contractions lead to vanishing operators for some of the color and Lorentz contractions.
Another subtlety is that  the tree decay may be strongly CKM suppressed so that the leading decay amplitude is the 1-loop one, see Fig. \ref{fig:b2-decay}.  
The decay width can thus be estimated as
\begin{equation}
\Gamma(\chi\rightarrow u\fl{i}u\fl{j}d\fl{b}d\fl{e}d^{c*}\fl{c}d^{c*}\fl{f})\simeq
\max\begin{cases}
\frac{\left(\cC^\ssc{ijbcef}_{(\ssc{mfv})}\right)^2}{8\pi}\left(\frac{1}{16\pi^2}\right)^4\left(\frac{m_\chi}{\Lambda}\right)^{12}m_\chi,\\
\left(\frac{1}{16\pi^2}\ckm\fl{ix}\ckm\fl{ye}\right)^2\frac{\left(\cC^\ssc{yjbcxf}_{(\ssc{mfv})}\right)^2}{8\pi}\left(\frac{1}{16\pi^2}\right)^4\left(\frac{m_\chi}{\Lambda}\right)^{12}m_\chi.
\end{cases}\label{eq:decay-width}
\end{equation}
where the first (second) line gives the NDA estimates for the tree level (1-loop) dominated decay width.
The $W$ emitted from the left-handed quark lines coming from the effective decay vertex gives the additional CKM factors in the second line.

An example where the leading decay amplitude is generated at 1-loop is the $B=2$ ADM with $m_\chi=3.3$ GeV,  discussed in Sec. \ref{metastability}.
 Decays into final states with one charm, bottom or top quark are kinematically forbidden. For instance, the lightest $B=2$ final states with one valence charm quark are $\Lambda^+_c+\Sigma^-$ and $n+\Sigma_c^0$. The first has the valence quark content $\sim udc+dds$ and the rest mass $m_{\Lambda^+_c}+m_{\Sigma^-}=3.48$ GeV, while the valence quark content of the second is$\sim udd +ddc$ and its rest mass $m_{n}+m_{\Sigma^0_c}=3.4$ GeV.  In contrast, the decays to $\Xi^0$ ($\sim uds$) or $\Lambda^0$ ($\sim uds$) baryons are allowed. 
 Eq.~\eqref{eq:decay-width} gives
\begin{align}
\Gamma^\ssc{mfv}_\ssc{tree}\left(\chi\rightarrow\Lambda^0\Lambda^0\right)&\simeq\left(y_s V_{ub}^2\right)^2\frac{1}{8\pi}\left(\frac{1}{16\pi^2}\right)^4\left(\frac{m_\chi}{\Lambda}\right)^{12}m_\chi,\\
\Gamma^\ssc{mfv}_\ssc{loop}\left(\chi\rightarrow\Lambda^0\Lambda^0\right)&\simeq\left(\frac{y_b V_{ub}V_{ts}}{16\pi^2}\right)^2\frac{1}{8\pi}\left(\frac{1}{16\pi^2}\right)^4\left(\frac{m_\chi}{\Lambda}\right)^{12}m_\chi,
\end{align}
with the same estimate, within our precision, for the $\chi \to \Xi^0, \Xi^0$ or $\chi \to \Lambda^0, \Lambda^0$ decays. Note that in the 1-loop amplitude the partonic transition at the decay vertex, $\chi \to udb+tds$, carries no CKM suppression. Furthermore, the $y_s$ Yukawa insertion in the tree level amplitude is replaced by $y_b$. The $b$ and $t$ quark lines then convert to $u$ and $s$ quark lines via $W$ exchange, as shown in Fig. \ref{fig:b2-decay}. The smaller CKM and Yukawa suppressions compensate the loop factor so that the 1-loop amplitude dominates, with the NDA estimate  $\Gamma^\ssc{mfv}_\ssc{loop}/\Gamma^\ssc{mfv}_\ssc{tree}\sim {\mathcal O}(10)$.

This procedure can be repeated for different DM masses, arriving at the dominant decay modes as a function of $m_\chi$. The results are listed in Table~\ref{tab:mfv-pdw}, where we give the kinematical thresholds (1st column) for a number of decay channels (4th column), along with the corresponding partonic transitions (3rd column) and the decay vertex transitions (2nd column). The latter two differ for the loop processes, cf. Fig. \ref{fig:b2-decay}. The total decay width for given $m_\chi$ is then the sum of partial decay widths, $\Gamma_i$, (5th column) for the decay channels that are kinematically allowed. For convenience we also give the decay times, $\tau_i$, (6th column) that correspond to individual partial decay widths. Note that in the calculation of the partial decay widths we neglect the phase space suppression, while the quoted $\Gamma_i$ in Table~\ref{tab:mfv-pdw} are obtained from the NDA estimates \eqref{eq:decay-width} with $m_\chi$ at the kinematical threshold, and setting $\Lambda=1$TeV.

\begin{table}[t!]
\begin{tabular}{ccccccc}\hline\hline
Thr. [GeV]~&~Decay vertex~ &~Partonic~& ~Final State~ & ~~~~$\Gamma_i$ [GeV]~~~~ &~~~~$\tau_i$ [s]~~~~& ~~Process\\\hline
2.06 & $\chi\rightarrow cudbsd $ & $\chi\rightarrow udd\,uds$ & $n+\Lambda^0$ & $1.34\times 10^{-60}$ & $4.91\times 10^{35}$ & Loop\\
2.23 & $\chi\rightarrow uusdsd $ & $\chi\rightarrow uds\,uds$ & $\Lambda^0+\Lambda^0$ & $3.74\times 10^{-55}$ & $1.76\times 10^{30}$ & Tree\\
2.43 & $\chi\rightarrow cusbsd $ & $\chi\rightarrow uds\,uss$ & $\Lambda^0+\Xi^0$ & $4.25\times 10^{-57}$ & $1.55\times 10^{32}$ & Loop\\
3.48 & $\chi\rightarrow ccdssd $ & $\chi\rightarrow udc\,dds$ & $\Lambda^+_c+\Sigma^-$ & $1.23\times 10^{-55}$ & $5.33\times 10^{30}$ & Loop\\
3.61 & $\chi\rightarrow ucsdsd $ & $\chi\rightarrow udc\,dss$ & $\Lambda^+_c+\Xi^-$ & $1.87\times 10^{-50}$ & $3.52\times 10^{25}$ & Tree\\
3.81 & $\chi\rightarrow ccsdsd $ & $\chi\rightarrow uds\,ssc$ & $\Lambda^0+\Omega^0_c$ & $1.42\times 10^{-52}$ & $4.62\times 10^{27}$ & Loop\\
3.79 & $\chi\rightarrow ccsdsd $ & $\chi\rightarrow usc\,dss$ & $\Xi^+_c+\Xi^-$ & $1.33\times 10^{-52}$ & $4.96\times 10^{27}$ & Loop\\
4.63 & $\chi\rightarrow ccdbsd $ & $\chi\rightarrow dcc\,dds$ & $\Xi^+_{cc}+\Sigma^-$ & $1.01\times 10^{-52}$ & $6.51\times 10^{27}$ & Loop\\
4.93 & $\chi\rightarrow ccsdsd $ & $\chi\rightarrow ddc\,dsc$ & $\Sigma^0_c+\Xi^0_c$ & $1.04\times 10^{-46}$ & $6.33\times 10^{21}$ & Tree\\
5.17 & $\chi\rightarrow ccsbsd $ & $\chi\rightarrow dsc\,ssc$ & $\Xi^0_c+\Omega^0_c$ & $4.14\times 10^{-52}$ & $1.59\times 10^{27}$ & Loop\\
6.56 & $\chi\rightarrow cudsbd $ & $\chi\rightarrow udd\,udb$ & $n+\Lambda^0_b$ & $7.25\times 10^{-54}$ & $9.08\times 10^{28}$ & Loop\\
6.73 & $\chi\rightarrow uudbsd $ & $\chi\rightarrow udd\,usb$ & $n+\Xi^0_b$ & $1.87\times 10^{-45}$ & $3.53\times 10^{20}$ & Tree\\
6.94 & $\chi\rightarrow uusbsd $ & $\chi\rightarrow uss\,udb$ & $\Xi^0+\Lambda^0_b $ & $5.18\times 10^{-44}$ & $1.27\times 10^{19}$ & Tree\\
7.10 & $\chi\rightarrow cusdbs $ & $\chi\rightarrow uss\,usb$ & $\Xi^0+\Xi^0_b$ & $3.49\times 10^{-52}$ & $1.89\times 10^{27}$ & Loop\\
8.07 & $\chi\rightarrow ucbdbd $ & $\chi\rightarrow udb\,ddc$ & $\Lambda^0_b+\Sigma^0_c$ & $2.16\times 10^{-51}$ & $3.05\times 10^{26}$ & Loop\\
8.09 & $\chi\rightarrow ucbdsd $ & $\chi\rightarrow udb\,dsc$ & $\Lambda^0_b+\Xi^0_c$ & $2.00\times 10^{-42}$ & $3.30\times 10^{17}$ & Tree\\
8.31 & $\chi\rightarrow ucbssd $ & $\chi\rightarrow udb\,ssc$ & $\Lambda^0_b+\Omega^0_c$ & $5.33\times 10^{-41}$ & $1.24\times 10^{16}$ & Tree\\
8.48 & $\chi\rightarrow ucbsbs $ & $\chi\rightarrow usb\,ssc$ & $\Xi^0_b+\Omega^0_c$ & $7.66\times 10^{-50}$ & $8.59\times 10^{24}$ & Loop\\
11.24 & $\chi\rightarrow uubdbd $ & $\chi\rightarrow udb\,udb$ & $\Lambda^0_b+\Lambda^0_b$ & $2.24\times 10^{-42}$ & $2.93\times 10^{17}$ & Tree\\
11.41 & $\chi\rightarrow uubbsd $ & $\chi\rightarrow udb\,usb$ & $\Lambda^0_b+\Xi^0_b$ & $9.86\times 10^{-38}$ & $6.68\times 10^{12}$ & Tree\\
11.58 & $\chi\rightarrow uubsbs $ & $\chi\rightarrow usb\,usb$ & $\Xi^0_b+\Xi^0_b$ & $2.94\times 10^{-42}$ & $2.24\times 10^{17}$ & Tree\\
\hline\hline
\end{tabular}
\caption{Partial decay widths, $\Gamma_i$, and related decay times, $\tau_i=1/\Gamma_i$, for representative decay channels above kinematical thresholds (1st column) assuming the MFV flavor breaking ansatz. The EFT scale is set to $\Lambda=1$ TeV. The last column denotes whether the dominant amplitude is tree level or 1-loop, while the 2nd and the 3rd columns give the decay vertex transition and the partonic transition after the potential $W$ exchange, respectively.}
\label{tab:mfv-pdw}
\end{table}

In the case of FN flavor breaking the leading tree level and loop induced decay widths for $B=2$, $m_\chi=3.3$ GeV DM have the NDA estimates of
\begin{align}
\Gamma^\ssc{fn}_\ssc{tree}\left(\chi\rightarrow\Lambda^0\Lambda^0\right)&\simeq \left(\lambda^2\right)^2\frac{1}{8\pi}\left(\frac{1}{16\pi^2}\right)^4\left(\frac{m_\chi}{\Lambda}\right)^{12}m_\chi,\\
\Gamma^\ssc{fn}_\ssc{loop}\left(\chi\rightarrow\Lambda^0\Lambda^0\right)&\simeq \left(\frac{\lambda^4}{16\pi^2}\right)^2\frac{1}{8\pi}\left(\frac{1}{16\pi^2}\right)^4\left(\frac{m_\chi}{\Lambda}\right)^{12}m_\chi.
\end{align}
In this case the tree level decay dominates over the loop induced decay by four orders of magnitude. The dominance of the tree level decay amplitude over the 1-loop decay amplitude holds also, if the DM mass is varied. This can be traced to the following 
difference between the MFV and FN ans\"{a}tze. In the MFV case the Levi-Civita tensors enforce that two quark flavors in the effective decay vertex need to be from the third generation. This can be changed either by using the $V_{\rm CKM}$ misalignment or through a loop transition. In FN flavor structure ansatz, on the other hand, the flavor indices need not be antisymmetric. 

\section{Loop functions in neutral meson mixing}
\label{app:meson-mixing}
Here we list the analytical form of the loop functions $F(x), F^F(x_1,x_2), G(x_1,x_2), G^F(x_1,x_2,_x3)$ and $H(x)$, $H^F(x_1,x_2)$ that appear in the 1-loop expressions for the Wilson coefficients in the neutral meson mixing, Sec. \ref{sec:flavor}.
The mediator loop functions  with mass degenerate quarks in the loop are given by
\beq
F(x)=x\,H(x), \qquad H(x)=\frac{1}{(1-x)^3} \left[1-x^2+2 x \log (x)\right],
\eeq
while for two different quarks running in the loop they are
\begin{align}
F^F(x_i,x_j)&=\frac{x_i x_j}{\left(1-x_i\right) \left(1-x_j\right)}
	+\left[\frac{x_i x_j\log\left(x_i\right)}{\left(1-x_i\right)^2\left(x_i-x_j\right)}
	+x_i\leftrightarrow x_j\right],
\\
H^F(x_i,x_j)&=\frac{1}{\left(1-x_i\right) \left(1-x_j\right)}+\left[\frac{x_i^2 \log
   \left(x_i\right)}{\left(1-x_i\right)^2 \left(x_i-x_j\right)}+x_i\leftrightarrow x_j\right].
\end{align}
The loop functions for the mediator-$W$ loops are
\begin{align}
\begin{split}
G(x_w,x)&=\frac{x}{x_w}\Bigg[\frac{x_w+x}{(x-1) \left(x-x_w\right)}-\frac{x \left(2 x x_w-x_w^2-2 x_w+x^2\right) \log (x)}{(x-1)^2
   \left(x-x_w\right){}^2}\\
   &\qquad\qquad\qquad+\frac{2 x x_w \log\left(x_w\right)}{\left(x-x_w\right){}^2 \left(x_w-1\right)}\Bigg],
   \end{split}
   \\
   \begin{split}
G^F(x_w,x_i,x_j)&=\frac{x_ix_j}{x_w}\Bigg[\left\{
	\frac{\left(x_i+x_w\right) \log \left(x_i\right)}{\left(1-x_i\right)
   \left(x_i-x_j\right) \left(x_w-x_i\right)}+x_i\leftrightarrow x_j\right\}\\
   &\qquad\qquad\qquad-\frac{2 x_w\log \left(x_w\right)}{\left(1-x_w\right) \left(x_w-x_i\right) \left(x_w-x_j\right)}\Bigg].
\end{split}
\end{align}

\bibliographystyle{apsrev4-1}

\begin{thebibliography}{82}%
\makeatletter
\providecommand \@ifxundefined [1]{%
 \@ifx{#1\undefined}
}%
\providecommand \@ifnum [1]{%
 \ifnum #1\expandafter \@firstoftwo
 \else \expandafter \@secondoftwo
 \fi
}%
\providecommand \@ifx [1]{%
 \ifx #1\expandafter \@firstoftwo
 \else \expandafter \@secondoftwo
 \fi
}%
\providecommand \natexlab [1]{#1}%
\providecommand \enquote  [1]{``#1''}%
\providecommand \bibnamefont  [1]{#1}%
\providecommand \bibfnamefont [1]{#1}%
\providecommand \citenamefont [1]{#1}%
\providecommand \href@noop [0]{\@secondoftwo}%
\providecommand \href [0]{\begingroup \@sanitize@url \@href}%
\providecommand \@href[1]{\@@startlink{#1}\@@href}%
\providecommand \@@href[1]{\endgroup#1\@@endlink}%
\providecommand \@sanitize@url [0]{\catcode `\\12\catcode `\$12\catcode
  `\&12\catcode `\#12\catcode `\^12\catcode `\_12\catcode `\%12\relax}%
\providecommand \@@startlink[1]{}%
\providecommand \@@endlink[0]{}%
\providecommand \url  [0]{\begingroup\@sanitize@url \@url }%
\providecommand \@url [1]{\endgroup\@href {#1}{\urlprefix }}%
\providecommand \urlprefix  [0]{URL }%
\providecommand \Eprint [0]{\href }%
\providecommand \doibase [0]{http://dx.doi.org/}%
\providecommand \selectlanguage [0]{\@gobble}%
\providecommand \bibinfo  [0]{\@secondoftwo}%
\providecommand \bibfield  [0]{\@secondoftwo}%
\providecommand \translation [1]{[#1]}%
\providecommand \BibitemOpen [0]{}%
\providecommand \bibitemStop [0]{}%
\providecommand \bibitemNoStop [0]{.\EOS\space}%
\providecommand \EOS [0]{\spacefactor3000\relax}%
\providecommand \BibitemShut  [1]{\csname bibitem#1\endcsname}%
\let\auto@bib@innerbib\@empty
\bibitem [{\citenamefont {Hambye}(2011)}]{Hambye2011}%
  \BibitemOpen
  \bibfield  {author} {\bibinfo {author} {\bibfnamefont {T.}~\bibnamefont
  {Hambye}},\ }\href@noop {} {\bibfield  {journal} {\bibinfo  {journal} {PoS}\
  }\textbf {\bibinfo {volume} {IDM2010}},\ \bibinfo {pages} {098} (\bibinfo
  {year} {2011})},\ \Eprint {http://arxiv.org/abs/1012.4587} {arXiv:1012.4587
  [hep-ph]} \BibitemShut {NoStop}%
\bibitem [{\citenamefont {Ackerman}\ \emph {et~al.}(2009)\citenamefont
  {Ackerman}, \citenamefont {Buckley}, \citenamefont {Carroll},\ and\
  \citenamefont {Kamionkowski}}]{Ackerman:mha}%
  \BibitemOpen
  \bibfield  {author} {\bibinfo {author} {\bibfnamefont {L.}~\bibnamefont
  {Ackerman}}, \bibinfo {author} {\bibfnamefont {M.~R.}\ \bibnamefont
  {Buckley}}, \bibinfo {author} {\bibfnamefont {S.~M.}\ \bibnamefont
  {Carroll}}, \ and\ \bibinfo {author} {\bibfnamefont {M.}~\bibnamefont
  {Kamionkowski}},\ }\href {\doibase 10.1103/PhysRevD.79.023519,
  10.1142/9789814293792_0021} {\bibfield  {journal} {\bibinfo  {journal}
  {Phys.Rev.}\ }\textbf {\bibinfo {volume} {D79}},\ \bibinfo {pages} {023519}
  (\bibinfo {year} {2009})},\ \Eprint {http://arxiv.org/abs/0810.5126}
  {arXiv:0810.5126 [hep-ph]} \BibitemShut {NoStop}%
\bibitem [{\citenamefont {Feng}\ \emph {et~al.}(2008)\citenamefont {Feng},
  \citenamefont {Tu},\ and\ \citenamefont {Yu}}]{Feng:2008mu}%
  \BibitemOpen
  \bibfield  {author} {\bibinfo {author} {\bibfnamefont {J.~L.}\ \bibnamefont
  {Feng}}, \bibinfo {author} {\bibfnamefont {H.}~\bibnamefont {Tu}}, \ and\
  \bibinfo {author} {\bibfnamefont {H.-B.}\ \bibnamefont {Yu}},\ }\href
  {\doibase 10.1088/1475-7516/2008/10/043} {\bibfield  {journal} {\bibinfo
  {journal} {JCAP}\ }\textbf {\bibinfo {volume} {0810}},\ \bibinfo {pages}
  {043} (\bibinfo {year} {2008})},\ \Eprint {http://arxiv.org/abs/0808.2318}
  {arXiv:0808.2318 [hep-ph]} \BibitemShut {NoStop}%
\bibitem [{\citenamefont {Feng}\ \emph {et~al.}(2009)\citenamefont {Feng},
  \citenamefont {Kaplinghat}, \citenamefont {Tu},\ and\ \citenamefont
  {Yu}}]{Feng:2009mn}%
  \BibitemOpen
  \bibfield  {author} {\bibinfo {author} {\bibfnamefont {J.~L.}\ \bibnamefont
  {Feng}}, \bibinfo {author} {\bibfnamefont {M.}~\bibnamefont {Kaplinghat}},
  \bibinfo {author} {\bibfnamefont {H.}~\bibnamefont {Tu}}, \ and\ \bibinfo
  {author} {\bibfnamefont {H.-B.}\ \bibnamefont {Yu}},\ }\href {\doibase
  10.1088/1475-7516/2009/07/004} {\bibfield  {journal} {\bibinfo  {journal}
  {JCAP}\ }\textbf {\bibinfo {volume} {0907}},\ \bibinfo {pages} {004}
  (\bibinfo {year} {2009})},\ \Eprint {http://arxiv.org/abs/0905.3039}
  {arXiv:0905.3039 [hep-ph]} \BibitemShut {NoStop}%
\bibitem [{\citenamefont {Farrar}\ and\ \citenamefont
  {Fayet}(1978)}]{Farrar:1978xj}%
  \BibitemOpen
  \bibfield  {author} {\bibinfo {author} {\bibfnamefont {G.~R.}\ \bibnamefont
  {Farrar}}\ and\ \bibinfo {author} {\bibfnamefont {P.}~\bibnamefont {Fayet}},\
  }\href {\doibase 10.1016/0370-2693(78)90858-4} {\bibfield  {journal}
  {\bibinfo  {journal} {Phys.Lett.}\ }\textbf {\bibinfo {volume} {B76}},\
  \bibinfo {pages} {575} (\bibinfo {year} {1978})}\BibitemShut {NoStop}%
\bibitem [{\citenamefont {Dimopoulos}\ \emph {et~al.}(1982)\citenamefont
  {Dimopoulos}, \citenamefont {Raby},\ and\ \citenamefont
  {Wilczek}}]{Dimopoulos:1981dw}%
  \BibitemOpen
  \bibfield  {author} {\bibinfo {author} {\bibfnamefont {S.}~\bibnamefont
  {Dimopoulos}}, \bibinfo {author} {\bibfnamefont {S.}~\bibnamefont {Raby}}, \
  and\ \bibinfo {author} {\bibfnamefont {F.}~\bibnamefont {Wilczek}},\ }\href
  {\doibase 10.1016/0370-2693(82)90313-6} {\bibfield  {journal} {\bibinfo
  {journal} {Phys.Lett.}\ }\textbf {\bibinfo {volume} {B112}},\ \bibinfo
  {pages} {133} (\bibinfo {year} {1982})}\BibitemShut {NoStop}%
\bibitem [{\citenamefont {Farrar}\ and\ \citenamefont
  {Weinberg}(1983)}]{Farrar:1982te}%
  \BibitemOpen
  \bibfield  {author} {\bibinfo {author} {\bibfnamefont {G.~R.}\ \bibnamefont
  {Farrar}}\ and\ \bibinfo {author} {\bibfnamefont {S.}~\bibnamefont
  {Weinberg}},\ }\href {\doibase 10.1103/PhysRevD.27.2732} {\bibfield
  {journal} {\bibinfo  {journal} {Phys.Rev.}\ }\textbf {\bibinfo {volume}
  {D27}},\ \bibinfo {pages} {2732} (\bibinfo {year} {1983})}\BibitemShut
  {NoStop}%
\bibitem [{\citenamefont {Kadastik}\ \emph {et~al.}(2010)\citenamefont
  {Kadastik}, \citenamefont {Kannike},\ and\ \citenamefont
  {Raidal}}]{Kadastik:2009dj}%
  \BibitemOpen
  \bibfield  {author} {\bibinfo {author} {\bibfnamefont {M.}~\bibnamefont
  {Kadastik}}, \bibinfo {author} {\bibfnamefont {K.}~\bibnamefont {Kannike}}, \
  and\ \bibinfo {author} {\bibfnamefont {M.}~\bibnamefont {Raidal}},\ }\href
  {\doibase 10.1103/PhysRevD.81.015002} {\bibfield  {journal} {\bibinfo
  {journal} {Phys.Rev.}\ }\textbf {\bibinfo {volume} {D81}},\ \bibinfo {pages}
  {015002} (\bibinfo {year} {2010})},\ \Eprint {http://arxiv.org/abs/0903.2475}
  {arXiv:0903.2475 [hep-ph]} \BibitemShut {NoStop}%
\bibitem [{\citenamefont {Kadastik}\ \emph {et~al.}(2009)\citenamefont
  {Kadastik}, \citenamefont {Kannike},\ and\ \citenamefont
  {Raidal}}]{Kadastik:2009cu}%
  \BibitemOpen
  \bibfield  {author} {\bibinfo {author} {\bibfnamefont {M.}~\bibnamefont
  {Kadastik}}, \bibinfo {author} {\bibfnamefont {K.}~\bibnamefont {Kannike}}, \
  and\ \bibinfo {author} {\bibfnamefont {M.}~\bibnamefont {Raidal}},\ }\href
  {\doibase 10.1103/PhysRevD.80.085020, 10.1103/PhysRevD.81.029903} {\bibfield
  {journal} {\bibinfo  {journal} {Phys.Rev.}\ }\textbf {\bibinfo {volume}
  {D80}},\ \bibinfo {pages} {085020} (\bibinfo {year} {2009})},\ \Eprint
  {http://arxiv.org/abs/0907.1894} {arXiv:0907.1894 [hep-ph]} \BibitemShut
  {NoStop}%
\bibitem [{\citenamefont {Frigerio}\ and\ \citenamefont
  {Hambye}(2010)}]{Frigerio:2009wf}%
  \BibitemOpen
  \bibfield  {author} {\bibinfo {author} {\bibfnamefont {M.}~\bibnamefont
  {Frigerio}}\ and\ \bibinfo {author} {\bibfnamefont {T.}~\bibnamefont
  {Hambye}},\ }\href {\doibase 10.1103/PhysRevD.81.075002} {\bibfield
  {journal} {\bibinfo  {journal} {Phys.Rev.}\ }\textbf {\bibinfo {volume}
  {D81}},\ \bibinfo {pages} {075002} (\bibinfo {year} {2010})},\ \Eprint
  {http://arxiv.org/abs/0912.1545} {arXiv:0912.1545 [hep-ph]} \BibitemShut
  {NoStop}%
\bibitem [{\citenamefont {Cirelli}\ and\ \citenamefont
  {Strumia}(2009)}]{Cirelli:2009uv}%
  \BibitemOpen
  \bibfield  {author} {\bibinfo {author} {\bibfnamefont {M.}~\bibnamefont
  {Cirelli}}\ and\ \bibinfo {author} {\bibfnamefont {A.}~\bibnamefont
  {Strumia}},\ }\href {\doibase 10.1088/1367-2630/11/10/105005} {\bibfield
  {journal} {\bibinfo  {journal} {New J.Phys.}\ }\textbf {\bibinfo {volume}
  {11}},\ \bibinfo {pages} {105005} (\bibinfo {year} {2009})},\ \Eprint
  {http://arxiv.org/abs/0903.3381} {arXiv:0903.3381 [hep-ph]} \BibitemShut
  {NoStop}%
\bibitem [{\citenamefont {Cirelli}\ \emph {et~al.}(2006)\citenamefont
  {Cirelli}, \citenamefont {Fornengo},\ and\ \citenamefont
  {Strumia}}]{Cirelli:2005uq}%
  \BibitemOpen
  \bibfield  {author} {\bibinfo {author} {\bibfnamefont {M.}~\bibnamefont
  {Cirelli}}, \bibinfo {author} {\bibfnamefont {N.}~\bibnamefont {Fornengo}}, \
  and\ \bibinfo {author} {\bibfnamefont {A.}~\bibnamefont {Strumia}},\ }\href
  {\doibase 10.1016/j.nuclphysb.2006.07.012} {\bibfield  {journal} {\bibinfo
  {journal} {Nucl.Phys.}\ }\textbf {\bibinfo {volume} {B753}},\ \bibinfo
  {pages} {178} (\bibinfo {year} {2006})},\ \Eprint
  {http://arxiv.org/abs/hep-ph/0512090} {arXiv:hep-ph/0512090 [hep-ph]}
  \BibitemShut {NoStop}%
\bibitem [{\citenamefont {Hambye}(2009)}]{Hambye:2008bq}%
  \BibitemOpen
  \bibfield  {author} {\bibinfo {author} {\bibfnamefont {T.}~\bibnamefont
  {Hambye}},\ }\href {\doibase 10.1088/1126-6708/2009/01/028} {\bibfield
  {journal} {\bibinfo  {journal} {JHEP}\ }\textbf {\bibinfo {volume} {0901}},\
  \bibinfo {pages} {028} (\bibinfo {year} {2009})},\ \Eprint
  {http://arxiv.org/abs/0811.0172} {arXiv:0811.0172 [hep-ph]} \BibitemShut
  {NoStop}%
\bibitem [{\citenamefont {Bai}\ and\ \citenamefont {Hill}(2010)}]{Bai:2010qg}%
  \BibitemOpen
  \bibfield  {author} {\bibinfo {author} {\bibfnamefont {Y.}~\bibnamefont
  {Bai}}\ and\ \bibinfo {author} {\bibfnamefont {R.~J.}\ \bibnamefont {Hill}},\
  }\href {\doibase 10.1103/PhysRevD.82.111701} {\bibfield  {journal} {\bibinfo
  {journal} {Phys.Rev.}\ }\textbf {\bibinfo {volume} {D82}},\ \bibinfo {pages}
  {111701} (\bibinfo {year} {2010})},\ \Eprint {http://arxiv.org/abs/1005.0008}
  {arXiv:1005.0008 [hep-ph]} \BibitemShut {NoStop}%
\bibitem [{\citenamefont {Kaplan}\ \emph {et~al.}(2009)\citenamefont {Kaplan},
  \citenamefont {Luty},\ and\ \citenamefont {Zurek}}]{Kaplan:2009ag}%
  \BibitemOpen
  \bibfield  {author} {\bibinfo {author} {\bibfnamefont {D.~E.}\ \bibnamefont
  {Kaplan}}, \bibinfo {author} {\bibfnamefont {M.~A.}\ \bibnamefont {Luty}}, \
  and\ \bibinfo {author} {\bibfnamefont {K.~M.}\ \bibnamefont {Zurek}},\ }\href
  {\doibase 10.1103/PhysRevD.79.115016} {\bibfield  {journal} {\bibinfo
  {journal} {Phys.Rev.}\ }\textbf {\bibinfo {volume} {D79}},\ \bibinfo {pages}
  {115016} (\bibinfo {year} {2009})},\ \Eprint {http://arxiv.org/abs/0901.4117}
  {arXiv:0901.4117 [hep-ph]} \BibitemShut {NoStop}%
\bibitem [{\citenamefont {Zurek}(2014)}]{Zurek:2013wia}%
  \BibitemOpen
  \bibfield  {author} {\bibinfo {author} {\bibfnamefont {K.~M.}\ \bibnamefont
  {Zurek}},\ }\href {\doibase 10.1016/j.physrep.2013.12.001} {\bibfield
  {journal} {\bibinfo  {journal} {Phys.Rept.}\ }\textbf {\bibinfo {volume}
  {537}},\ \bibinfo {pages} {91} (\bibinfo {year} {2014})},\ \Eprint
  {http://arxiv.org/abs/1308.0338} {arXiv:1308.0338 [hep-ph]} \BibitemShut
  {NoStop}%
\bibitem [{\citenamefont {Batell}\ \emph {et~al.}(2011)\citenamefont {Batell},
  \citenamefont {Pradler},\ and\ \citenamefont {Spannowsky}}]{Batell2011}%
  \BibitemOpen
  \bibfield  {author} {\bibinfo {author} {\bibfnamefont {B.}~\bibnamefont
  {Batell}}, \bibinfo {author} {\bibfnamefont {J.}~\bibnamefont {Pradler}}, \
  and\ \bibinfo {author} {\bibfnamefont {M.}~\bibnamefont {Spannowsky}},\
  }\href {\doibase 10.1007/JHEP08(2011)038} {\bibfield  {journal} {\bibinfo
  {journal} {JHEP}\ }\textbf {\bibinfo {volume} {1108}},\ \bibinfo {pages}
  {038} (\bibinfo {year} {2011})},\ \Eprint {http://arxiv.org/abs/1105.1781}
  {arXiv:1105.1781 [hep-ph]} \BibitemShut {NoStop}%
\bibitem [{\citenamefont {Kile}\ and\ \citenamefont {Soni}(2011)}]{Kile2011}%
  \BibitemOpen
  \bibfield  {author} {\bibinfo {author} {\bibfnamefont {J.}~\bibnamefont
  {Kile}}\ and\ \bibinfo {author} {\bibfnamefont {A.}~\bibnamefont {Soni}},\
  }\href {\doibase 10.1103/PhysRevD.84.035016} {\bibfield  {journal} {\bibinfo
  {journal} {Phys.Rev.}\ }\textbf {\bibinfo {volume} {D84}},\ \bibinfo {pages}
  {035016} (\bibinfo {year} {2011})},\ \Eprint {http://arxiv.org/abs/1104.5239}
  {arXiv:1104.5239 [hep-ph]} \BibitemShut {NoStop}%
\bibitem [{\citenamefont {Agrawal}\ \emph
  {et~al.}(2012{\natexlab{a}})\citenamefont {Agrawal}, \citenamefont
  {Blanchet}, \citenamefont {Chacko},\ and\ \citenamefont
  {Kilic}}]{Agrawal2012}%
  \BibitemOpen
  \bibfield  {author} {\bibinfo {author} {\bibfnamefont {P.}~\bibnamefont
  {Agrawal}}, \bibinfo {author} {\bibfnamefont {S.}~\bibnamefont {Blanchet}},
  \bibinfo {author} {\bibfnamefont {Z.}~\bibnamefont {Chacko}}, \ and\ \bibinfo
  {author} {\bibfnamefont {C.}~\bibnamefont {Kilic}},\ }\href {\doibase
  10.1103/PhysRevD.86.055002} {\bibfield  {journal} {\bibinfo  {journal}
  {Phys.Rev.}\ }\textbf {\bibinfo {volume} {D86}},\ \bibinfo {pages} {055002}
  (\bibinfo {year} {2012}{\natexlab{a}})},\ \Eprint
  {http://arxiv.org/abs/1109.3516} {arXiv:1109.3516 [hep-ph]} \BibitemShut
  {NoStop}%
\bibitem [{\citenamefont {Davoudiasl}\ \emph {et~al.}(2011)\citenamefont
  {Davoudiasl}, \citenamefont {Morrissey}, \citenamefont {Sigurdson},\ and\
  \citenamefont {Tulin}}]{Davoudiasl2011}%
  \BibitemOpen
  \bibfield  {author} {\bibinfo {author} {\bibfnamefont {H.}~\bibnamefont
  {Davoudiasl}}, \bibinfo {author} {\bibfnamefont {D.~E.}\ \bibnamefont
  {Morrissey}}, \bibinfo {author} {\bibfnamefont {K.}~\bibnamefont
  {Sigurdson}}, \ and\ \bibinfo {author} {\bibfnamefont {S.}~\bibnamefont
  {Tulin}},\ }\href {\doibase 10.1103/PhysRevD.84.096008} {\bibfield  {journal}
  {\bibinfo  {journal} {Phys.Rev.}\ }\textbf {\bibinfo {volume} {D84}},\
  \bibinfo {pages} {096008} (\bibinfo {year} {2011})},\ \Eprint
  {http://arxiv.org/abs/1106.4320} {arXiv:1106.4320 [hep-ph]} \BibitemShut
  {NoStop}%
\bibitem [{\citenamefont {Lopez-Honorez}\ and\ \citenamefont
  {Merlo}(2013)}]{Lopez-Honorez:2013wla}%
  \BibitemOpen
  \bibfield  {author} {\bibinfo {author} {\bibfnamefont {L.}~\bibnamefont
  {Lopez-Honorez}}\ and\ \bibinfo {author} {\bibfnamefont {L.}~\bibnamefont
  {Merlo}},\ }\href {\doibase 10.1016/j.physletb.2013.04.015} {\bibfield
  {journal} {\bibinfo  {journal} {Phys.Lett.}\ }\textbf {\bibinfo {volume}
  {B722}},\ \bibinfo {pages} {135} (\bibinfo {year} {2013})},\ \Eprint
  {http://arxiv.org/abs/1303.1087} {arXiv:1303.1087 [hep-ph]} \BibitemShut
  {NoStop}%
\bibitem [{\citenamefont {Agrawal}\ \emph
  {et~al.}(2014{\natexlab{a}})\citenamefont {Agrawal}, \citenamefont {Blanke},\
  and\ \citenamefont {Gemmler}}]{Agrawal:2014aoa}%
  \BibitemOpen
  \bibfield  {author} {\bibinfo {author} {\bibfnamefont {P.}~\bibnamefont
  {Agrawal}}, \bibinfo {author} {\bibfnamefont {M.}~\bibnamefont {Blanke}}, \
  and\ \bibinfo {author} {\bibfnamefont {K.}~\bibnamefont {Gemmler}},\
  }\href@noop {} {\  (\bibinfo {year} {2014}{\natexlab{a}})},\ \Eprint
  {http://arxiv.org/abs/1405.6709} {arXiv:1405.6709 [hep-ph]} \BibitemShut
  {NoStop}%
\bibitem [{\citenamefont {Batell}\ \emph {et~al.}(2014)\citenamefont {Batell},
  \citenamefont {Lin},\ and\ \citenamefont {Wang}}]{Batell:2013zwa}%
  \BibitemOpen
  \bibfield  {author} {\bibinfo {author} {\bibfnamefont {B.}~\bibnamefont
  {Batell}}, \bibinfo {author} {\bibfnamefont {T.}~\bibnamefont {Lin}}, \ and\
  \bibinfo {author} {\bibfnamefont {L.-T.}\ \bibnamefont {Wang}},\ }\href
  {\doibase 10.1007/JHEP01(2014)075} {\bibfield  {journal} {\bibinfo  {journal}
  {JHEP}\ }\textbf {\bibinfo {volume} {1401}},\ \bibinfo {pages} {075}
  (\bibinfo {year} {2014})},\ \Eprint {http://arxiv.org/abs/1309.4462}
  {arXiv:1309.4462 [hep-ph]} \BibitemShut {NoStop}%
\bibitem [{\citenamefont {Agrawal}\ \emph
  {et~al.}(2014{\natexlab{b}})\citenamefont {Agrawal}, \citenamefont {Batell},
  \citenamefont {Hooper},\ and\ \citenamefont {Lin}}]{Agrawal:2014una}%
  \BibitemOpen
  \bibfield  {author} {\bibinfo {author} {\bibfnamefont {P.}~\bibnamefont
  {Agrawal}}, \bibinfo {author} {\bibfnamefont {B.}~\bibnamefont {Batell}},
  \bibinfo {author} {\bibfnamefont {D.}~\bibnamefont {Hooper}}, \ and\ \bibinfo
  {author} {\bibfnamefont {T.}~\bibnamefont {Lin}},\ }\href@noop {} {\
  (\bibinfo {year} {2014}{\natexlab{b}})},\ \Eprint
  {http://arxiv.org/abs/1404.1373} {arXiv:1404.1373 [hep-ph]} \BibitemShut
  {NoStop}%
\bibitem [{\citenamefont {Haisch}\ \emph {et~al.}(2014)\citenamefont {Haisch},
  \citenamefont {Hibbs},\ and\ \citenamefont {Re}}]{Haisch:2013fla}%
  \BibitemOpen
  \bibfield  {author} {\bibinfo {author} {\bibfnamefont {U.}~\bibnamefont
  {Haisch}}, \bibinfo {author} {\bibfnamefont {A.}~\bibnamefont {Hibbs}}, \
  and\ \bibinfo {author} {\bibfnamefont {E.}~\bibnamefont {Re}},\ }\href
  {\doibase 10.1103/PhysRevD.89.034009} {\bibfield  {journal} {\bibinfo
  {journal} {Phys.Rev.}\ }\textbf {\bibinfo {volume} {D89}},\ \bibinfo {pages}
  {034009} (\bibinfo {year} {2014})},\ \Eprint {http://arxiv.org/abs/1311.7131}
  {arXiv:1311.7131 [hep-ph]} \BibitemShut {NoStop}%
\bibitem [{\citenamefont {Kamenik}\ and\ \citenamefont
  {Zupan}(2011)}]{Kamenik:2011nb}%
  \BibitemOpen
  \bibfield  {author} {\bibinfo {author} {\bibfnamefont {J.~F.}\ \bibnamefont
  {Kamenik}}\ and\ \bibinfo {author} {\bibfnamefont {J.}~\bibnamefont
  {Zupan}},\ }\href {\doibase 10.1103/PhysRevD.84.111502} {\bibfield  {journal}
  {\bibinfo  {journal} {Phys.Rev.}\ }\textbf {\bibinfo {volume} {D84}},\
  \bibinfo {pages} {111502} (\bibinfo {year} {2011})},\ \Eprint
  {http://arxiv.org/abs/1107.0623} {arXiv:1107.0623 [hep-ph]} \BibitemShut
  {NoStop}%
\bibitem [{\citenamefont {Agrawal}\ \emph
  {et~al.}(2012{\natexlab{b}})\citenamefont {Agrawal}, \citenamefont
  {Blanchet}, \citenamefont {Chacko},\ and\ \citenamefont
  {Kilic}}]{Agrawal:2011ze}%
  \BibitemOpen
  \bibfield  {author} {\bibinfo {author} {\bibfnamefont {P.}~\bibnamefont
  {Agrawal}}, \bibinfo {author} {\bibfnamefont {S.}~\bibnamefont {Blanchet}},
  \bibinfo {author} {\bibfnamefont {Z.}~\bibnamefont {Chacko}}, \ and\ \bibinfo
  {author} {\bibfnamefont {C.}~\bibnamefont {Kilic}},\ }\href {\doibase
  10.1103/PhysRevD.86.055002} {\bibfield  {journal} {\bibinfo  {journal}
  {Phys.Rev.}\ }\textbf {\bibinfo {volume} {D86}},\ \bibinfo {pages} {055002}
  (\bibinfo {year} {2012}{\natexlab{b}})},\ \Eprint
  {http://arxiv.org/abs/1109.3516} {arXiv:1109.3516 [hep-ph]} \BibitemShut
  {NoStop}%
\bibitem [{\citenamefont {Merle}\ and\ \citenamefont
  {Niro}(2011)}]{Merle:2011yv}%
  \BibitemOpen
  \bibfield  {author} {\bibinfo {author} {\bibfnamefont {A.}~\bibnamefont
  {Merle}}\ and\ \bibinfo {author} {\bibfnamefont {V.}~\bibnamefont {Niro}},\
  }\href {\doibase 10.1088/1475-7516/2011/07/023} {\bibfield  {journal}
  {\bibinfo  {journal} {JCAP}\ }\textbf {\bibinfo {volume} {1107}},\ \bibinfo
  {pages} {023} (\bibinfo {year} {2011})},\ \Eprint
  {http://arxiv.org/abs/1105.5136} {arXiv:1105.5136 [hep-ph]} \BibitemShut
  {NoStop}%
\bibitem [{\citenamefont {Kajiyama}\ \emph {et~al.}(2012)\citenamefont
  {Kajiyama}, \citenamefont {Kannike},\ and\ \citenamefont
  {Raidal}}]{Kajiyama:2011gu}%
  \BibitemOpen
  \bibfield  {author} {\bibinfo {author} {\bibfnamefont {Y.}~\bibnamefont
  {Kajiyama}}, \bibinfo {author} {\bibfnamefont {K.}~\bibnamefont {Kannike}}, \
  and\ \bibinfo {author} {\bibfnamefont {M.}~\bibnamefont {Raidal}},\ }\href
  {\doibase 10.1103/PhysRevD.85.033008} {\bibfield  {journal} {\bibinfo
  {journal} {Phys.Rev.}\ }\textbf {\bibinfo {volume} {D85}},\ \bibinfo {pages}
  {033008} (\bibinfo {year} {2012})},\ \Eprint {http://arxiv.org/abs/1111.1270}
  {arXiv:1111.1270 [hep-ph]} \BibitemShut {NoStop}%
\bibitem [{\citenamefont {Lattanzi}\ \emph {et~al.}(2014)\citenamefont
  {Lattanzi}, \citenamefont {Lineros},\ and\ \citenamefont
  {Taoso}}]{Lattanzi:2014mia}%
  \BibitemOpen
  \bibfield  {author} {\bibinfo {author} {\bibfnamefont {M.}~\bibnamefont
  {Lattanzi}}, \bibinfo {author} {\bibfnamefont {R.~A.}\ \bibnamefont
  {Lineros}}, \ and\ \bibinfo {author} {\bibfnamefont {M.}~\bibnamefont
  {Taoso}},\ }\href@noop {} {\  (\bibinfo {year} {2014})},\ \Eprint
  {http://arxiv.org/abs/1406.0004} {arXiv:1406.0004 [hep-ph]} \BibitemShut
  {NoStop}%
\bibitem [{\citenamefont {Hirsch}\ \emph {et~al.}(2010)\citenamefont {Hirsch},
  \citenamefont {Morisi}, \citenamefont {Peinado},\ and\ \citenamefont
  {Valle}}]{Hirsch:2010ru}%
  \BibitemOpen
  \bibfield  {author} {\bibinfo {author} {\bibfnamefont {M.}~\bibnamefont
  {Hirsch}}, \bibinfo {author} {\bibfnamefont {S.}~\bibnamefont {Morisi}},
  \bibinfo {author} {\bibfnamefont {E.}~\bibnamefont {Peinado}}, \ and\
  \bibinfo {author} {\bibfnamefont {J.}~\bibnamefont {Valle}},\ }\href
  {\doibase 10.1103/PhysRevD.82.116003} {\bibfield  {journal} {\bibinfo
  {journal} {Phys.Rev.}\ }\textbf {\bibinfo {volume} {D82}},\ \bibinfo {pages}
  {116003} (\bibinfo {year} {2010})},\ \Eprint {http://arxiv.org/abs/1007.0871}
  {arXiv:1007.0871 [hep-ph]} \BibitemShut {NoStop}%
\bibitem [{\citenamefont {Boucenna}\ \emph {et~al.}(2012)\citenamefont
  {Boucenna}, \citenamefont {Morisi}, \citenamefont {Peinado}, \citenamefont
  {Shimizu},\ and\ \citenamefont {Valle}}]{Boucenna:2012qb}%
  \BibitemOpen
  \bibfield  {author} {\bibinfo {author} {\bibfnamefont {M.}~\bibnamefont
  {Boucenna}}, \bibinfo {author} {\bibfnamefont {S.}~\bibnamefont {Morisi}},
  \bibinfo {author} {\bibfnamefont {E.}~\bibnamefont {Peinado}}, \bibinfo
  {author} {\bibfnamefont {Y.}~\bibnamefont {Shimizu}}, \ and\ \bibinfo
  {author} {\bibfnamefont {J.}~\bibnamefont {Valle}},\ }\href {\doibase
  10.1103/PhysRevD.86.073008} {\bibfield  {journal} {\bibinfo  {journal}
  {Phys.Rev.}\ }\textbf {\bibinfo {volume} {D86}},\ \bibinfo {pages} {073008}
  (\bibinfo {year} {2012})},\ \Eprint {http://arxiv.org/abs/1204.4733}
  {arXiv:1204.4733 [hep-ph]} \BibitemShut {NoStop}%
\bibitem [{\citenamefont {Zhao}\ and\ \citenamefont
  {Zurek}(2014)}]{Zhao:2014nsa}%
  \BibitemOpen
  \bibfield  {author} {\bibinfo {author} {\bibfnamefont {Y.}~\bibnamefont
  {Zhao}}\ and\ \bibinfo {author} {\bibfnamefont {K.~M.}\ \bibnamefont
  {Zurek}},\ }\href {\doibase 10.1007/JHEP07(2014)017} {\bibfield  {journal}
  {\bibinfo  {journal} {JHEP}\ }\textbf {\bibinfo {volume} {1407}},\ \bibinfo
  {pages} {017} (\bibinfo {year} {2014})},\ \Eprint
  {http://arxiv.org/abs/1401.7664} {arXiv:1401.7664 [hep-ph]} \BibitemShut
  {NoStop}%
\bibitem [{\citenamefont {Feng}\ and\ \citenamefont
  {Kang}(2013)}]{Feng:2013vva}%
  \BibitemOpen
  \bibfield  {author} {\bibinfo {author} {\bibfnamefont {L.}~\bibnamefont
  {Feng}}\ and\ \bibinfo {author} {\bibfnamefont {Z.}~\bibnamefont {Kang}},\
  }\href {\doibase 10.1088/1475-7516/2013/10/008} {\bibfield  {journal}
  {\bibinfo  {journal} {JCAP}\ }\textbf {\bibinfo {volume} {1310}},\ \bibinfo
  {pages} {008} (\bibinfo {year} {2013})},\ \Eprint
  {http://arxiv.org/abs/1304.7492} {arXiv:1304.7492 [hep-ph]} \BibitemShut
  {NoStop}%
\bibitem [{\citenamefont {Masina}\ \emph {et~al.}(2012)\citenamefont {Masina},
  \citenamefont {Panci},\ and\ \citenamefont {Sannino}}]{Masina:2012hg}%
  \BibitemOpen
  \bibfield  {author} {\bibinfo {author} {\bibfnamefont {I.}~\bibnamefont
  {Masina}}, \bibinfo {author} {\bibfnamefont {P.}~\bibnamefont {Panci}}, \
  and\ \bibinfo {author} {\bibfnamefont {F.}~\bibnamefont {Sannino}},\ }\href
  {\doibase 10.1088/1475-7516/2012/12/002} {\bibfield  {journal} {\bibinfo
  {journal} {JCAP}\ }\textbf {\bibinfo {volume} {1212}},\ \bibinfo {pages}
  {002} (\bibinfo {year} {2012})},\ \Eprint {http://arxiv.org/abs/1205.5918}
  {arXiv:1205.5918 [astro-ph.CO]} \BibitemShut {NoStop}%
\bibitem [{\citenamefont {Masina}\ and\ \citenamefont
  {Sannino}(2011)}]{Masina:2011hu}%
  \BibitemOpen
  \bibfield  {author} {\bibinfo {author} {\bibfnamefont {I.}~\bibnamefont
  {Masina}}\ and\ \bibinfo {author} {\bibfnamefont {F.}~\bibnamefont
  {Sannino}},\ }\href {\doibase 10.1088/1475-7516/2011/09/021} {\bibfield
  {journal} {\bibinfo  {journal} {JCAP}\ }\textbf {\bibinfo {volume} {1109}},\
  \bibinfo {pages} {021} (\bibinfo {year} {2011})},\ \Eprint
  {http://arxiv.org/abs/1106.3353} {arXiv:1106.3353 [hep-ph]} \BibitemShut
  {NoStop}%
\bibitem [{\citenamefont {Farrar}\ and\ \citenamefont
  {Zaharijas}(2006)}]{Farrar:2005zd}%
  \BibitemOpen
  \bibfield  {author} {\bibinfo {author} {\bibfnamefont {G.~R.}\ \bibnamefont
  {Farrar}}\ and\ \bibinfo {author} {\bibfnamefont {G.}~\bibnamefont
  {Zaharijas}},\ }\href {\doibase 10.1103/PhysRevLett.96.041302} {\bibfield
  {journal} {\bibinfo  {journal} {Phys.Rev.Lett.}\ }\textbf {\bibinfo {volume}
  {96}},\ \bibinfo {pages} {041302} (\bibinfo {year} {2006})},\ \Eprint
  {http://arxiv.org/abs/hep-ph/0510079} {arXiv:hep-ph/0510079 [hep-ph]}
  \BibitemShut {NoStop}%
\bibitem [{\citenamefont {Kitano}\ and\ \citenamefont
  {Low}(2005)}]{Kitano:2004sv}%
  \BibitemOpen
  \bibfield  {author} {\bibinfo {author} {\bibfnamefont {R.}~\bibnamefont
  {Kitano}}\ and\ \bibinfo {author} {\bibfnamefont {I.}~\bibnamefont {Low}},\
  }\href {\doibase 10.1103/PhysRevD.71.023510} {\bibfield  {journal} {\bibinfo
  {journal} {Phys.Rev.}\ }\textbf {\bibinfo {volume} {D71}},\ \bibinfo {pages}
  {023510} (\bibinfo {year} {2005})},\ \Eprint
  {http://arxiv.org/abs/hep-ph/0411133} {arXiv:hep-ph/0411133 [hep-ph]}
  \BibitemShut {NoStop}%
\bibitem [{\citenamefont {Kitano}\ \emph {et~al.}(2008)\citenamefont {Kitano},
  \citenamefont {Murayama},\ and\ \citenamefont {Ratz}}]{Kitano:2008tk}%
  \BibitemOpen
  \bibfield  {author} {\bibinfo {author} {\bibfnamefont {R.}~\bibnamefont
  {Kitano}}, \bibinfo {author} {\bibfnamefont {H.}~\bibnamefont {Murayama}}, \
  and\ \bibinfo {author} {\bibfnamefont {M.}~\bibnamefont {Ratz}},\ }\href
  {\doibase 10.1016/j.physletb.2008.09.049} {\bibfield  {journal} {\bibinfo
  {journal} {Phys.Lett.}\ }\textbf {\bibinfo {volume} {B669}},\ \bibinfo
  {pages} {145} (\bibinfo {year} {2008})},\ \Eprint
  {http://arxiv.org/abs/0807.4313} {arXiv:0807.4313 [hep-ph]} \BibitemShut
  {NoStop}%
\bibitem [{\citenamefont {Fujii}\ and\ \citenamefont
  {Yanagida}(2002)}]{Fujii:2002aj}%
  \BibitemOpen
  \bibfield  {author} {\bibinfo {author} {\bibfnamefont {M.}~\bibnamefont
  {Fujii}}\ and\ \bibinfo {author} {\bibfnamefont {T.}~\bibnamefont
  {Yanagida}},\ }\href {\doibase 10.1016/S0370-2693(02)02341-9} {\bibfield
  {journal} {\bibinfo  {journal} {Phys.Lett.}\ }\textbf {\bibinfo {volume}
  {B542}},\ \bibinfo {pages} {80} (\bibinfo {year} {2002})},\ \Eprint
  {http://arxiv.org/abs/hep-ph/0206066} {arXiv:hep-ph/0206066 [hep-ph]}
  \BibitemShut {NoStop}%
\bibitem [{\citenamefont {Kaplan}(1992)}]{Kaplan:1991ah}%
  \BibitemOpen
  \bibfield  {author} {\bibinfo {author} {\bibfnamefont {D.~B.}\ \bibnamefont
  {Kaplan}},\ }\href {\doibase 10.1103/PhysRevLett.68.741} {\bibfield
  {journal} {\bibinfo  {journal} {Phys.Rev.Lett.}\ }\textbf {\bibinfo {volume}
  {68}},\ \bibinfo {pages} {741} (\bibinfo {year} {1992})}\BibitemShut
  {NoStop}%
\bibitem [{\citenamefont {Nussinov}(1985)}]{Nussinov:1985xr}%
  \BibitemOpen
  \bibfield  {author} {\bibinfo {author} {\bibfnamefont {S.}~\bibnamefont
  {Nussinov}},\ }\href {\doibase 10.1016/0370-2693(85)90689-6} {\bibfield
  {journal} {\bibinfo  {journal} {Phys.Lett.}\ }\textbf {\bibinfo {volume}
  {B165}},\ \bibinfo {pages} {55} (\bibinfo {year} {1985})}\BibitemShut
  {NoStop}%
\bibitem [{\citenamefont {Barr}(1991)}]{Barr:1991qn}%
  \BibitemOpen
  \bibfield  {author} {\bibinfo {author} {\bibfnamefont {S.~M.}\ \bibnamefont
  {Barr}},\ }\href {\doibase 10.1103/PhysRevD.44.3062} {\bibfield  {journal}
  {\bibinfo  {journal} {Phys.Rev.}\ }\textbf {\bibinfo {volume} {D44}},\
  \bibinfo {pages} {3062} (\bibinfo {year} {1991})}\BibitemShut {NoStop}%
\bibitem [{\citenamefont {Barr}\ \emph {et~al.}(1990)\citenamefont {Barr},
  \citenamefont {Chivukula},\ and\ \citenamefont {Farhi}}]{Barr:1990ca}%
  \BibitemOpen
  \bibfield  {author} {\bibinfo {author} {\bibfnamefont {S.~M.}\ \bibnamefont
  {Barr}}, \bibinfo {author} {\bibfnamefont {R.~S.}\ \bibnamefont {Chivukula}},
  \ and\ \bibinfo {author} {\bibfnamefont {E.}~\bibnamefont {Farhi}},\ }\href
  {\doibase 10.1016/0370-2693(90)91661-T} {\bibfield  {journal} {\bibinfo
  {journal} {Phys.Lett.}\ }\textbf {\bibinfo {volume} {B241}},\ \bibinfo
  {pages} {387} (\bibinfo {year} {1990})}\BibitemShut {NoStop}%
\bibitem [{\citenamefont {Gudnason}\ \emph {et~al.}(2006)\citenamefont
  {Gudnason}, \citenamefont {Kouvaris},\ and\ \citenamefont
  {Sannino}}]{Gudnason:2006ug}%
  \BibitemOpen
  \bibfield  {author} {\bibinfo {author} {\bibfnamefont {S.~B.}\ \bibnamefont
  {Gudnason}}, \bibinfo {author} {\bibfnamefont {C.}~\bibnamefont {Kouvaris}},
  \ and\ \bibinfo {author} {\bibfnamefont {F.}~\bibnamefont {Sannino}},\ }\href
  {\doibase 10.1103/PhysRevD.73.115003} {\bibfield  {journal} {\bibinfo
  {journal} {Phys.Rev.}\ }\textbf {\bibinfo {volume} {D73}},\ \bibinfo {pages}
  {115003} (\bibinfo {year} {2006})},\ \Eprint
  {http://arxiv.org/abs/hep-ph/0603014} {arXiv:hep-ph/0603014 [hep-ph]}
  \BibitemShut {NoStop}%
\bibitem [{\citenamefont {Beringer}\ \emph {et~al.}(2012)\citenamefont
  {Beringer} \emph {et~al.}}]{Beringer:1900zz}%
  \BibitemOpen
  \bibfield  {author} {\bibinfo {author} {\bibfnamefont {J.}~\bibnamefont
  {Beringer}} \emph {et~al.} (\bibinfo {collaboration} {Particle Data Group}),\
  }\href {\doibase 10.1103/PhysRevD.86.010001} {\bibfield  {journal} {\bibinfo
  {journal} {Phys.Rev.}\ }\textbf {\bibinfo {volume} {D86}},\ \bibinfo {pages}
  {010001} (\bibinfo {year} {2012})}\BibitemShut {NoStop}%
\bibitem [{\citenamefont {Ade}\ \emph {et~al.}(2013)\citenamefont {Ade} \emph
  {et~al.}}]{Ade2013a}%
  \BibitemOpen
  \bibfield  {author} {\bibinfo {author} {\bibfnamefont {P.}~\bibnamefont
  {Ade}} \emph {et~al.} (\bibinfo {collaboration} {Planck Collaboration}),\
  }\href@noop {} {\  (\bibinfo {year} {2013})},\ \Eprint
  {http://arxiv.org/abs/1303.5076} {arXiv:1303.5076 [astro-ph.CO]} \BibitemShut
  {NoStop}%
\bibitem [{\citenamefont {Falkowski}\ \emph {et~al.}(2011)\citenamefont
  {Falkowski}, \citenamefont {Ruderman},\ and\ \citenamefont
  {Volansky}}]{Falkowski:2011xh}%
  \BibitemOpen
  \bibfield  {author} {\bibinfo {author} {\bibfnamefont {A.}~\bibnamefont
  {Falkowski}}, \bibinfo {author} {\bibfnamefont {J.~T.}\ \bibnamefont
  {Ruderman}}, \ and\ \bibinfo {author} {\bibfnamefont {T.}~\bibnamefont
  {Volansky}},\ }\href {\doibase 10.1007/JHEP05(2011)106} {\bibfield  {journal}
  {\bibinfo  {journal} {JHEP}\ }\textbf {\bibinfo {volume} {1105}},\ \bibinfo
  {pages} {106} (\bibinfo {year} {2011})},\ \Eprint
  {http://arxiv.org/abs/1101.4936} {arXiv:1101.4936 [hep-ph]} \BibitemShut
  {NoStop}%
\bibitem [{\citenamefont {Cui}\ \emph {et~al.}(2011)\citenamefont {Cui},
  \citenamefont {Randall},\ and\ \citenamefont {Shuve}}]{Cui:2011qe}%
  \BibitemOpen
  \bibfield  {author} {\bibinfo {author} {\bibfnamefont {Y.}~\bibnamefont
  {Cui}}, \bibinfo {author} {\bibfnamefont {L.}~\bibnamefont {Randall}}, \ and\
  \bibinfo {author} {\bibfnamefont {B.}~\bibnamefont {Shuve}},\ }\href
  {\doibase 10.1007/JHEP08(2011)073} {\bibfield  {journal} {\bibinfo  {journal}
  {JHEP}\ }\textbf {\bibinfo {volume} {1108}},\ \bibinfo {pages} {073}
  (\bibinfo {year} {2011})},\ \Eprint {http://arxiv.org/abs/1106.4834}
  {arXiv:1106.4834 [hep-ph]} \BibitemShut {NoStop}%
\bibitem [{\citenamefont {Lin}\ \emph {et~al.}(2012)\citenamefont {Lin},
  \citenamefont {Yu},\ and\ \citenamefont {Zurek}}]{Lin:2011gj}%
  \BibitemOpen
  \bibfield  {author} {\bibinfo {author} {\bibfnamefont {T.}~\bibnamefont
  {Lin}}, \bibinfo {author} {\bibfnamefont {H.-B.}\ \bibnamefont {Yu}}, \ and\
  \bibinfo {author} {\bibfnamefont {K.~M.}\ \bibnamefont {Zurek}},\ }\href
  {\doibase 10.1103/PhysRevD.85.063503} {\bibfield  {journal} {\bibinfo
  {journal} {Phys.Rev.}\ }\textbf {\bibinfo {volume} {D85}},\ \bibinfo {pages}
  {063503} (\bibinfo {year} {2012})},\ \Eprint {http://arxiv.org/abs/1111.0293}
  {arXiv:1111.0293 [hep-ph]} \BibitemShut {NoStop}%
\bibitem [{\citenamefont {Cohen}\ \emph {et~al.}(2010)\citenamefont {Cohen},
  \citenamefont {Phalen}, \citenamefont {Pierce},\ and\ \citenamefont
  {Zurek}}]{Cohen:2010kn}%
  \BibitemOpen
  \bibfield  {author} {\bibinfo {author} {\bibfnamefont {T.}~\bibnamefont
  {Cohen}}, \bibinfo {author} {\bibfnamefont {D.~J.}\ \bibnamefont {Phalen}},
  \bibinfo {author} {\bibfnamefont {A.}~\bibnamefont {Pierce}}, \ and\ \bibinfo
  {author} {\bibfnamefont {K.~M.}\ \bibnamefont {Zurek}},\ }\href {\doibase
  10.1103/PhysRevD.82.056001} {\bibfield  {journal} {\bibinfo  {journal}
  {Phys.Rev.}\ }\textbf {\bibinfo {volume} {D82}},\ \bibinfo {pages} {056001}
  (\bibinfo {year} {2010})},\ \Eprint {http://arxiv.org/abs/1005.1655}
  {arXiv:1005.1655 [hep-ph]} \BibitemShut {NoStop}%
\bibitem [{\citenamefont {Blennow}\ \emph {et~al.}(2012)\citenamefont
  {Blennow}, \citenamefont {Fernandez-Martinez}, \citenamefont {Mena},
  \citenamefont {Redondo},\ and\ \citenamefont {Serra}}]{Blennow:2012de}%
  \BibitemOpen
  \bibfield  {author} {\bibinfo {author} {\bibfnamefont {M.}~\bibnamefont
  {Blennow}}, \bibinfo {author} {\bibfnamefont {E.}~\bibnamefont
  {Fernandez-Martinez}}, \bibinfo {author} {\bibfnamefont {O.}~\bibnamefont
  {Mena}}, \bibinfo {author} {\bibfnamefont {J.}~\bibnamefont {Redondo}}, \
  and\ \bibinfo {author} {\bibfnamefont {P.}~\bibnamefont {Serra}},\ }\href
  {\doibase 10.1088/1475-7516/2012/07/022} {\bibfield  {journal} {\bibinfo
  {journal} {JCAP}\ }\textbf {\bibinfo {volume} {1207}},\ \bibinfo {pages}
  {022} (\bibinfo {year} {2012})},\ \Eprint {http://arxiv.org/abs/1203.5803}
  {arXiv:1203.5803 [hep-ph]} \BibitemShut {NoStop}%
\bibitem [{\citenamefont {Chivukula}\ and\ \citenamefont
  {Georgi}(1987)}]{Chivukula:1987py}%
  \BibitemOpen
  \bibfield  {author} {\bibinfo {author} {\bibfnamefont {R.~S.}\ \bibnamefont
  {Chivukula}}\ and\ \bibinfo {author} {\bibfnamefont {H.}~\bibnamefont
  {Georgi}},\ }\href {\doibase 10.1016/0370-2693(87)90713-1} {\bibfield
  {journal} {\bibinfo  {journal} {Phys.Lett.}\ }\textbf {\bibinfo {volume}
  {B188}},\ \bibinfo {pages} {99} (\bibinfo {year} {1987})}\BibitemShut
  {NoStop}%
\bibitem [{\citenamefont {D'Ambrosio}\ \emph {et~al.}(2002)\citenamefont
  {D'Ambrosio}, \citenamefont {Giudice}, \citenamefont {Isidori},\ and\
  \citenamefont {Strumia}}]{D'Ambrosio:2002ex}%
  \BibitemOpen
  \bibfield  {author} {\bibinfo {author} {\bibfnamefont {G.}~\bibnamefont
  {D'Ambrosio}}, \bibinfo {author} {\bibfnamefont {G.}~\bibnamefont {Giudice}},
  \bibinfo {author} {\bibfnamefont {G.}~\bibnamefont {Isidori}}, \ and\
  \bibinfo {author} {\bibfnamefont {A.}~\bibnamefont {Strumia}},\ }\href
  {\doibase 10.1016/S0550-3213(02)00836-2} {\bibfield  {journal} {\bibinfo
  {journal} {Nucl.Phys.}\ }\textbf {\bibinfo {volume} {B645}},\ \bibinfo
  {pages} {155} (\bibinfo {year} {2002})},\ \Eprint
  {http://arxiv.org/abs/hep-ph/0207036} {arXiv:hep-ph/0207036 [hep-ph]}
  \BibitemShut {NoStop}%
\bibitem [{\citenamefont {Hall}\ and\ \citenamefont
  {Randall}(1990)}]{Hall:1990ac}%
  \BibitemOpen
  \bibfield  {author} {\bibinfo {author} {\bibfnamefont {L.}~\bibnamefont
  {Hall}}\ and\ \bibinfo {author} {\bibfnamefont {L.}~\bibnamefont {Randall}},\
  }\href {\doibase 10.1103/PhysRevLett.65.2939} {\bibfield  {journal} {\bibinfo
   {journal} {Phys.Rev.Lett.}\ }\textbf {\bibinfo {volume} {65}},\ \bibinfo
  {pages} {2939} (\bibinfo {year} {1990})}\BibitemShut {NoStop}%
\bibitem [{\citenamefont {Buras}(2003)}]{Buras:2003jf}%
  \BibitemOpen
  \bibfield  {author} {\bibinfo {author} {\bibfnamefont {A.~J.}\ \bibnamefont
  {Buras}},\ }\href@noop {} {\bibfield  {journal} {\bibinfo  {journal} {Acta
  Phys.Polon.}\ }\textbf {\bibinfo {volume} {B34}},\ \bibinfo {pages} {5615}
  (\bibinfo {year} {2003})},\ \Eprint {http://arxiv.org/abs/hep-ph/0310208}
  {arXiv:hep-ph/0310208 [hep-ph]} \BibitemShut {NoStop}%
\bibitem [{\citenamefont {Buras}\ \emph {et~al.}(2001)\citenamefont {Buras},
  \citenamefont {Gambino}, \citenamefont {Gorbahn}, \citenamefont {Jager},\
  and\ \citenamefont {Silvestrini}}]{Buras:2000dm}%
  \BibitemOpen
  \bibfield  {author} {\bibinfo {author} {\bibfnamefont {A.}~\bibnamefont
  {Buras}}, \bibinfo {author} {\bibfnamefont {P.}~\bibnamefont {Gambino}},
  \bibinfo {author} {\bibfnamefont {M.}~\bibnamefont {Gorbahn}}, \bibinfo
  {author} {\bibfnamefont {S.}~\bibnamefont {Jager}}, \ and\ \bibinfo {author}
  {\bibfnamefont {L.}~\bibnamefont {Silvestrini}},\ }\href {\doibase
  10.1016/S0370-2693(01)00061-2} {\bibfield  {journal} {\bibinfo  {journal}
  {Phys.Lett.}\ }\textbf {\bibinfo {volume} {B500}},\ \bibinfo {pages} {161}
  (\bibinfo {year} {2001})},\ \Eprint {http://arxiv.org/abs/hep-ph/0007085}
  {arXiv:hep-ph/0007085 [hep-ph]} \BibitemShut {NoStop}%
\bibitem [{\citenamefont {Froggatt}\ and\ \citenamefont
  {Nielsen}(1979)}]{Froggatt:1978nt}%
  \BibitemOpen
  \bibfield  {author} {\bibinfo {author} {\bibfnamefont {C.}~\bibnamefont
  {Froggatt}}\ and\ \bibinfo {author} {\bibfnamefont {H.~B.}\ \bibnamefont
  {Nielsen}},\ }\href {\doibase 10.1016/0550-3213(79)90316-X} {\bibfield
  {journal} {\bibinfo  {journal} {Nucl.Phys.}\ }\textbf {\bibinfo {volume}
  {B147}},\ \bibinfo {pages} {277} (\bibinfo {year} {1979})}\BibitemShut
  {NoStop}%
\bibitem [{\citenamefont {Leurer}\ \emph {et~al.}(1994)\citenamefont {Leurer},
  \citenamefont {Nir},\ and\ \citenamefont {Seiberg}}]{Leurer1994}%
  \BibitemOpen
  \bibfield  {author} {\bibinfo {author} {\bibfnamefont {M.}~\bibnamefont
  {Leurer}}, \bibinfo {author} {\bibfnamefont {Y.}~\bibnamefont {Nir}}, \ and\
  \bibinfo {author} {\bibfnamefont {N.}~\bibnamefont {Seiberg}},\ }\href
  {\doibase 10.1016/0550-3213(94)90074-4} {\bibfield  {journal} {\bibinfo
  {journal} {Nucl.Phys.}\ }\textbf {\bibinfo {volume} {B420}},\ \bibinfo
  {pages} {468} (\bibinfo {year} {1994})},\ \Eprint
  {http://arxiv.org/abs/hep-ph/9310320} {arXiv:hep-ph/9310320 [hep-ph]}
  \BibitemShut {NoStop}%
\bibitem [{\citenamefont {Ackermann}\ \emph
  {et~al.}(2012{\natexlab{a}})\citenamefont {Ackermann} \emph
  {et~al.}}]{Ackermann2012}%
  \BibitemOpen
  \bibfield  {author} {\bibinfo {author} {\bibfnamefont {M.}~\bibnamefont
  {Ackermann}} \emph {et~al.} (\bibinfo {collaboration} {LAT collaboration}),\
  }\href {\doibase 10.1088/0004-637X/761/2/91} {\bibfield  {journal} {\bibinfo
  {journal} {Astrophys.J.}\ }\textbf {\bibinfo {volume} {761}},\ \bibinfo
  {pages} {91} (\bibinfo {year} {2012}{\natexlab{a}})},\ \Eprint
  {http://arxiv.org/abs/1205.6474} {arXiv:1205.6474 [astro-ph.CO]} \BibitemShut
  {NoStop}%
\bibitem [{\citenamefont {Aguilar}\ \emph {et~al.}(2013)\citenamefont {Aguilar}
  \emph {et~al.}}]{Aguilar:2013qda}%
  \BibitemOpen
  \bibfield  {author} {\bibinfo {author} {\bibfnamefont {M.}~\bibnamefont
  {Aguilar}} \emph {et~al.} (\bibinfo {collaboration} {AMS Collaboration}),\
  }\href {\doibase 10.1103/PhysRevLett.110.141102} {\bibfield  {journal}
  {\bibinfo  {journal} {Phys.Rev.Lett.}\ }\textbf {\bibinfo {volume} {110}},\
  \bibinfo {pages} {141102} (\bibinfo {year} {2013})}\BibitemShut {NoStop}%
\bibitem [{\citenamefont {Ibarra}\ \emph {et~al.}(2013)\citenamefont {Ibarra},
  \citenamefont {Lamperstorfer},\ and\ \citenamefont {Silk}}]{Ibarra:2013zia}%
  \BibitemOpen
  \bibfield  {author} {\bibinfo {author} {\bibfnamefont {A.}~\bibnamefont
  {Ibarra}}, \bibinfo {author} {\bibfnamefont {A.~S.}\ \bibnamefont
  {Lamperstorfer}}, \ and\ \bibinfo {author} {\bibfnamefont {J.}~\bibnamefont
  {Silk}},\ }\href@noop {} {\  (\bibinfo {year} {2013})},\ \Eprint
  {http://arxiv.org/abs/1309.2570} {arXiv:1309.2570 [hep-ph]} \BibitemShut
  {NoStop}%
\bibitem [{\citenamefont {Desai}\ \emph {et~al.}(2004)\citenamefont {Desai}
  \emph {et~al.}}]{Desai2004}%
  \BibitemOpen
  \bibfield  {author} {\bibinfo {author} {\bibfnamefont {S.}~\bibnamefont
  {Desai}} \emph {et~al.} (\bibinfo {collaboration} {Super-Kamiokande
  Collaboration}),\ }\href {\doibase 10.1103/PhysRevD.70.083523,
  10.1103/PhysRevD.70.109901} {\bibfield  {journal} {\bibinfo  {journal}
  {Phys.Rev.}\ }\textbf {\bibinfo {volume} {D70}},\ \bibinfo {pages} {083523}
  (\bibinfo {year} {2004})},\ \Eprint {http://arxiv.org/abs/hep-ex/0404025}
  {arXiv:hep-ex/0404025 [hep-ex]} \BibitemShut {NoStop}%
\bibitem [{\citenamefont {Covi}\ \emph {et~al.}(2010)\citenamefont {Covi},
  \citenamefont {Grefe}, \citenamefont {Ibarra},\ and\ \citenamefont
  {Tran}}]{Covi2010}%
  \BibitemOpen
  \bibfield  {author} {\bibinfo {author} {\bibfnamefont {L.}~\bibnamefont
  {Covi}}, \bibinfo {author} {\bibfnamefont {M.}~\bibnamefont {Grefe}},
  \bibinfo {author} {\bibfnamefont {A.}~\bibnamefont {Ibarra}}, \ and\ \bibinfo
  {author} {\bibfnamefont {D.}~\bibnamefont {Tran}},\ }\href {\doibase
  10.1088/1475-7516/2010/04/017} {\bibfield  {journal} {\bibinfo  {journal}
  {JCAP}\ }\textbf {\bibinfo {volume} {1004}},\ \bibinfo {pages} {017}
  (\bibinfo {year} {2010})},\ \Eprint {http://arxiv.org/abs/0912.3521}
  {arXiv:0912.3521 [hep-ph]} \BibitemShut {NoStop}%
\bibitem [{1089675()}]{FermiLAT:2012aa}%
  \BibitemOpen
  \bibfield  {author} {1089675,\ }\href {\doibase 10.1088/0004-637X/750/1/3}
  {\bibfield  {journal} {\bibinfo  {journal} {Astrophys.J.}\ }\textbf {\bibinfo
  {volume} {750}},\ \bibinfo {pages} {3} (\bibinfo {year} {2012})},\ \Eprint
  {http://arxiv.org/abs/1202.4039} {arXiv:1202.4039 [astro-ph.HE]} \BibitemShut
  {NoStop}%
\bibitem [{\citenamefont {Ackermann}\ \emph
  {et~al.}(2012{\natexlab{b}})\citenamefont {Ackermann} \emph
  {et~al.}}]{Ackermann:2012qk}%
  \BibitemOpen
  \bibfield  {author} {\bibinfo {author} {\bibfnamefont {M.}~\bibnamefont
  {Ackermann}} \emph {et~al.} (\bibinfo {collaboration} {LAT Collaboration}),\
  }\href {\doibase 10.1103/PhysRevD.86.022002} {\bibfield  {journal} {\bibinfo
  {journal} {Phys.Rev.}\ }\textbf {\bibinfo {volume} {D86}},\ \bibinfo {pages}
  {022002} (\bibinfo {year} {2012}{\natexlab{b}})},\ \Eprint
  {http://arxiv.org/abs/1205.2739} {arXiv:1205.2739 [astro-ph.HE]} \BibitemShut
  {NoStop}%
\bibitem [{\citenamefont {Abdo}\ \emph {et~al.}(2010)\citenamefont {Abdo} \emph
  {et~al.}}]{Abdo:2010nz}%
  \BibitemOpen
  \bibfield  {author} {\bibinfo {author} {\bibfnamefont {A.}~\bibnamefont
  {Abdo}} \emph {et~al.} (\bibinfo {collaboration} {Fermi-LAT collaboration}),\
  }\href {\doibase 10.1103/PhysRevLett.104.101101} {\bibfield  {journal}
  {\bibinfo  {journal} {Phys.Rev.Lett.}\ }\textbf {\bibinfo {volume} {104}},\
  \bibinfo {pages} {101101} (\bibinfo {year} {2010})},\ \Eprint
  {http://arxiv.org/abs/1002.3603} {arXiv:1002.3603 [astro-ph.HE]} \BibitemShut
  {NoStop}%
\bibitem [{\citenamefont {Han}\ \emph {et~al.}(2010)\citenamefont {Han},
  \citenamefont {Lewis},\ and\ \citenamefont {McElmurry}}]{Han:2009ya}%
  \BibitemOpen
  \bibfield  {author} {\bibinfo {author} {\bibfnamefont {T.}~\bibnamefont
  {Han}}, \bibinfo {author} {\bibfnamefont {I.}~\bibnamefont {Lewis}}, \ and\
  \bibinfo {author} {\bibfnamefont {T.}~\bibnamefont {McElmurry}},\ }\href
  {\doibase 10.1007/JHEP01(2010)123} {\bibfield  {journal} {\bibinfo  {journal}
  {JHEP}\ }\textbf {\bibinfo {volume} {1001}},\ \bibinfo {pages} {123}
  (\bibinfo {year} {2010})},\ \Eprint {http://arxiv.org/abs/0909.2666}
  {arXiv:0909.2666 [hep-ph]} \BibitemShut {NoStop}%
\bibitem [{\citenamefont {Giudice}\ \emph {et~al.}(2011)\citenamefont
  {Giudice}, \citenamefont {Gripaios},\ and\ \citenamefont
  {Sundrum}}]{Giudice:2011ak}%
  \BibitemOpen
  \bibfield  {author} {\bibinfo {author} {\bibfnamefont {G.~F.}\ \bibnamefont
  {Giudice}}, \bibinfo {author} {\bibfnamefont {B.}~\bibnamefont {Gripaios}}, \
  and\ \bibinfo {author} {\bibfnamefont {R.}~\bibnamefont {Sundrum}},\ }\href
  {\doibase 10.1007/JHEP08(2011)055} {\bibfield  {journal} {\bibinfo  {journal}
  {JHEP}\ }\textbf {\bibinfo {volume} {1108}},\ \bibinfo {pages} {055}
  (\bibinfo {year} {2011})},\ \Eprint {http://arxiv.org/abs/1105.3161}
  {arXiv:1105.3161 [hep-ph]} \BibitemShut {NoStop}%
\bibitem [{\citenamefont {Grinstein}\ \emph {et~al.}(2011)\citenamefont
  {Grinstein}, \citenamefont {Kagan}, \citenamefont {Zupan},\ and\
  \citenamefont {Trott}}]{Grinstein:2011dz}%
  \BibitemOpen
  \bibfield  {author} {\bibinfo {author} {\bibfnamefont {B.}~\bibnamefont
  {Grinstein}}, \bibinfo {author} {\bibfnamefont {A.~L.}\ \bibnamefont
  {Kagan}}, \bibinfo {author} {\bibfnamefont {J.}~\bibnamefont {Zupan}}, \ and\
  \bibinfo {author} {\bibfnamefont {M.}~\bibnamefont {Trott}},\ }\href
  {\doibase 10.1007/JHEP10(2011)072} {\bibfield  {journal} {\bibinfo  {journal}
  {JHEP}\ }\textbf {\bibinfo {volume} {1110}},\ \bibinfo {pages} {072}
  (\bibinfo {year} {2011})},\ \Eprint {http://arxiv.org/abs/1108.4027}
  {arXiv:1108.4027 [hep-ph]} \BibitemShut {NoStop}%
\bibitem [{\citenamefont {Bona}\ \emph {et~al.}(2008)\citenamefont {Bona} \emph
  {et~al.}}]{Bona:2007vi}%
  \BibitemOpen
  \bibfield  {author} {\bibinfo {author} {\bibfnamefont {M.}~\bibnamefont
  {Bona}} \emph {et~al.} (\bibinfo {collaboration} {UTfit Collaboration}),\
  }\href {\doibase 10.1088/1126-6708/2008/03/049} {\bibfield  {journal}
  {\bibinfo  {journal} {JHEP}\ }\textbf {\bibinfo {volume} {0803}},\ \bibinfo
  {pages} {049} (\bibinfo {year} {2008})},\ \Eprint
  {http://arxiv.org/abs/0707.0636} {arXiv:0707.0636 [hep-ph]} \BibitemShut
  {NoStop}%
\bibitem [{\citenamefont {Charles}\ \emph {et~al.}(2014)\citenamefont
  {Charles}, \citenamefont {Descotes-Genon}, \citenamefont {Ligeti},
  \citenamefont {Monteil}, \citenamefont {Papucci} \emph
  {et~al.}}]{Charles:2013aka}%
  \BibitemOpen
  \bibfield  {author} {\bibinfo {author} {\bibfnamefont {J.}~\bibnamefont
  {Charles}}, \bibinfo {author} {\bibfnamefont {S.}~\bibnamefont
  {Descotes-Genon}}, \bibinfo {author} {\bibfnamefont {Z.}~\bibnamefont
  {Ligeti}}, \bibinfo {author} {\bibfnamefont {S.}~\bibnamefont {Monteil}},
  \bibinfo {author} {\bibfnamefont {M.}~\bibnamefont {Papucci}},  \emph
  {et~al.},\ }\href {\doibase 10.1103/PhysRevD.89.033016} {\bibfield  {journal}
  {\bibinfo  {journal} {Phys.Rev.}\ }\textbf {\bibinfo {volume} {D89}},\
  \bibinfo {pages} {033016} (\bibinfo {year} {2014})},\ \Eprint
  {http://arxiv.org/abs/1309.2293} {arXiv:1309.2293 [hep-ph]} \BibitemShut
  {NoStop}%
\bibitem [{\citenamefont {Chatrchyan}\ \emph {et~al.}(2013)\citenamefont
  {Chatrchyan} \emph {et~al.}}]{Chatrchyan:2013izb}%
  \BibitemOpen
  \bibfield  {author} {\bibinfo {author} {\bibfnamefont {S.}~\bibnamefont
  {Chatrchyan}} \emph {et~al.} (\bibinfo {collaboration} {CMS Collaboration}),\
  }\href {\doibase 10.1103/PhysRevLett.110.141802} {\bibfield  {journal}
  {\bibinfo  {journal} {Phys.Rev.Lett.}\ }\textbf {\bibinfo {volume} {110}},\
  \bibinfo {pages} {141802} (\bibinfo {year} {2013})},\ \Eprint
  {http://arxiv.org/abs/1302.0531} {arXiv:1302.0531 [hep-ex]} \BibitemShut
  {NoStop}%
\bibitem [{\citenamefont {Collaboration}(2014)}]{CMS:2014nia}%
  \BibitemOpen
  \bibfield  {author} {\bibinfo {author} {\bibfnamefont {C.}~\bibnamefont
  {Collaboration}} (\bibinfo {collaboration} {CMS Collaboration}),\ }\href@noop
  {} {\  (\bibinfo {year} {2014})}\BibitemShut {NoStop}%
\bibitem [{\citenamefont {ATLAS}\ and\ \citenamefont
  {Collaborations}(2013)}]{ATLASandCMSCollaborations:2013ofa}%
  \BibitemOpen
  \bibfield  {author} {\bibinfo {author} {\bibnamefont {ATLAS}}\ and\ \bibinfo
  {author} {\bibfnamefont {C.}~\bibnamefont {Collaborations}} (\bibinfo
  {collaboration} {ATLAS and CMS Collaborations}),\ }\href@noop {} {\
  (\bibinfo {year} {2013})}\BibitemShut {NoStop}%
\bibitem [{\citenamefont {Alwall}\ \emph {et~al.}(2014)\citenamefont {Alwall},
  \citenamefont {Frederix}, \citenamefont {Frixione}, \citenamefont {Hirschi},
  \citenamefont {Maltoni} \emph {et~al.}}]{Alwall:2014hca}%
  \BibitemOpen
  \bibfield  {author} {\bibinfo {author} {\bibfnamefont {J.}~\bibnamefont
  {Alwall}}, \bibinfo {author} {\bibfnamefont {R.}~\bibnamefont {Frederix}},
  \bibinfo {author} {\bibfnamefont {S.}~\bibnamefont {Frixione}}, \bibinfo
  {author} {\bibfnamefont {V.}~\bibnamefont {Hirschi}}, \bibinfo {author}
  {\bibfnamefont {F.}~\bibnamefont {Maltoni}},  \emph {et~al.},\ }\href
  {\doibase 10.1007/JHEP07(2014)079} {\bibfield  {journal} {\bibinfo  {journal}
  {JHEP}\ }\textbf {\bibinfo {volume} {1407}},\ \bibinfo {pages} {079}
  (\bibinfo {year} {2014})},\ \Eprint {http://arxiv.org/abs/1405.0301}
  {arXiv:1405.0301 [hep-ph]} \BibitemShut {NoStop}%
\bibitem [{\citenamefont {Alloul}\ \emph {et~al.}(2014)\citenamefont {Alloul},
  \citenamefont {Christensen}, \citenamefont {Degrande}, \citenamefont {Duhr},\
  and\ \citenamefont {Fuks}}]{Alloul:2013bka}%
  \BibitemOpen
  \bibfield  {author} {\bibinfo {author} {\bibfnamefont {A.}~\bibnamefont
  {Alloul}}, \bibinfo {author} {\bibfnamefont {N.~D.}\ \bibnamefont
  {Christensen}}, \bibinfo {author} {\bibfnamefont {C.}~\bibnamefont
  {Degrande}}, \bibinfo {author} {\bibfnamefont {C.}~\bibnamefont {Duhr}}, \
  and\ \bibinfo {author} {\bibfnamefont {B.}~\bibnamefont {Fuks}},\ }\href
  {\doibase 10.1016/j.cpc.2014.04.012} {\bibfield  {journal} {\bibinfo
  {journal} {Comput.Phys.Commun.}\ }\textbf {\bibinfo {volume} {185}},\
  \bibinfo {pages} {2250} (\bibinfo {year} {2014})},\ \Eprint
  {http://arxiv.org/abs/1310.1921} {arXiv:1310.1921 [hep-ph]} \BibitemShut
  {NoStop}%
\bibitem [{\citenamefont {Dreiner}\ \emph {et~al.}(2010)\citenamefont
  {Dreiner}, \citenamefont {Haber},\ and\ \citenamefont
  {Martin}}]{Dreiner:2008tw}%
  \BibitemOpen
  \bibfield  {author} {\bibinfo {author} {\bibfnamefont {H.~K.}\ \bibnamefont
  {Dreiner}}, \bibinfo {author} {\bibfnamefont {H.~E.}\ \bibnamefont {Haber}},
  \ and\ \bibinfo {author} {\bibfnamefont {S.~P.}\ \bibnamefont {Martin}},\
  }\href {\doibase 10.1016/j.physrep.2010.05.002} {\bibfield  {journal}
  {\bibinfo  {journal} {Phys.Rept.}\ }\textbf {\bibinfo {volume} {494}},\
  \bibinfo {pages} {1} (\bibinfo {year} {2010})},\ \Eprint
  {http://arxiv.org/abs/0812.1594} {arXiv:0812.1594 [hep-ph]} \BibitemShut
  {NoStop}%
\bibitem [{\citenamefont {D'Onofrio}\ \emph {et~al.}(2012)\citenamefont
  {D'Onofrio}, \citenamefont {Rummukainen},\ and\ \citenamefont
  {Tranberg}}]{D'Onofrio:2012ni}%
  \BibitemOpen
  \bibfield  {author} {\bibinfo {author} {\bibfnamefont {M.}~\bibnamefont
  {D'Onofrio}}, \bibinfo {author} {\bibfnamefont {K.}~\bibnamefont
  {Rummukainen}}, \ and\ \bibinfo {author} {\bibfnamefont {A.}~\bibnamefont
  {Tranberg}},\ }\href@noop {} {\bibfield  {journal} {\bibinfo  {journal}
  {PoS}\ }\textbf {\bibinfo {volume} {LATTICE2012}},\ \bibinfo {pages} {055}
  (\bibinfo {year} {2012})},\ \Eprint {http://arxiv.org/abs/1212.3206}
  {arXiv:1212.3206} \BibitemShut {NoStop}%
\bibitem [{\citenamefont {Harvey}\ and\ \citenamefont
  {Turner}(1990)}]{Harvey:1990qw}%
  \BibitemOpen
  \bibfield  {author} {\bibinfo {author} {\bibfnamefont {J.~A.}\ \bibnamefont
  {Harvey}}\ and\ \bibinfo {author} {\bibfnamefont {M.~S.}\ \bibnamefont
  {Turner}},\ }\href {\doibase 10.1103/PhysRevD.42.3344} {\bibfield  {journal}
  {\bibinfo  {journal} {Phys.Rev.}\ }\textbf {\bibinfo {volume} {D42}},\
  \bibinfo {pages} {3344} (\bibinfo {year} {1990})}\BibitemShut {NoStop}%
\bibitem [{\citenamefont {Feldstein}\ and\ \citenamefont
  {Fitzpatrick}(2010)}]{Feldstein2010}%
  \BibitemOpen
  \bibfield  {author} {\bibinfo {author} {\bibfnamefont {B.}~\bibnamefont
  {Feldstein}}\ and\ \bibinfo {author} {\bibfnamefont {A.~L.}\ \bibnamefont
  {Fitzpatrick}},\ }\href {\doibase 10.1088/1475-7516/2010/09/005} {\bibfield
  {journal} {\bibinfo  {journal} {JCAP}\ }\textbf {\bibinfo {volume} {1009}},\
  \bibinfo {pages} {005} (\bibinfo {year} {2010})},\ \Eprint
  {http://arxiv.org/abs/1003.5662} {arXiv:1003.5662 [hep-ph]} \BibitemShut
  {NoStop}%
\bibitem [{\citenamefont {Burnier}\ \emph {et~al.}(2006)\citenamefont
  {Burnier}, \citenamefont {Laine},\ and\ \citenamefont
  {Shaposhnikov}}]{Burnier:2005hp}%
  \BibitemOpen
  \bibfield  {author} {\bibinfo {author} {\bibfnamefont {Y.}~\bibnamefont
  {Burnier}}, \bibinfo {author} {\bibfnamefont {M.}~\bibnamefont {Laine}}, \
  and\ \bibinfo {author} {\bibfnamefont {M.}~\bibnamefont {Shaposhnikov}},\
  }\href {\doibase 10.1088/1475-7516/2006/02/007} {\bibfield  {journal}
  {\bibinfo  {journal} {JCAP}\ }\textbf {\bibinfo {volume} {0602}},\ \bibinfo
  {pages} {007} (\bibinfo {year} {2006})},\ \Eprint
  {http://arxiv.org/abs/hep-ph/0511246} {arXiv:hep-ph/0511246 [hep-ph]}
  \BibitemShut {NoStop}%
\end{thebibliography}

\end{document}